\documentclass[11pt]{article}
\usepackage{fullpage}
\usepackage{psfrag,epsf}
\usepackage{enumerate}
\usepackage{fancyhdr}
\usepackage[colorlinks,citecolor=blue,urlcolor=blue,linkcolor=blue,bookmarks=false]{hyperref}
\usepackage{amsfonts,epsfig,graphicx}
\usepackage{afterpage}
\usepackage{amsmath,amssymb,amsthm} 
\usepackage[T1]{fontenc} 
\usepackage{epsf} 
\usepackage{graphics} 
\usepackage[round]{natbib}
\setcitestyle{aysep={}} 
\usepackage{psfrag,xspace,subcaption}
\usepackage{color,etoolbox}
\usepackage{sectsty}
\usepackage[protrusion=true,expansion=true]{microtype}
\usepackage{booktabs}
\usepackage{multirow}
\usepackage{enumitem}
\usepackage[titletoc]{appendix}
\usepackage{titlesec}

\let\hat\widehat
\let\tilde\widetilde

\newcommand{\RNum}[1]{\uppercase\expandafter{\romannumeral #1\relax}}

\renewcommand{\P}{\mbox{$\mathbb{P}$}}
\newcommand{\E}{\mbox{$\mathbb{E}$}}

\let\hat\widehat
\let\tilde\widetilde

\usepackage[ruled]{algorithm2e}

\newcommand{\vecbn}{{\bf b_n}}
\newcommand{\veca}{{\bf a}}
\newcommand{\vecmu}{{\boldsymbol \mu}}
\newcommand{\riftspace}{{\sc Rift~}}
\newcommand{\riftspaces}{{\sc Rift}s~}
\newcommand{\riftnospace}{{\sc Rift}}
\newcommand{\riftnospaces}{{\sc Rift}s}
\newcommand{\mriftspace}{{\sc M-Rift~}}
\newcommand{\mriftnospace}{{\sc M-Rift}}

\newtheorem{theorem}{Theorem}
\newtheorem{newcounterthree}{New Counter}

\newtheorem{lemma}[newcounterthree]{Lemma}

\newcommand{\blind}{0}

\begin{document}

\def\spacingset#1{\renewcommand{\baselinestretch}%
{#1}\small\normalsize} \spacingset{1}


\if0\blind
{
  \title{\bf Gaussian Mixture Clustering Using Relative Tests of Fit}
  \author{Purvasha Chakravarti 
     \and Sivaraman Balakrishnan \and  Larry Wasserman \hspace{.2cm} \\
    Department of Statistics and Data Science, \\
    Carnegie Mellon University, \\
    Pittsburgh, PA, 15213. \\
    \href{mailto:pchakrav@stat.cmu.edu,siva@stat.cmu.edu,larry@stat.cmu.edu}{\{pchakrav, siva, larry\}@stat.cmu.edu} }
  \maketitle
} \fi

\if1\blind
{
  \bigskip
  \bigskip
  \bigskip
  \begin{center}
    {\LARGE\bf Gaussian Mixture Clustering Using Relative Tests of Fit}
\end{center}
  \medskip
} \fi

\begin{abstract}
\noindent We consider clustering based on significance tests for Gaussian Mixture Models (GMMs). Our starting point is the SigClust method developed by
Liu et al. (2008), which introduces a test based on 
the k-means objective (with k = 2) to decide whether the
data should be split into two clusters. When applied recursively, this test
yields a method for hierarchical clustering that is equipped with a significance guarantee. 
We study the limiting distribution and power of this approach in some examples 
and show that there are large regions of the parameter space where
the power is low. We then introduce a new test based on the idea of \emph{relative fit}.
Unlike prior work, we test for whether a mixture of Gaussians
provides a better fit relative to a single Gaussian,
without assuming that either model is correct.
The proposed test has a simple critical value and provides provable error control.
One version of our test provides exact, finite sample control of the type I error. 
We show how our tests can be used for hierarchical clustering as well as in a sequential manner for model selection. We conclude with an extensive simulation study and a 
cluster analysis of a gene expression dataset. 
\end{abstract}



\section{Introduction}
\label{section::Intro}
Gaussian mixture models (GMMs) are a commonly used tool
for clustering. A major challenge in using GMMs for clustering is
in adequately answering inferential questions regarding the 
number of mixture components or the number of clusters to use in data analysis. This task typically
requires hypothesis testing or model selection.
However, deriving rigorous tests for GMMs is notoriously difficult
since the usual regularity conditions fail for mixture models~\citep{ghosh1984asymptotic,dacunha1999testing,gassiat2002likelihood,
mclachlan2004finite,mclachlan2014number,
chen2017finite,gu2017testing}.

In this direction, \citet{liu2008statistical} 
proposed an approach called SigClust.
Their method starts by fitting a multivariate Gaussian to the data.
Then a significance test based on $k$-means clustering, with $k=2$, is applied.
If the test rejects, then the data is split into two clusters.
This test can be applied recursively leading to a top-down
hierarchical clustering \citep{kimes2017statistical}.
This approach roughly attempts to distinguish clusters which are actually present in the data from the natural sampling variability. The method
is appealing because it is simple and because,
as we further elaborate on in the sequel, it
provides certain rigorous error control guarantees.

In this paper we 
study the power of SigClust
and show that there are large regions of the parameter space where the method has poor power.
A natural way to fix this would be 
to use another statistic designed to distinguish
``a Gaussian'' versus ``a mixture of two Gaussians''
such as the generalized likelihood ratio test.
However, such an approach has two problems:
first, as mentioned above, mixture models are irregular and the limiting distribution
of the likelihood ratio test
(and other familiar tests) is intractable.
Second, such tests assume that one of the models
(Gaussian or mixture of Gaussians) is correct.
Instead from a practical standpoint, 
for the purposes of clustering, we only regard these models as approximations.

So we consider a different approach.
We test
whether one model is closer to the true distribution than the other
without assuming either model is true.
We call this a test of {\em relative fit}.
Our test is based on data splitting.
Half the data are used to fit the models and the other half are used
to construct the test. The result is a test
with a simple limiting distribution which makes it easy to determine
an appropriate cutoff for it.
In fact, we provide several versions of the test.
One version provides exact type I error control
without requiring any asymptotic approximations.

Following \citet{kimes2017statistical}, we also 
apply the test recursively to obtain a hierarchical clustering of the data
with significance guarantees.
We develop a bottom-up version of mixture clustering
which can be regarded as a linkage clustering procedure where
we first over-fit a mixture and subsequently combine elements of the mixture.
We also construct a sequential, non-hierarchical version, of the approach. 
We call our procedure \riftspace (Relative Information Fit Test).

Throughout this paper we assume that the dimension $d$ is fixed and the sample size $n$ is increasing.
In contrast, \citet{liu2008statistical}
and \citet{kimes2017statistical}
focus on the large $d$, fixed $n$ case which requires
dealing with challenging issues such as estimating the covariance matrix
in high dimensions (see also \cite{vogt2017clustering}).
However, because of the challenges of high dimensional estimation,
these prior works only establish results about power in very specific cases.
In contrast, we provide a more detailed understanding
of the power in the fixed-$d$ case.

\subsection{Related Work}
Estimating the number of clusters has been 
approached in many ways
\citep{Bock1985,milligan1985examination,mclachlan2004finite}. A common approach is to find the optimal number of
clusters by optimizing a criterion function, examples of which are the
Hartigan index \citep{hartigan1975clustering}, the silhouette
statistic \citep{rousseeuw1987silhouettes} or the gap statistic \citep{tibshirani2001estimating}.

Another approach to estimating the number of clusters
is to assess the statistical significance
of the clusters. \citet{mcshane2002methods} proposed a method to calculate
p-values by assuming that the cluster structure
lies in the first three principal components of the
data. \citet{tibshirani2005cluster} use resampling techniques to
quantify the prediction strength of different clusters and
\citet{suzuki2006pvclust} assess the significance of hierarchical
clustering using bootstrapping procedures. More recently,
\citet{maitra2012bootstrapping} proposed a distribution-free bootstrap
procedure which assumes that the data in a cluster is sampled from a
spherically symmetric, compact and unimodal distribution.
\citet{engelman1969percentage}
considered the maximal F-ratio that compares between group dispersions
with within group dispersions.
\citet{lee1979multivariate} proposed a subsequent multivariate
version and a robust version was recently proposed by
\citet{garcia2009robust}. Another example is a statistical test
proposed by \citet{vogt2017clustering}.
They develop a fairly general significance test
but it relies on assuming that the number of covariates tends to infinity
and that the clusters are, in a certain sense, well-separated
(i.e. can be consistently estimated as the number of features increases).

Alternatively, 
and closer to our approach,
Gaussian mixture models can be used for cluster
analysis. See for instance, the works
\cite{fraley2002model,mclachlan2004finite,mclachlan2014number} 
for overviews. 
There is much prior
work for testing the order
of a Gaussian mixture. For example, the works
\cite{ghosh1984asymptotic,hartigan1985failure} used the likelihood ratio test with the
null hypothesis that the order is one. \citet{hartigan1985failure}
explored the impact of nonregularity of the mixture models and
\citet{ghosh1984asymptotic} used a separation condition in order to
find the asymptotic distribution of the likelihood ratio test
statistic.

Since finite normal mixture models are irregular and the limiting
distribution of the likelihood ratio test statistic is difficult to derive,
deriving a general theory for testing the order of a mixture is
hard. Instead most of the algorithms test for homogeneity in the
data. The works \cite{charnigo2004testing,liu2004asymptotics,chen2009hypothesis} 
among others, are examples of this
approach. More recently, \citet{li2010testing} and
\citet{chen2012inference} constructed a new likelihood-based
expectation-maximization (EM) test for the order of finite mixture
models that uses a penalty function on the variance to obtain a
bounded penalized likelihood.
Further developments can be found in
the works \cite{dacunha1999testing,gassiat2002likelihood,
chen2017finite,gu2017testing}.
Our approach differs in three ways:
we use a test that avoids the irregularities,
it avoids assuming that the mixture model is correct
and it is valid for multivariate mixtures.
We only treat the mixture model as an approximate working model.

\citet{liu2008statistical} proposed a Monte Carlo based algorithm (SigClust) that
defines a cluster as data generated from a single multivariate
Gaussian distribution. 
The distribution of the test statistic under the null hypothesis
SigClust depends on the eigenvalues of the null covariance
matrix. \citet{huang2015statistical} proposed a soft-thresholding
method that provides an estimate of these eigenvalues, and this
soft-thresholding method leads to a
modified version of SigClust that is better suited to high-dimensional problems.

\subsection{Outline}
In Section \ref{section::sigclust} we review the SigClust
procedure and we derive its power in some cases. We show that SigClust can have poor 
power against certain alternatives. 
This section also
has results on the geometric properties
of $k$-means clustering in a special case, which is a prelude to finding the power.
In Section \ref{section::rift} we describe our new procedure.
We also describe several other tests that are used for comparison.
In Section \ref{section::hier} we show how to use our new tests in a hierarchical framework.
Section \ref{section::sequential} describes a sequential testing version of our approach 
which can be used for model selection for the GMM. 
We consider some examples in Section \ref{section::examples}, and analyze a gene expression
dataset in Section~\ref{sec:gene}.
Finally, concluding remarks are in Section \ref{section::conclusion}.
We defer the technical details of most proofs to the Appendix. 

\subsection{Notation}
Throughout this paper 
we use $\|\cdot\|$ to 
denote the Euclidean norm, i.e. for \mbox{$\mathbf{x} \in \mathbb{R}^d$, $\|\mathbf{x}\| := \sqrt{ \sum_{i=1}^d x_i^2}$.}
We use the symbols $\overset{p}{\to}$ and $\rightsquigarrow$ to denote the 
standard stochastic convergence concepts of convergence in probability and in distribution respectively.

\section{Setup and the SigClust Procedure}
\label{section::sigclust}
We let $\{X_1, X_2, \ldots, X_n\} \sim \mathbb{P}$ be i.i.d. observations from some
distribution with probability measure $\mathbb{P}$ on $\mathbb{R}^d$. 
We recall the $k$-means clustering algorithm which chooses
cluster centers $\vecbn = (b_{n1},\ldots, b_{nk}) \in \mathbb{R}^{d \times k}$ 
to minimize the within-cluster sum of squares,
\begin{equation}
W_n(\veca) = \frac{1}{n} \sum_{i = 1}^n \min_{1 \leq j \leq k} \|X_i - a_j\|^2
\end{equation}
as a function of $\veca  = (a_1, \ldots, a_k) \in \mathbb{R}^{d \times k}$.
For each center
$a_j$, we can also associate a convex polyhedron $A_j$ which contains
all points in $\mathbb{R}^d$ closer to $a_j$ than to any other
center. 
The sets
$\{A_1,\ldots, A_k\}$
are the Voronoi tessellation of $\mathbb{R}^d$.
The tessellation defines the clustering.
We also define,
\begin{align*}
W(\veca) = \mathbb{E}[W_n(\veca )],
\end{align*}
and we let
$\vecmu = (\mu_1, \ldots, \mu_k) \in \mathbb{R}^{d \times k}$ denote the minimizer of
$W(\veca)$. When the minimizers are not unique we let $\vecbn$ and $\vecmu$ denote arbitrary minimizers
of $W_n(\veca)$ and $W(\veca)$ respectively.

\subsection{SigClust}

In this section, we describe the SigClust procedure. 
\citet{liu2008statistical} define a cluster as a population sampled
from a single Gaussian distribution.
To capture the non-Gaussianity due to the presence of multiple clusters,
\citet{liu2008statistical} consider a
test statistic based on
$k$-means, with $k=2$.
Specifically,
define $T_n$
to be the ratio between the within-class sum of squares and the total sum of
squares as,
\begin{align*}
T_n = T_n(\vecbn) = \frac{W_n(\vecbn)}{\frac{1}{n}\sum_{i = 1}^n 
\|X_i - \overline{X}\|^2} = 
\frac{\sum_{i = 1}^n \min_{1 \leq j \leq 2} \|X_i - b_{nj}\|^2}{\sum_{i = 1}^n \|X_i - \overline{X}\|^2},
\end{align*}
where $\vecbn = (b_{n1}, b_{n2})$ is the vector of optimal
centers chosen by the $2$-means clustering algorithm and
$\overline{X}$ is the sample mean of the data. 
We note in passing that in their extension of this method to
hierarchical clustering, \citet{kimes2017statistical} 
also consider other statistics that arise in hierarchical clustering.

Roughly, we reject the null for small values of this statistic. 
In order to estimate the p-value we use a version of the parametric bootstrap.
The (estimated) p-value is an estimate of
$\mathbb{P}_{\hat\mu,\hat\Sigma}(T_n^* < T_n)$ where
$T_n^*$ is computed on the bootstrap samples from $\mathbb{P}_{\hat\mu,\hat\Sigma}$,  
where $\mathbb{P}_{\hat\mu,\hat\Sigma}=N(\hat\mu,\hat\Sigma)$ and
where $\hat\mu = \overline{X}$ and
$\hat\Sigma$ is the sample covariance matrix.
We note that in the high dimensional case, as discussed earlier, \citet{liu2008statistical} use a regularized estimator of $\Sigma$. 

\subsection{Limiting Distribution of SigClust under the Null}
In order to analytically understand the SigClust procedure and to develop results regarding its power
we first find the limiting distribution
of the test statistic under the null in a simplified setup.

We focus in this and subsequent sections on the case when under the null hypothesis, we obtain samples
from $\{X_1,\ldots,X_n\} \sim N(0, \Sigma)$ where $\Sigma$ is a diagonal matrix. 
We assume that the two leading eigenvalues are distinct which ensures that, under the null, the k-means objective at the population-level has a unique optimal solution whose optimal value is tractable to analyze in closed-form. 
For notational convenience, we will assume that, $\sigma_1^2 > \sigma_2^2 \geq \sigma_3^3 \ldots \geq \sigma_d^2 > 0.$

Our results extend in a straightforward way to the general non-spherical, axis-aligned case with minor modifications. These results in turn are easily generalized to the 
 non-spherical, not necessarily axis-aligned, case by noting the invariance of the test statistic to orthonormal rotations under the null.
The spherical case is more challenging since the population optimal k-means solution is not unique and the limiting distribution is more complicated. To illustrate some of the difficulties, we derive the limiting distribution of the SigClust
 statistic, under the null, for the two-dimensional case in Appendix~\ref{sec:spherical}, but do not consider the power of the test in that setting.

Recall,
that $\vecmu$ denotes the (unique) population optimal $k$-means solution, and we use $\{A_1,A_2\}$ to denote
the corresponding Voronoi partition. 
Our results build on the following result from \citet{Pollard1982} and \citet{Bock1985}:
\begin{lemma}[Corollary 6.2 of \cite{Bock1985}]
\label{lem:bock}
The minimum within cluster sum of squares $W_n(\vecbn)$ has an asymptotically normal distribution given by,
\begin{align*}
\sqrt{n}(W_n(\vecbn) - W(\vecmu))  \rightsquigarrow N(0, \tau^2), \ \ \text{as} \ \ n \to \infty,
\end{align*}
where 
\[
\tau^2 := 
\sum_{i = 1}^2 \mathbb{P}(A_i) \mathbb{E}\left[ \|X - \mu_i\|^4 | X \in A_i \right] - \left[W(\vecmu) \right]^2.
\]
\end{lemma}
To analyze the power of the SigClust procedure, and to better understand its limiting distribution, we need to calculate 
$W(\vecmu), \tau^2$ and the mass of the Voronoi cells. It is easy to verify that under the null the probability of each of the Voronoi cells corresponding to $\vecmu$ is 1/2. In Appendix~\ref{app:claims}, we establish the following claims:
\begin{align}
\label{eqn:muopt}
W(\vecmu) &= \sum_{i = 1}^d \sigma_i^2 - \frac{2 \sigma_1^2}{\pi}\\
\label{eqn:tausq}
\tau^2 &= 2 \sum_{i = 1}^d \sigma_i^4 - \frac{16 \sigma_1^4}{\pi^2}.
\end{align}
As a consequence of these calculations, we obtain the limiting distribution of the SigClust statistic under the null:

\begin{theorem}
\label{thm:null}
For $W(\vecmu)$ and $\tau^2$ defined in~\eqref{eqn:muopt} and~\eqref{eqn:tausq} we have that,
\begin{align*}
\sqrt{n} \left(T_n(\vecbn) - \frac{W(\vecmu)}{\sum_{i = 1}^d \sigma_i^2}\right) \rightsquigarrow 
N\left(0, \left[\frac{\tau}{\sum_{i = 1}^d \sigma_i^2}\right]^2\right), \ \ \text{as} \ \ n \to \infty.
\end{align*}
\end{theorem}

\noindent {\bf Remark: } 
Leveraging this result, we are able to
characterize the rejection region of the test and in Theorem \ref{thm::power1}
we analyze its power.
The proof of 
Theorem~\ref{thm:null} is quite long and technical.
Most of the work is done in
the Appendix.
Here is a brief proof that leverages
Lemma~\ref{lem:bock} which contains most of the technical details.

\begin{proof}
From Lemma~\ref{lem:bock} we have that,
\[\sqrt{n}(W_n(\vecbn) - W(\vecmu))  \rightsquigarrow N(0, \tau^2), \ \ \text{as} \ \ n \to \infty.\]
Furthermore by the Weak Law of Large Numbers we have that,
\[S^2 = \frac{1}{n}\sum_{i = 1}^n \|X_i - \overline{X}\|^2 \overset{p}{\to} \sum_{i = 1}^d \sigma_i^2. \]
Putting these together yields the desired claim. 
\end{proof}

\subsection{Geometry of $k$-means under the alternative}
\label{sec::alternate}
Our goal is to find special cases where we are able to explicitly 
calculate SigClust's power and understand cases in which it
has high power and cases where it has low power. 
In order to find the power, we first need to understand the 
behaviour of $2$-means clustering under the alternative. 
In particular, we need to understand what the optimal split is and 
what the optimal within sum of squares is, if the data was indeed 
generated from the alternative.

We focus on the case when the data, under the alternative, is generated from a 
mixture of two Gaussian
distributions 
of the form
\begin{align}
\label{eqn:mix}
\{X_1, \ldots, X_n\} \sim \frac{1}{2} N(-\theta_1, D) + \frac{1}{2} N(\theta_1, D),
\end{align} 
where $\theta_1 = (a/2, 0, \ldots, 0) \in \mathbb{R}^d$, $a$ is a non-zero constant and 
$D$ is a diagonal matrix,
\begin{align*}
D = \left[ \begin{matrix} \sigma_1^2 & 0 & 0 & \ldots & 0 \\
0 & \sigma_2^2 & 0 & \ldots & 0 \\
& & \vdots \\
0 & 0 & 0 & \ldots & \sigma_d^2
\end{matrix} \right].
\end{align*}
In this section, we will consider cases where $\sigma_1^2, \sigma_2^2 > \sigma_3^2 \geq \ldots \geq \sigma_d^2,$ allowing in some cases $\sigma_2^2$ to be larger than $\sigma_1^2$.
We treat the case when $a > 0$ and $0 < \sigma_d^2 \leq \ldots \leq \sigma_2^2, \sigma_1^2 < \infty,$
are fixed (do not vary with the sample-size).

For technical reasons, we make a small modification to $2-$means clustering. We consider $2-$means clustering with symmertric centers. That is, we consider $\mathbf{b_n}^{(0)} $ that minimizes the within-cluster sum of squares,
\begin{equation}
\label{eqn::symmsampleWSS}
W_n^{(0)}(t) := W_n(t, -t) = \frac{1}{n}\sum_{i = 1}^n \min \{||X_i -  t||^2, ||X_i +  t||^2\},
\end{equation}
as a function of $t \in \mathbb{R}^d.$ 

We also introduce notation for the optimal split by considering a symmetric population version of the $2$-means clustering for the following theorems and lemmas. We define the following terms to be used in the lemmas. Let
\[\mu^*= \left( \begin{array}{c}
\mu_1^*\\
\mu_2^*
\end{array}\right) = \left( \begin{array}{c}
\mu_1^*\\
-\mu_1^*
\end{array}\right),\] 
where $\mu_1^*$ and $-\mu_1^*$ denote the optimum cluster centers that minimize the within sum of squares when symmetric $2$-means clustering is performed on the data. The corresponding minimum within sum of squares is denoted by $W(\mu^*)$. That is,  
\[ W^{(0)}(\mu_1^*) := W(\mu^*) = \inf_{t \in \mathbb{R}^d} E\left[\min \{ || X - t ||^2,  || X + t ||^2 \right]  = E\left[\min \{ || X - \mu_1^* ||^2,  || X + \mu_1^* ||^2\} \right].\]

We conjecture that this symmetric assumption has no practical effect on SigClust, since the samples are drawn from a symmetric distribution and in practice the optimum $2$-means cluster centers are close to being symmetric. Moreover, to consider the limiting distribution of $W_n(\mathbf{b_n})$, given by Theorem 6.4 (b) of \citet[pp.~101]{Bock1985}, we need the population optimal centers to be unique. This is guaranteed only if the population optimal centers are symmetric about the origin, since if $(\mu_1^*, \mu_2^*)$ minimizes the population within sum of squares, then due to the symmetry of the distribution, $(-\mu_1^*, -\mu_2^*)$ also minimizes the population within sum of squares. Therefore for the minimizer to be unique, $\mu_2^* = - \mu_1^*$.

Therefore we state a result analogous to Theorem 6.4 (b) of \citet[pp.~101]{Bock1985} for symmetric $2$-means clustering for our population as follows:
\begin{theorem}
Let the data be generated from $\frac{1}{2} N(-\theta_1, D) +
\frac{1}{2} N(\theta_1, D)$, as defined above, and $\mathbf{b_n}^{(0)}, \mu_1^*, \mu^*, W_n^{(0)}(t)$ and $W^{(0)}(\mu_1^*)$ are as defined above. Suppose
\begin{itemize}
\item[(i)] the vector $ \mu_1^*$ that minimizes $W^{(0)}(\mu_1^*)$ is unique upto relabeling of its coordinates;
\item[(ii)] the matrix $G$ is positive definite, where $G$ as defined in \citet{Pollard1982} (as $\Gamma$) is a matrix made up of $d \times
d$ matrices of the form,
\begin{equation}
\label{eqn::Gamma}
G_{ij} = \left\lbrace \begin{array}{c c}
2\mathbb{P}(A_i) \mathbf{I}_d - 2  r_{i j}^{-1} \int_{M_{i j}} f(x) (x - \mu_i^*) (x - \mu_i^*)^T \ d\sigma(x) & \mbox{ for } \ \ i = j\\
-2 r_{ij}^{-1}  \int_{M_{i j}} f(x) (x - \mu_i^*) (x - \mu_j^*)^T \ d\sigma(x) & \mbox{ for } \ \ i \neq j,
\end{array} \right.
\end{equation}
for $i, j \in \{ 1, 2 \}$ where $r_{ij} = \| \mu_i^* - \mu_j^*\|$, $f(\cdot)$ is the corresponding density function, $\sigma(\cdot)$ is the $(d - 1)$ dimensional Lebesgue measure, $A_1$ is the convex polyhedron that contains all points in $\mathbb{R}^d$ that are closer to $\mu_1^*$ compared to $-\mu_1^*$ and $A_2$ is vice-versa and $M_{i j}$ denotes the face common to $A_i$ and $A_j$, and $\mathbf{I}_d$ denotes the $d \times d$ identity matrix.
\end{itemize} 
Then as $n \to \infty$, 
\[\sqrt{n}( W_n^{(0)}(\mathbf{b_n}^{(0)}) - W(\mu^*)) \rightsquigarrow N(0, \tau^{*2}),\] 
where 
\[ 
\tau^{*2} = \sum_{i = 1}^2 P(A_i) E[||X- E[X| X \in A_i]||^4 | X \in A_i] - [W(\mu^*)]^2.
\]
\label{thm::newPollard}
\end{theorem}

Since $\mu_1^*$ and $-\mu_1^*$ denote the optimum cluster centers, the corresponding optimal separating hyperplane passes through the origin. We denote the corresponding optimal separating hyperplane by 
\[\mathcal{H}(b^*) = \left\lbrace y \in \mathbb{R}^d : b^{*T} y = 0  \right\rbrace, \ \text{ where } \ \sum_{i = 1}^d b_i^{*2} = 1.\]
Then the corresponding within sum of squares can be written as:
\begin{align*}
W(b^*) := W(\mu^*) 
&= \inf_{b \in \mathbb{R}^d}  \left\lbrace P(b^T X > 0) E[||X - E[X| b^T X > 0]||^2 | b^T X > 0]  \right.\\
&\left. \ \ \ \ \ \ \ \ +  P(b^T X < 0) E[||X- E[X| b^T X < 0]||^2 | b^T X < 0] \right\rbrace\\
&= \inf_{b \in \mathbb{R}^d} E[||X- E[X| b^T X > 0]||^2 | b^T X > 0], \ \ \ \ \ (\text{Since, } - X \overset{d}{=} X)\\
&= E[||X- E[X| b^{*T} X > 0]||^2 | b^{*T} X > 0].
\end{align*}

The following theorem gives the optimal separating hyperplane and the optimal within sum of squares under the alternative.

\begin{theorem}
For data generated from $\frac{1}{2} N(-\theta_1, D) +
\frac{1}{2} N(\theta_1, D)$, where $\theta_1 = (a/2, 0, \ldots, 0)
\in \mathbb{R}^d$, $a > 0$ is fixed and $D$ is a diagonal
matrix with elements $D_{jj} = \sigma_j^2$, such that $\sigma_1^2, \sigma_2^2 > \sigma_3^2 \geq \ldots \geq \sigma_d^2$ are fixed. 
\begin{enumerate}
\item When
\begin{align}
\label{eqn::firstdimsplit}
\sigma_2^2 < \sigma_1^2 + \frac{a^2}{4}, 
\end{align}
the unique optimal separating hyperplane which gives the minimum within sum of squares $W(b^*)$ is given by $\mathcal{H}(b) = \{y \in \mathbb{R}^d: y_1 = 0\}$, that is, the unique optimal $b^*$ is such that $b_1^* = 1$ and $b_i^* = 0$ for every $i \neq 1$. The corresponding optimal within sum of squares is given by
\begin{equation}
\label{eqn:w1mu}
W(\mu^*) = W(b^*) = \sum_{j = 1}^d \sigma_j^2 +  \frac{a^2}{4}  - \left( \sqrt{\frac{2}{\pi}} \ \sigma_1 \ e^{-\frac{a^2}{8 \sigma_1^2}} + \frac{a}{2} \ P \left(|Z| < \frac{a}{2 \sigma_1}\right) \right)^{2}.
\end{equation}
\item When 
\begin{align}
\label{eqn::seconddimsplit}
\sigma_2^2 > \max\left\lbrace \frac{2 \sigma_1^4 +  \frac{a^4}{16} + \frac{a^2}{2}\sqrt{\sigma_1^4 + \frac{a^4}{64}}}{2 \sigma_1^2}, \frac{\pi}{2}\left( \sqrt{\frac{2}{\pi}} \ \sigma_1 \ e^{-\frac{a^2}{8 \ \sigma_1^2}} + \frac{a}{2} \ P \left(|Z| < \frac{a}{2 \ \sigma_1}\right) \right)^{2} \right\rbrace, 
\end{align}
the unique optimal separating hyperplane which gives the minimum within sum of squares $W(b^*)$  is given by $\mathcal{H}(b) = \{y \in \mathbb{R}^d: y_2 = 0\}$, that is, the unique optimal $b^*$ is such that $b_2^* = 1$ and $b_i^* = 0$ for every $i \neq 2$. The corresponding optimal within sum of squares is given by
\begin{equation}
\label{eqn:w1mu2}
W(\mu^*) = W(b^*) = \sum_{j = 1}^d \sigma_j^2 +  \frac{a^2}{4}  - \frac{2}{\pi} \sigma_2^2.
\end{equation}
\end{enumerate}
\label{thm::optimalsplit}
\end{theorem}

In simpler words, the theorem implies that when the condition in~\eqref{eqn::firstdimsplit} holds, i.e. when the variance along the second covariate is small, the optimal symmetric $2$-means split at the population-level splits the data along the first covariate. On the other hand when the condition in~\eqref{eqn::seconddimsplit} holds, i.e. when the variance along the second covariate is large, the optimal symmetric $2$-means split at the population-level is along the second covariate. 

We conjecture that even for $2$-means clustering without the symmetric assumption, as long as the data is generated from  $\frac{1}{2} N(-\theta_1, D) + \frac{1}{2} N(\theta_1, D)$, the above statement holds. That is, when the condition in~\eqref{eqn::firstdimsplit} holds, the optimal $2$-means split at the population-level is along the first covariate and on the other hand when the condition in~\eqref{eqn::seconddimsplit} holds, the optimal $2$-means split at the population-level is along the second covariate. 

Additionally we also have the following lemma:
\begin{lemma}
\label{lemma::Gamma alternate}
In both the cases mentioned in Theorem \ref{thm::optimalsplit}, the matrix $G$ given by equation (\ref{eqn::Gamma}) is positive definite.
\end{lemma}

Therefore Theorem \ref{thm::optimalsplit} and the above Lemma \ref{lemma::Gamma alternate} combined together with Theorem \ref{thm::newPollard} give the limiting distribution under the alternative.

\subsection{Power}

In this section we derive the asymptotic power of
the test 
using the previous results on the limiting distribution. 
Since in the previous section we assumed using a symmetric $2-$means clustering we now consider the test statistic for the symmetric $2-$means clustering. We define
\begin{equation}
T_n^{(0)} := T_n^{(0)}\left(\vecbn^{(0)}\right) = \frac{W_n^{(0)}(\vecbn^{(0)})}{\frac{1}{n} \sum_{i = 1}^n \| X_i - \overline{X} \|^2}.
\end{equation}
Let
\begin{align*}
{\rm Power}_n(a) = \mathbb{P}(T_n^{(0)} > t_{\alpha,n}),
\end{align*}
denote the power of the test
where 
$t_{\alpha,n}$ denotes the $\alpha$-level critical value. Building once again on the result
in Lemma~\ref{lem:bock} and additionally on Theorem \ref{thm::newPollard}, we show the following result:

\begin{theorem}
\label{thm::power1}
Suppose that samples are generated according to the model described in~\eqref{eqn:mix} and let $Z \sim N(0,1)$ then: 
\begin{enumerate}
\item {\bf Consistent: } If,
\begin{align}
\label{eqn:consistent}
\sigma_2^2 < \sigma_1^2 + \frac{a^2}{4},
\end{align}
then SigClust is consistent, i.e. ${\rm Power}_n(a) \to 1$
as
$n \to \infty$. 
\item {\bf Inconsistent: } On the other hand if,
\begin{align}
\label{eqn:inconsistent}
 \sigma_2^2 > \max\left\lbrace \frac{2 \sigma_1^4 +  \frac{a^4}{16} + \frac{a^2}{2}\sqrt{\sigma_1^4 + \frac{a^4}{64}}}{2 \sigma_1^2}, \frac{\pi}{2}\left( \sqrt{\frac{2}{\pi}} \ \sigma_1 \ \exp\left({-\frac{a^2}{8 \sigma_1^2}}\right) + \frac{a}{2} \ P \left(|Z| < \frac{a}{2 \ \sigma_1}\right) \right)^{2} \right\rbrace
\end{align}
then SigClust is inconsistent, i.e. ${\rm Power}_n(a) < 1$
as
$n \to \infty$.
\end{enumerate}
\end{theorem}

\noindent {\bf Remarks: }
\begin{enumerate}
\item In order to roughly understand the result, as we show more precisely in the Appendix
for small values of $a > 0$:
\begin{align*}
\frac{\pi}{2}\Big( \sqrt{\frac{2}{\pi}} \sigma_1 \exp\left({-\frac{a^2}{8 \sigma_1^2}}\right) + \frac{a}{2} \mathbb{P}\Big(|Z| < \frac{a}{2 \sigma_1}\Big) \Big)^{2} \approx \sigma_1^2 + \frac{a^2}{4}, 
\end{align*}
where we use $\approx$ to mean equal up to a small error of size roughly $a^4/\sigma_1^2$. As a consequence, in our setup we see that 
when the variance of the second covariate is sufficiently large SigClust has no power in detecting departures from Gaussianity along the first covariate. 

\item We observe a phase-transition in the power of SigClust, and we provide a precise characterization of this phase-transition.
We highlight that the low power of SigClust is a persistent phenomenon, i.e. there is a large, non-vanishing part of the parameter space
where \emph{the test is not consistent}. 
We see  that
the power of SigClust is very sensitive to the
particular values of the variances in the matrix $D$. 
In the next section we consider alternative 
tests based on relative-fit that address these drawbacks of SigClust. 

\item The proof of this result is quite technical and we defer the details to~Appendix~\ref{sec:power}. At a high-level, the proof follows from Theorem \ref{thm::optimalsplit} which characterizes the optimal 2-means split at the population-level, and uses it to study the distribution of the test statistic under the alternate.
We then leverage our previous characterization of the distribution of the test statistic under the null to study the power of SigClust. 

\item Despite the technical nature of the proof, the intuition behind the phase-transition is quite natural. As shown in Theorem \ref{thm::optimalsplit}, when the condition in~\eqref{eqn:consistent} holds, the optimal $2$-means split at the population-level 
splits the data along the first covariate and as a result the test is able to detect the non-Gaussianity of the first covariate. 
On the other hand when the condition in~\eqref{eqn:inconsistent} holds, the optimal $2$-means split at the population-level is along the second covariate and the resulting test is asymptotically inconsistent. 

\item Finally, we note in passing that in the case when 
\begin{align*}
\frac{\pi}{2}\Big( \sqrt{\frac{2}{\pi}} \sigma_1 \exp\left({-\frac{a^2}{8 \sigma_1^2}}\right) + \frac{a}{2} \mathbb{P}\Big(|Z| < \frac{a}{2 \sigma_1}\Big) \Big)^{2} = \sigma_2^2,
\end{align*}
the $2$-means solution is no longer unique, and we are unable to use our techniques to characterize the power of the test. However, we conjecture that SigClust remains inconsistent even in this case. 
\end{enumerate}

\section{A Test For Relative Fit of Mixtures}
\label{section::rift}

A natural way to improve the low power of SigClust
is to formally test 
for whether the data are generated from a Gaussian 
versus a mixture of Gaussians.
There is a long history of research on this problem;
see, for example,
\cite{dacunha1999testing,gassiat2002likelihood,chen2017finite,gu2017testing}
and references therein.
As we mentioned
earlier, the mixture model is irregular and
there has been little success in deriving a practical, simple test with
valid type I error control.
Furthermore, and more importantly, such tests ignore the fact
that we are only using the parametric model as an approximation;
we don't expect that the true distribution is exactly
Gaussian or a mixture of Gaussians.
This motivates our new approach
where we test the relative fit of the models
without assuming that either model is correct.
Also, our test is valid for multivariate mixtures whereas
many of the existing tests are for the univariate case.

\subsection{The Basic Test}

Let ${\cal P}_1$
denote the set of multivariate Gaussians and let
${\cal P}_2$
denote the set of mixtures of two multivariate Gaussians.
We are given a sample
$X_1,\ldots, X_{2n} \sim P$
but we do not assume that $P$ is necessarily in either
${\cal P}_1$ or ${\cal P}_2$.
Note that, for notational simplicity, we denote the total sample size by $2n$.

We randomly split the data into two halves ${\cal D}_1$ and 
${\cal D}_2$.
Assume each has size $n$. 
Using
${\cal D}_1$,
fit a Gaussian
$\hat p_1$ 
and a mixture of two Gaussians 
$\hat p_2$.
Any consistent estimation procedure can be used;
in our examples we use the Expectation Maximization (EM) algorithm. Understanding precise conditions under which EM yields a global maximizer is an area of active research \citep{balakrishnan2017statistical} but we do not pursue this further in this paper. 

Instead of testing 
$H_0: P\in {\cal P}_1$ versus
$H_1: P\in {\cal P}_2$
we test whether
$\hat p_2$ 
is a significantly better fit for the data than
$\hat p_1$.
This is a different hypothesis from the usual one, but, arguably, it is more relevant since it is
$\hat p_1$ or $\hat p_2$ that will be used for clustering.
Furthermore, this
does not require that
the true distribution be in either
${\cal P}_1$ or 
${\cal P}_2$.

To formalize the test,
let 
\begin{equation}
\Gamma  = K(p, \hat p_1) - K(p, \hat p_2),
\end{equation}
where $K(p,q) = \int p \log (p/q)$ is the Kullback-Leibler distance and $p$ is the true density. 
Note that $\Gamma$ is a random variable.
Formally, we will test, conditional on ${\cal D}_1$,
\begin{equation}
H_0 : \Gamma \leq 0 \ \ \ {\rm versus}\ \ \  H_1 :  \Gamma > 0.
\end{equation}
Since $\Gamma$ is a random variable, these are random hypotheses.
Let
\begin{equation}
\label{eqn::gammahat}
\hat \Gamma = \frac{1}{n}\sum_{i\in {\cal D}_2} R_i
\end{equation}
where
$R_i = \log\left({\hat p_2(X_i)}/{\hat p_1(X_i)}\right).$
Below, we show that, conditionally on ${\cal D}_1$,
\begin{align*}
\sqrt{n}(\hat\Gamma - \Gamma) \rightsquigarrow N(0,\tau^2)
\end{align*}
where
$\tau^2 \equiv \tau^2 ({\cal D}_1) =
\mathbb{E}[R_i^2] - \Gamma^2.$
The quantity $\tau^2$ can be estimated by
$\hat\tau^2 = \frac{1}{n}\sum_{i\in {\cal D}_2} (R_i - \overline{R})^2.$
We reject $H_0$ if 
\begin{align*}
\hat \Gamma > \frac{z_\alpha \hat \tau}{\sqrt{n}},
\end{align*}
and we refer to this as the \riftspace (Relative Information Fit Test).
For technical reasons,
we make a small modification to the test statistic.
We replace $R_i$
with $\tilde R_i = R_i + \delta Z_i$
where $Z_1,\ldots, Z_n \sim N(0,1)$
and $\delta$ is some small positive number, for example,
$\delta = 0.00001$.
This has no practical effect on the test and is only needed for the theory.

For the following result, let the fitted Gaussian density be given by
$\hat p_1= N(\hat{\mu}, \hat\Sigma)$ 
and the fitted mixture of two Gaussians be given by
$\hat p_2 = 
\hat\alpha \hat{f}_1 + (1 - \hat\alpha) \hat{f}_2$, where $\hat{f}_1 = 
N(\hat{\mu}_1, \hat\Sigma_1)$ and $\hat{f}_2 = N(\hat{\mu}_2, \hat\Sigma_2)$.
For technical reasons,
we restrict the parameter estimates  to lie in a compact set.
Formally, we assume that each $\hat\mu_i$ 
is restricted to lie in a compact set $\mathcal{A}$ and 
that the eigenvalues of $\hat\Sigma$ and $\hat\Sigma_i$ lie in some interval $[c_1, c_2]$ for $i = 1,2$, where $c_1, c_2 > 0$. 
As a consequence of data splitting, the test of relative fit has a simple limiting distribution
unlike the usual tests for mixtures which have intractable limits.

\begin{theorem}
\label{thm::BE1}
Let $Z\sim N(0,\tau^2)$
where 
$\tau^2 =  \mathbb{E}[( \tilde R_i  - \Gamma )^2 | {\cal D}_1].$ 
Then, under $H_0$
\begin{equation}
\sup_t \Bigl|\mathbb{P}(\sqrt{n}(\hat\Gamma - \Gamma)\leq t\,|\, {\cal D}_1) - 
\mathbb{P}(Z \leq t)\Bigr| \leq \frac{C}{\sqrt{n}}
\end{equation}
where 
$C = \frac{C_0 }{\delta^3} \left[8C_1^3 + 
\delta \left( 12 C_1^2 \sqrt{\frac{2}{\pi}} + 6 C_1\delta + 
2 \sqrt{\frac{2}{\pi}} \delta^2  \right)\right]$, 
$C_0=33/4$ and $C_1$ is a constant.
\end{theorem}
\noindent {\bf Remark: } It is also possible to consider the normalized version of the statistic $\widehat{\Gamma}$. Formally, under the conditions of the above result conditional on $\mathcal{D}_1$:
\begin{align*}
\sup_t \Bigl|P\Big(\sqrt{n}\Big(\frac{\hat\Gamma}{\hat \tau} - \frac{\Gamma}{\tau}\Big)\leq t \Big) - 
\mathbb{P}(Z \leq t)\Bigr| \leq \frac{C}{\sqrt{n}}
\end{align*} 
where $Z \sim N(0,1)$. 
We note that since the constant $C$ does not depend on $\mathcal{D}_1$ this result also holds unconditionally.

We now turn our attention to the power of \riftnospace. 
Suppose that we consider a distribution $p$ such that,
\begin{align}
\label{eqn:gammastar}
\Gamma^* = \inf_{p_1 \in \mathcal{P}_1} K(p, p_1) - \inf_{p_2 \in \mathcal{P}_2} K(p, p_2) > 0, 
\end{align}
i.e. $p$ is a distribution for which the class of mixtures of two Gaussians provides a strictly better fit than a single Gaussian. Then we have the following result characterizing the power of \riftnospace:
\begin{theorem}
\label{thm::goodpower}
Suppose that $\Gamma^*$ in~\eqref{eqn:gammastar} is strictly positive, then \riftspace is asymptotically consistent, i.e. as $n\to\infty$.
\begin{align*}
\text{Power}_n(\textsc{\riftnospace}) = \mathbb{P}(\hat\Gamma > z_\alpha\hat\tau/\sqrt{n})\to 1.
\end{align*}
\end{theorem}
\noindent {\bf Remark: } A consequence of this result is that 
\riftspace is consistent against any fixed distribution 
$p\in {\cal P}_2 \backslash {\cal P}_1$.
In other words, the power deficiency of SigClust observed
in Theorem \ref{thm::power1} does not happen for our test.

\subsection{Variants of \riftnospace}
In this section we introduce and study a few variants of \riftspace that can be advantageous
in various applications. 

\vspace{.2cm}

\noindent {\bf A Robust, Exact Test.}  The Kullback-Leibler (KL) distance
between two densities
$p$ and $q$ is
$K(p,q) = \mathbb{E}_p [W]$
where
$W = \log (p(X)/q(X))$.
This distance can be sensitive to the tail of the distribution of $W$.
For this reason we also consider a robustified version of the KL distance,
namely,
$\tilde K(p,q) = {\rm Median}_P [W]$,
that is, the median of $W$ under $p$ (we will assume for convenience that the median is unique).
In this case, the sample median of
$W_1,\ldots, W_n$ is a consistent estimator of
$\tilde K(p,q)$,
where $W_i = \log \left( p(X_i)/q(X_i)\right)$.

For relative fit we define
\begin{equation}
\tilde{\Gamma} = {\rm Median}_p [R]
\end{equation}
where
$R = \log \hat p_2(X)/\hat p_1(X)$.
A point estimate is the sample median based on ${\cal D}_2$.
To test
$H_0: \tilde\Gamma \leq 0$ versus
$H_1: \tilde\Gamma >0$ we use
the sign test.
Hence, under $H_0$,
$\mathbb{P}({\rm rejecting\ }H_0)\leq \alpha$.
We will refer to this as median-\riftspace or \mriftnospace.
This approach has two advantages: it is robust and 
it does not require any asymptotic approximations.

\vspace{.2cm}

\noindent {\bf $\ell_2$ Version.}
The test does not have to be based on Kullback-Leibler distance.
We can also use the $\ell_2$ distance as we now explain.
Define the $\ell_2$-relative fit by
$\Theta = \int (p - \hat p_1)^2 - \int (p - \hat p_2)^2.$
We test, conditional on ${\cal D}_1$,
$$
H_0 : \Theta \leq 0 \ \ \ {\rm versus}\ \ \  H_1 : \Theta > 0.
$$
To estimate $\Theta$, note that we can write
$\Theta = \int \hat p_1^2 -\int \hat p_2^2 - 2 \int p (\hat p_1 - \hat p_2)$
which can be estimated by
$$
\hat\Theta = \int \hat p_1^2 -\int \hat p_2^2 - \frac{2}{n}\sum_{i\in {\cal D}_2} U_i
$$
where
$U_i = \hat p_1 (X_i) - \hat p_2 (X_i).$
To evaluate the integrals, we use importance sampling.
We sample $Y_1,\ldots, Y_N\sim g$
from a convenient density $g$ (such as a $t$-distribution) and then use
$$
\int \hat p_1^2 \approx \frac{1}{N}\sum_j \frac{\hat p_1^2(Y_j)}{g(Y_j)},\ \ \ 
\int \hat p_2^2 \approx \frac{1}{N}\sum_j \frac{\hat p_2^2(Y_j)}{g(Y_j)}.
$$
Again, for technical reasons,
we make a small modification to the test statistic.
We replace $U_i$
with $\tilde U_i = U_i + \delta Z_i$
where $Z_1,\ldots, Z_n \sim N(0,1)$
and $\delta$ is some tiny positive number, for example,
$\delta = 0.00001$.
Again this has no practical effect on the test and is only needed for the theory.
Recall that the Gaussian density is given by
$\hat p_1= N(\hat{\mu}, \hat\Sigma)$ 
and the mixture of two Gaussians is given by
$\hat p_2 = 
\alpha \hat{f}_1 + (1 - \alpha) \hat{f}_2$, 
where $\hat{f}_1 = N(\hat{\mu}_1, \hat\Sigma_1)$ and $\hat{f}_2 = N(\hat{\mu}_2, \hat\Sigma_2)$.
Once again, we assume that $\hat\mu_i$ 
are restricted to lie in a compact set $\mathcal{A}$ 
and that the eigenvalues of $\hat\Sigma$ and $\hat\Sigma_i$ lie in the interval $[c_1, c_2]$ for $c_1, c_2 > 0$ 
and for $i = 1,2$.

\begin{theorem}
\label{thm::BE2}

Let $Z\sim N(0,a^2)$ where $a^2 = \text{var}(\tilde{U}_i)$.
Then, under $H_0$, 
\begin{equation}
\sup_t|\mathbb{P}(\sqrt{n}(\hat\Theta - \Theta)\leq t | \mathcal{D}_1) - \mathbb{P}(Z \leq t)| \leq \frac{\tilde C}{\sqrt{n}},
\end{equation}
where $\tilde C = \frac{C_0 }{\delta^3} 
\left[8C_2^3 + 
\delta \left( 12 C_2^2 \sqrt{\frac{2}{\pi}} + 
6 C_2\delta + 2 \sqrt{\frac{2}{\pi}} \delta^2  \right)\right]$, 
$C_0=33/4$ and $C_2$ is a constant.
\end{theorem}

\subsection{Aside: A Test for Mixtures}

Our focus is on the relative fit as described in the previous section.
However, it is possible to modify our test so that
it tests the more traditional hypotheses
$$
H_0: P \in {\cal P}_1\ \ \ {\rm versus}\ \ \ H_1: P \in {\cal P}_2
$$
where ${\cal P}_1$ are Normals and
${\cal P}_2$ are the mixtures of two Normals.
There is currently no available test that is simple, asymptotically valid and has easily computable critical values
in the multivariate case.
But we can use our split test for this hypothesis
if we modify the test using the idea of  
\citet{ghosh1984asymptotic}
where we force the fit under the alternative to be bounded
away from the null.
When combined with data splitting, this results in a valid test.
Specifically, when we fit $H_1$, we will constrain the fitted density $\hat p_2$ to satisfy
$K(p,\hat p_2) > \Delta$ for all $p\in {\cal P}_1$. 
Here, $\Delta$ is any small, positive constant.

\begin{theorem}
\label{thm::valid}
If $P\in {\cal P}_1$ then
$\mathbb{P}(\hat\Gamma > z_\alpha \hat\tau/\sqrt{n})=\alpha + o(1)$.
\end{theorem}

\noindent Hence, combining data splitting with the Ghosh-Sen separation idea
yields an asymptotically valid test for mixtures
with a simple critical value.
To the best of our knowledge, this is the first such test.

\subsection{Truncation}

If we use \riftspace for top-down hierarchical clustering, as described later in Section \ref{section::hier}, then
after the first split,
the null hypothesis will be a truncated Normal rather than a Normal
since the test is now applied to the data in a cluster.
Instead of comparing the fit of a Normal $\hat{p}_1$ and a fit of a  
mixture of two Normals $\hat{p}_2$, we need 
to compare the fit of a truncated normal to a fit of a 
truncated mixture of two Normals. 
We can use exactly the same test except that
$\hat p_j$ should be replaced with
$\hat p_j/\hat P_j(S)$
where $S$ denotes the subset of $\mathbb{R}^d$ corresponding
to the cluster being tested.
We can estimate $P_j(S)$ as follows.
First, generate $Z_1, Z_2, \ldots, Z_m \sim \hat P_j$ for some large $m$.
Then set
$\hat P_j(S) = \frac{1}{m} \sum_{i = 1}^m \mathbb{I}(Z_i \in S)$.
Then replace
$\hat p_j$ with
with $\hat p_j/\hat P_j(S)$ 
in the test.

\subsection{Other Tests}

Another way to decide whether to split the Normal is to use a goodness-of-fit test for Normality.
In this section we describe two such tests.
Note that such tests can only be used for the first split
in the clustering problem.
We include them in our study because they are simple
and they provide a point of comparison. We also note that it is possible to use tests for goodness-of-fit 
with minimax-optimal power against neighborhoods defined in particular metrics, based on binning and the $\chi^2$-test, but these tests are complex and have tuning parameters that need to be carefully chosen.

\vspace{0.2cm}

\noindent {\bf 1. Mardia's multivariate Kurtosis test.}
\cite{mardia1974applications} proposed using the Kurtosis measure
to test for normality. If ${X}$ is a
$d$-dimensional random (column) vector with expectation $\mu =
\mathbb{E}[{X}]$ and non-singular covariance matrix $\Sigma = 
\mathbb{E}[({X}- \mu)({X} - \mu)^T]$, \cite{mardia1970measures} defines the
multivariate Kurtosis as
\[
\beta_2 = \mathbb{E}\left[\left\lbrace ({X}- \mu)^T \Sigma^{-1} ({X} - \mu)\right\rbrace^2\right]. 
\]

The proposed test uses the Kurtosis measure to test for multivariate
normality. If $X_1,\ldots, X_n \in \mathbb{R}^d$ are independent
observations from any multivariate normal distribution, then the
sample analogue of Kurtosis is given by,
\[
b_{2,d} = 
\frac{1}{n} \sum_{j = 1}^n 
\left\lbrace \left(X_j - \overline{X}\right)^T S_n^{-1}\left(X_j - \overline{X}\right)\right\rbrace^2, 
\]
where 
\[ 
\overline{X} = 
\frac{1}{n} \sum_{i = 1}^n X_j, \ \ 
S_n = \frac{1}{n} \sum_{j = 1}^n \left(X_j - \overline{X}\right) \left(X_j - \overline{X}\right)^T
\]
are the sample mean vector and the sample covariance
matrix. \cite{mardia1970measures} shows that $b_{2,d}$ has a
distribution under the null hypothesis, $H_0$, given by
\[ 
\frac{\sqrt{n}\left( b_{2,d} - d(d+2)\right)}{\sqrt{8 d (d + 2)}} \overset{d}{\to} N(0, 1)  
\]
as $n \to \infty$. So we reject the null hypothesis for both large and
small values of $b_{2,d}$. This multivariate normality test is
consistent if, and only if,
\[
\mathbb{E}\left[\left\lbrace 
\left( X_i - \mu\right)^T\Sigma^{-1}\left( X_i - \mu\right)\right\rbrace^2 \right] \neq d(d + 2).
\]
For detecting clusters, the method starts by fitting a multivariate Gaussian to the
data. 
We then perform the multivariate normality test using the Kurtosis measure and if
the test gets rejected then the data is split into two clusters. 
We reject $H_0$ at level $\alpha$ if 
$$
\left\vert \frac{\sqrt{n}\left( b_{2,d} - d(d+2)\right)}{\sqrt{8 d (d + 2)}} \right\vert > z_{\alpha/2}.
$$

\vspace{0.2cm}

\noindent {\bf 2. NN Test.}
Nearest neighbor (NN) goodness of fit tests
were developed by
\citet{bickel1983sums} and \citet{zhou1993goodness}.
Let $X_1, \ldots , X_n \in \mathbb{R}^d$ be samples from $P$
with a density function $p(x)$.
We want to test
$H_0: P = P_0$
where $P_0$ has density $p_0$.

In the clustering framework, we consider the null hypothesis that the
data is drawn from a single multivariate Gaussian distribution. That
is, we consider $p_0$ to be the multivariate Gaussian distribution,
with some mean $\mu$ and covariance matrix $\Sigma$. 
To implement these tests,
we first split
the data into two halves $\mathcal{D}_1$ and $\mathcal{D}_2$ and use
$\mathcal{D}_1$ in order to estimate the $\mu$ and $\Sigma$. 
Therefore in our setting,
$P_0 = N(\hat\mu,\hat\Sigma)$ is the estimated null.

Let
$R_i = \min_{j \neq i} \|X_i - X_j\|$.
The first version of this test
uses
\[W_i = \exp\left({-n D_i}\right) := \exp\left({-n p_0(X_i)V(R_i)}\right)\]
where $V(r) = K_d r^d$
is the volume of a ball of radius $r$ and
$D_i = p_0(X_i)V(R_i)$.
Under $H_0$, the $W_i$'s
are approximately Uniform on $[0,1]$
and hence we can use the Kolmogorov-Smirnov test.

For the second version, we consider the test proposed by \citet{zhou1993goodness} that uses
$$
T_n^* = \frac{1}{\sqrt{n}} \sum_{i = 1}^n \left[h(n D_i) - \mathbb{E}_0[h(n D_i)]\right]
$$
where
$h$ is a bounded function on $[0,\infty)$.
The authors show that
$\sqrt{n} \ T_n^* \overset{d}{\to} N(0, \sigma^2(h))$
which is independent of the null distribution $P_0$, where $\sigma^2(h)$ only depends on the function $h$.

We consider $h(x) =
\exp(-x)$ and calculate the test statistic in terms of $W_i$ as
$$
T_n^* = 
\frac{1}{\sqrt{n}} \sum_{i = 1}^n 
\left[\exp(-n D_i) - \mathbb{E}_0[\exp(-n D_i)]\right] = 
\frac{1}{\sqrt{n}} \sum_{i = 1}^n \left[W_i - \mathbb{E}_0[W_i]\right]. 
$$
Since under the null distribution $P_0$, $ W_i \approx U(0,1)$, $\mathbb{E}_0[W_i] = 0.5$. 
Therefore, we reject $H_0$ at level $\alpha$ if
$$
\left\vert \frac{\sqrt{n} \ T_n^*}{\hat{\sigma}(h)} \right\vert > z_{\alpha/2}
$$
where $\hat{\sigma}^2(h)$ is the estimated variance of the $W_i$'s.

\section{Hierarchical Clustering}
\label{section::hier}

To propose a hierarchical version of \riftnospace, we apply the procedure
recursively.
Figures~\ref{fig::top-down} and \ref{TopDownFlowchart}
in the Appendix
describe the details for the top-down approach and 
Figure~\ref{fig::bottom-up} in the Appendix describes the details for the bottom-up approach.
The final clustering is given by the leaf nodes of the tree derived by the algorithms.

In each case we begin by splitting the data
into two halves
${\cal D}_1$ and 
${\cal D}_2$.
The first half ${\cal D}_1$
is used to estimate the parameters and to recursively split the clusters forming a cluster tree.
The second half ${\cal D}_2$ is used to conduct the significance tests.
In the top-down approach, the tests are applied from the top of the tree downwards
and we stop when $H_0$ is not rejected.
In the bottom-up approach we start at the bottom of the tree and combine leaves until the test rejects.

\section{A Sequential Approach}
\label{section::sequential}

\riftspace can also be used 
in a sequential model selection framework.
Using ${\cal D}_1$ we fit a mixture of $k$ Gaussians
for $k=1,2,\ldots, K_n$
where $K_n$ can be chosen to be quite large, for example, $K_n = \sqrt{n}$.
Now,
using ${\cal D}_2$,
we choose $k$ by testing a series of hypotheses.
For $j=1,2,\ldots,$ we test the null that
$\hat p_j$ fits better than any $\hat p_s$ for $s>j$.
Formally, we test
$$
H_{0j}: = K(p, \hat p_j) - K(p, \hat p_s) \leq 0\ \ \ {\rm for \ all\ }s>j
$$
versus
$$
H_{1j}: = K(p, \hat p_j) - K(p, \hat p_s) > 0\ \ \ {\rm for \ some\ }s>j.
$$
We reject $H_{0j}$ if
\begin{equation}
\max_s \hat\Gamma_{js} > \frac{ z_{\alpha/m_j}\hat \tau_{js}}{\sqrt{n}}
\end{equation}
where
$m_j = K_n - j$,
$\hat \Gamma_{js} = \frac{1}{n}\sum_{i\in {\cal D}_2} R_i$,
$R_i = \log\left( {\hat p_s(X_i)}/{\hat p_j(X_i)}\right)$
and
$\hat\tau^2_{js} = \frac{1}{n}\sum_{i\in {\cal D}_2} (R_i - \overline{R})^2.$

Let $\hat k$ be the first value of $j$ for which $H_{0j}$ is not rejected.
We then use $\hat p_{\hat k}$ to define the clusters.
Notice that, unlike procedures like AIC or BIC,
this method provides a valid, asymptotic, type I error control.

\begin{lemma}
Under $H_{0j}$,
\begin{equation}
\limsup_{n\to\infty} \mathbb{P}( {\rm rejecting\ }H_{0j}) \leq \alpha.
\end{equation}
\end{lemma}
\noindent This follows from the results in Section 3.
Of course, $\Gamma$ can be replaced with the $\ell_2$ version or the median version.

\section{Simulations}
\label{section::examples}
In this section we compare SigClust and the \riftspace variants we proposed through 
a variety of simulations. In Section~\ref{sec:simnorm} we investigate the asymptotic normality
of the \riftspace statistic defined in~\eqref{eqn::gammahat} under the null. In Section~\ref{sec:simtwogauss} 
we
compare the power of various tests for detecting and splitting a mixture of two Gaussians.
Finally, in Sections~\ref{sec:simhierarchical} and~\ref{sec:simsequential} we study hierarchical clustering using the \riftspace statistic and evaluate model selection using the sequential \riftspace procedure. 

\subsection{Asymptotic Normality of the \riftspace Test Statistic}
\label{sec:simnorm}
In this section we check if the distribution of the \riftspace test statistic is indeed Normal
as claimed in Theorem \ref{thm::BE1}. We explore four simulated data
sets and use Q-Q plots to check for Normality. For
the four examples, we generate data from the following distributions:
\begin{enumerate}
\item $0.5 N(\mu, \mathbf{I}_d) + 0.5 N(-\mu, \mathbf{I}_d),$ with 
$d = 2$, $n = 1000$ and $\mu = (2, 0)$.
\item A mixture of two uniform distributions over rectangles given by, $0.5 \ \text{Unif}([-2, -1] \times [0, 1]) + 0.5 \ \text{Unif}([2, 3] \times [0, 1]),$ with 
$d = 2$ and $n = 1000$.
\item $0.5 N(\mu, \mathbf{I}_d) + 0.5 N(-\mu, \mathbf{I}_d),$ with 
$d = 1000$, $n = 1000$ and $\mu = (10, 0, \ldots, 0)$.
\item A single Gaussian distribution, $N({\bf 0}, \Sigma)$, where $\Sigma_{11} = 100$ and $\Sigma_{jj} = 1$ for $j = 2, \ldots, d$.
\end{enumerate}
 
\begin{figure}[h]
\centering
\includegraphics[width=0.24\linewidth]{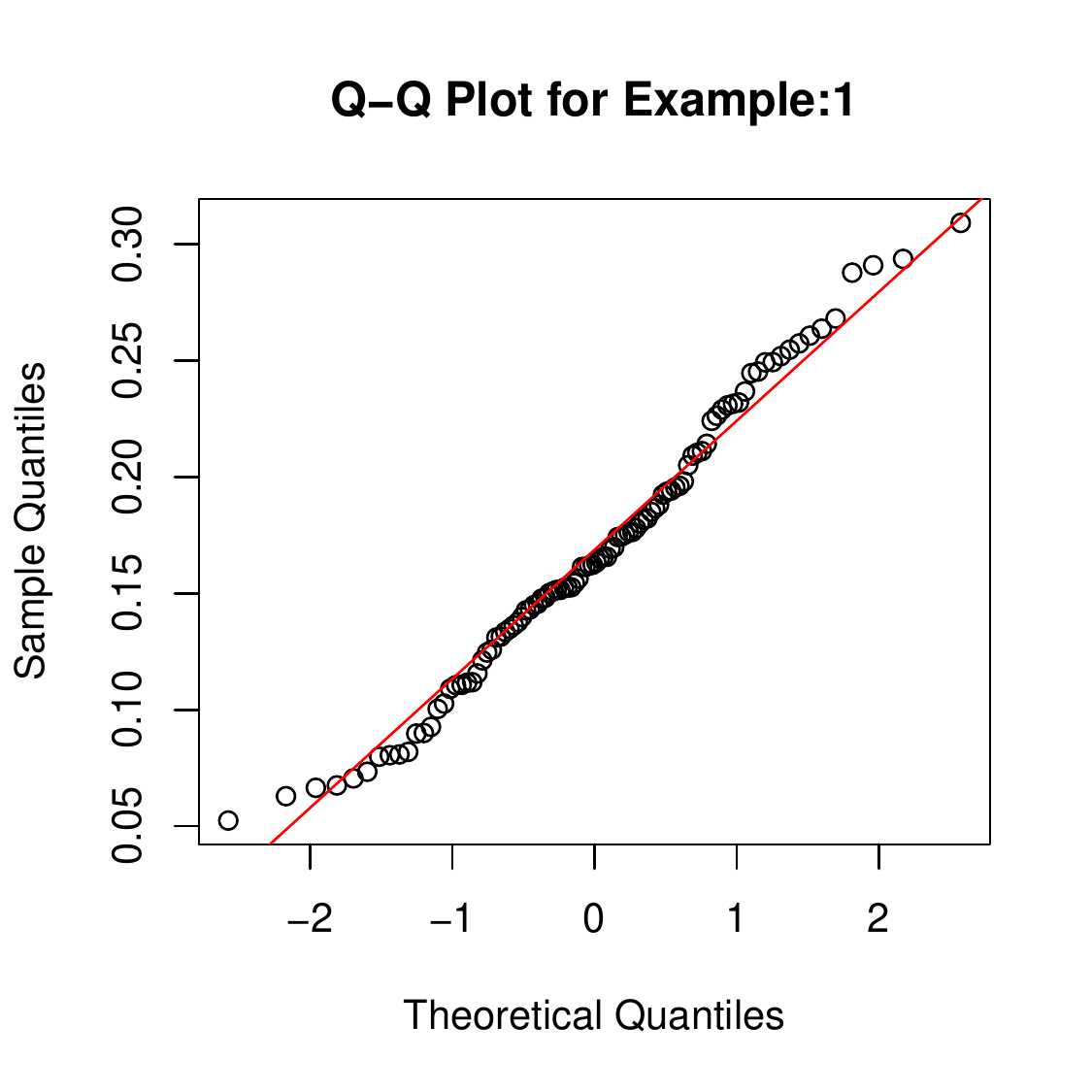}
\includegraphics[width=0.24\linewidth]{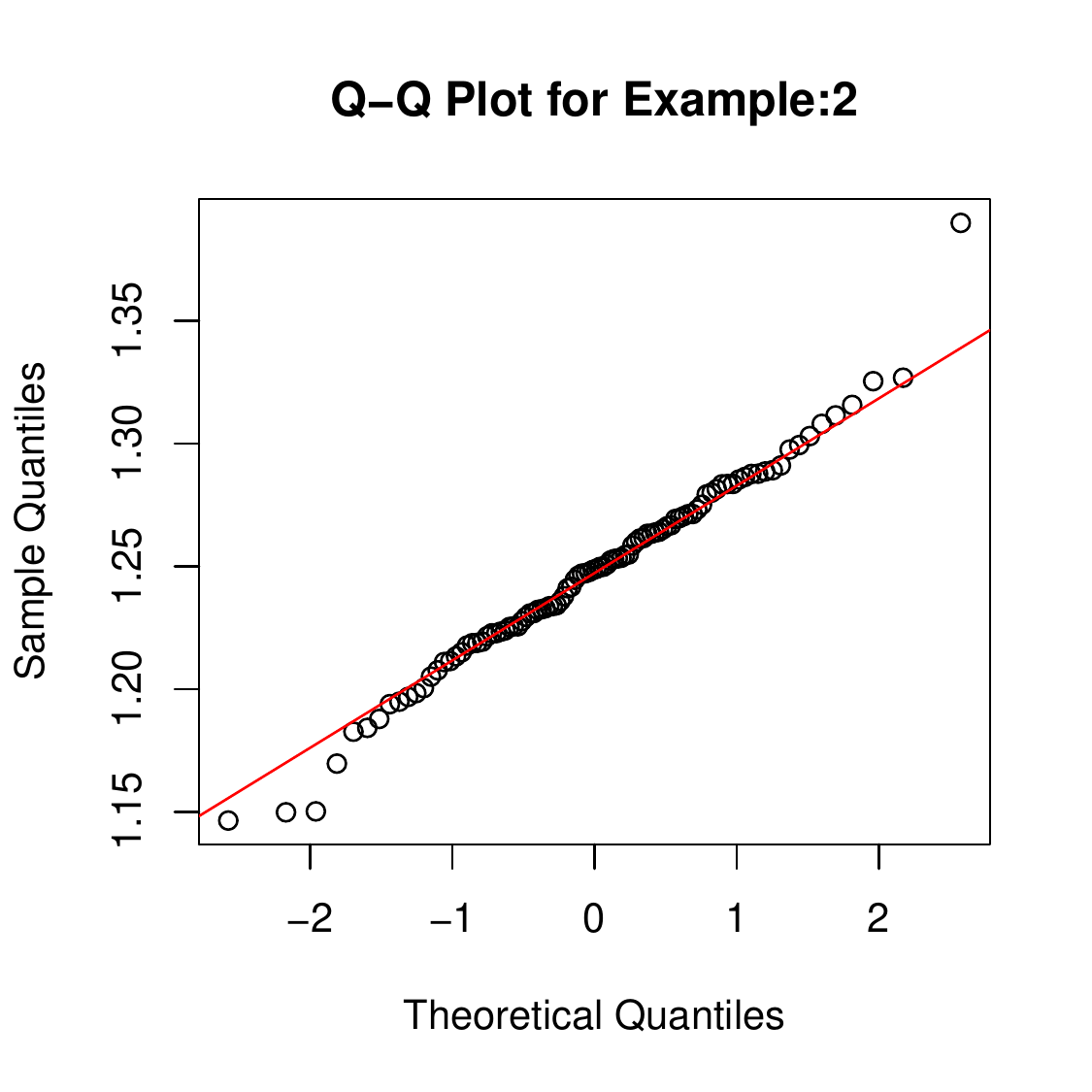}
\includegraphics[width=0.24\linewidth]{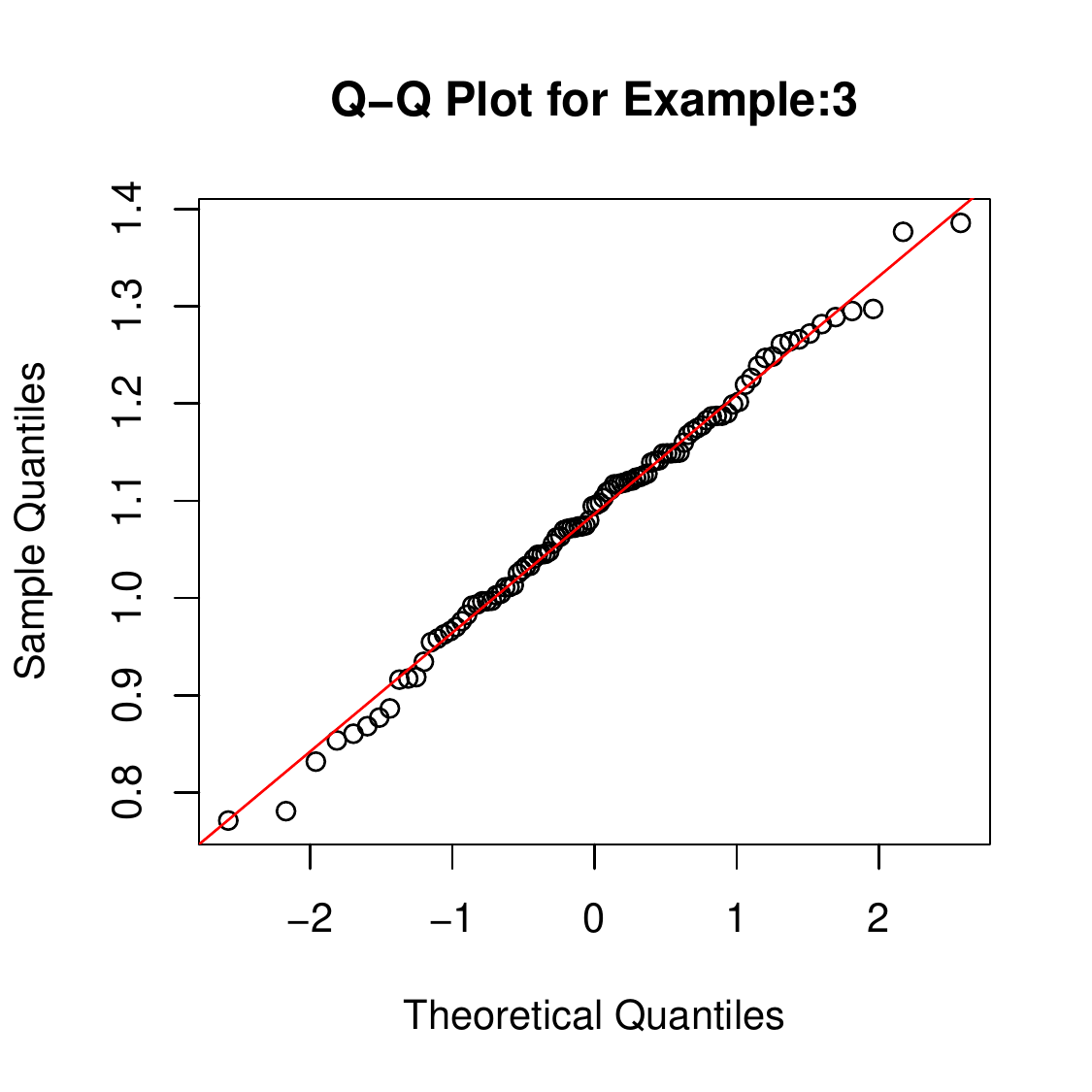}
\includegraphics[width=0.24\linewidth]{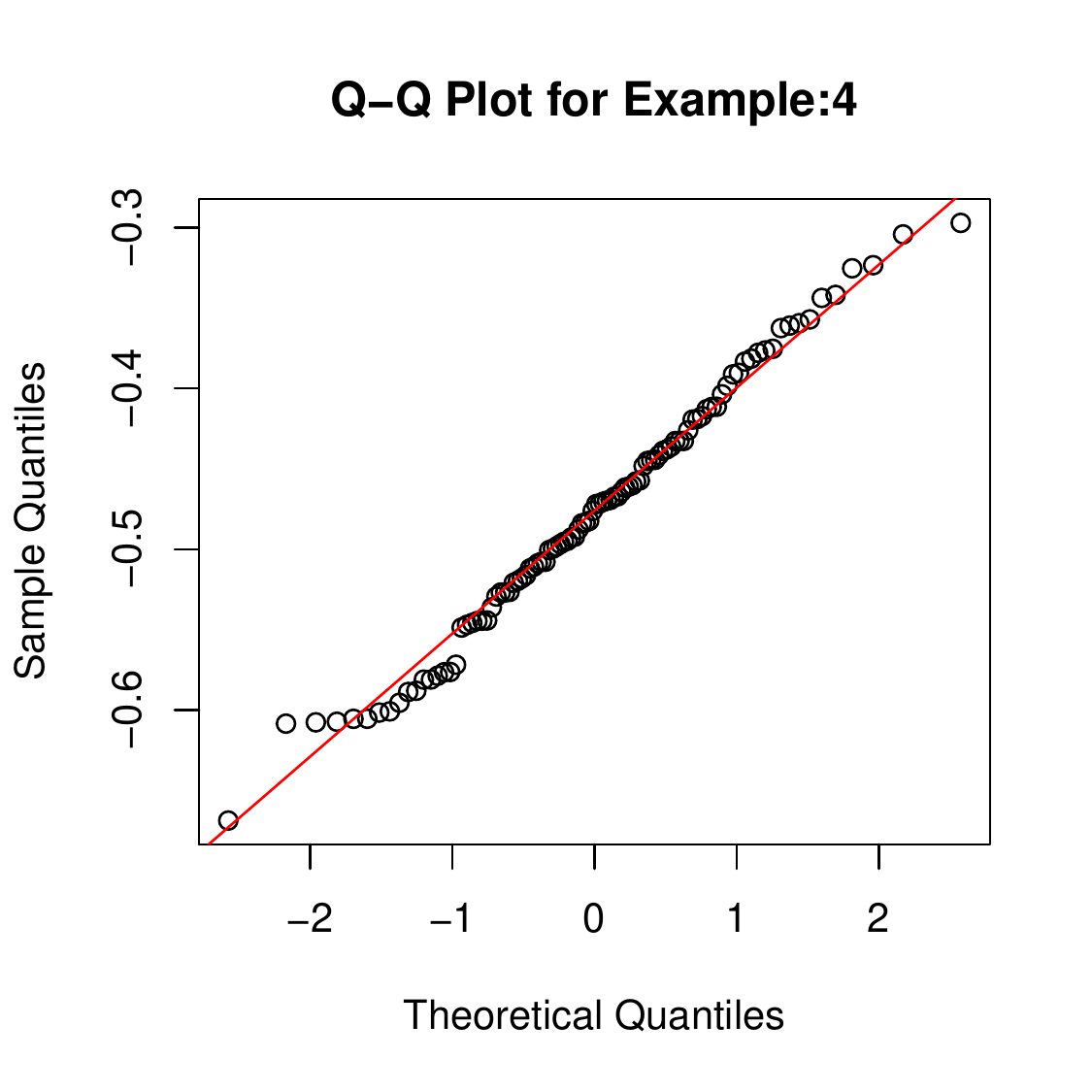}
\caption{Q-Q plots to check Normality of the \riftspace test statistic.}
\label{QQPlots}
\end{figure}
\noindent The test statistic, $\hat{\Gamma}$ defined in~\eqref{eqn::gammahat} is computed for $100$ simulations
in each of these cases. Figure~\ref{QQPlots} provides the Q-Q plots of
the test statistic. We notice that all of them are close to Normal, confirming the result in
Theorem \ref{thm::BE1}.

\subsection{Comparing the Different Tests for Mixtures of Two Gaussians}
\label{sec:simtwogauss}

We first consider data generated from a collection of mixture of two
Gaussians, $0.5 N(\mu, \mathbf{I}_d) + 0.5 N(-\mu, \mathbf{I}_d),$ where
$\mu = (a, 0, \ldots, 0)$, with varying distances (varying $a$)
between their means. We compare the power of
the different tests in detecting the two clusters. Specifically, we
compare the number of times the tests correctly reject the null
hypothesis that the data is generated from a single (Gaussian) cluster.

\begin{figure}[h]
\centering
\includegraphics[width=0.49\linewidth]{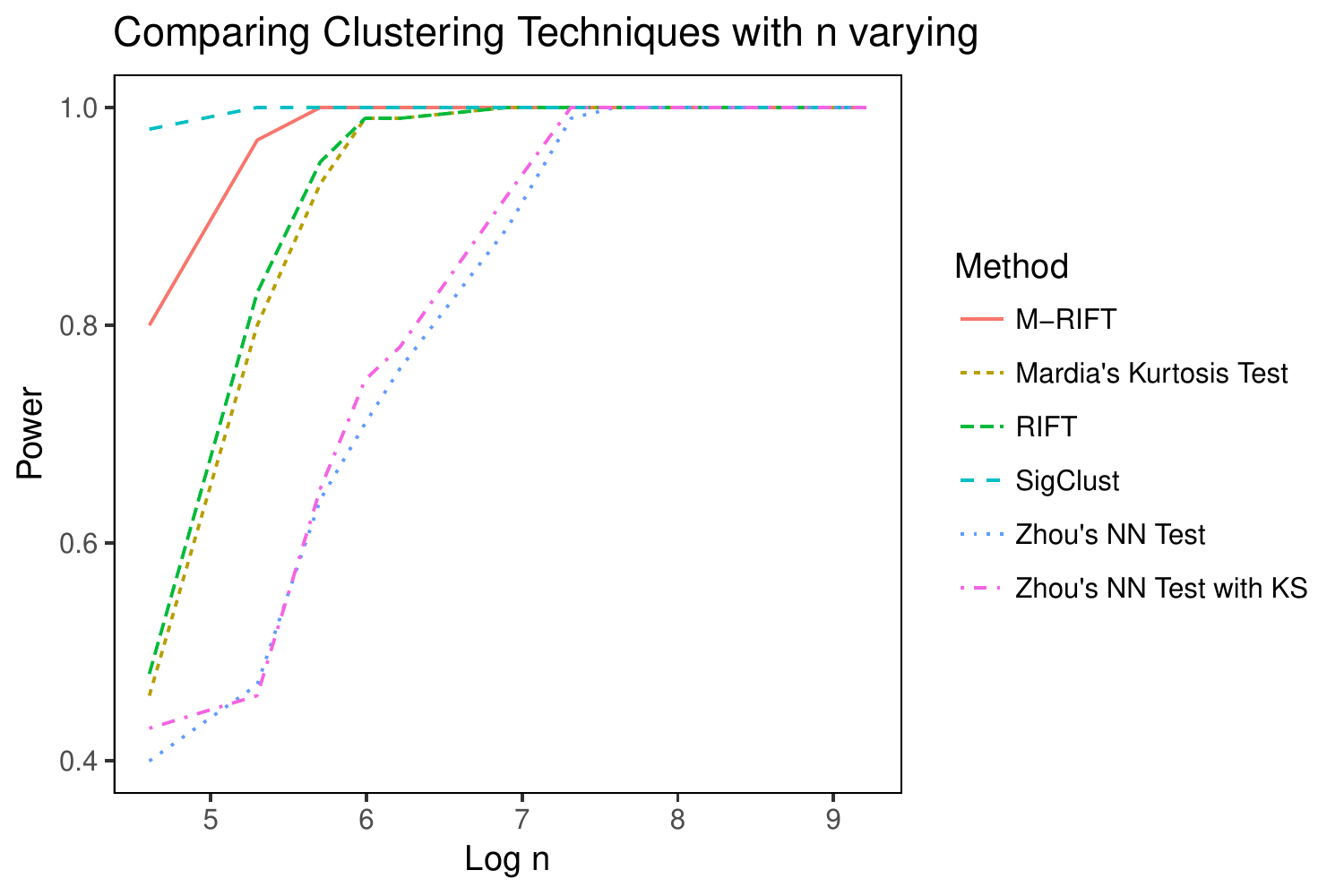}
\includegraphics[width=0.49\linewidth]{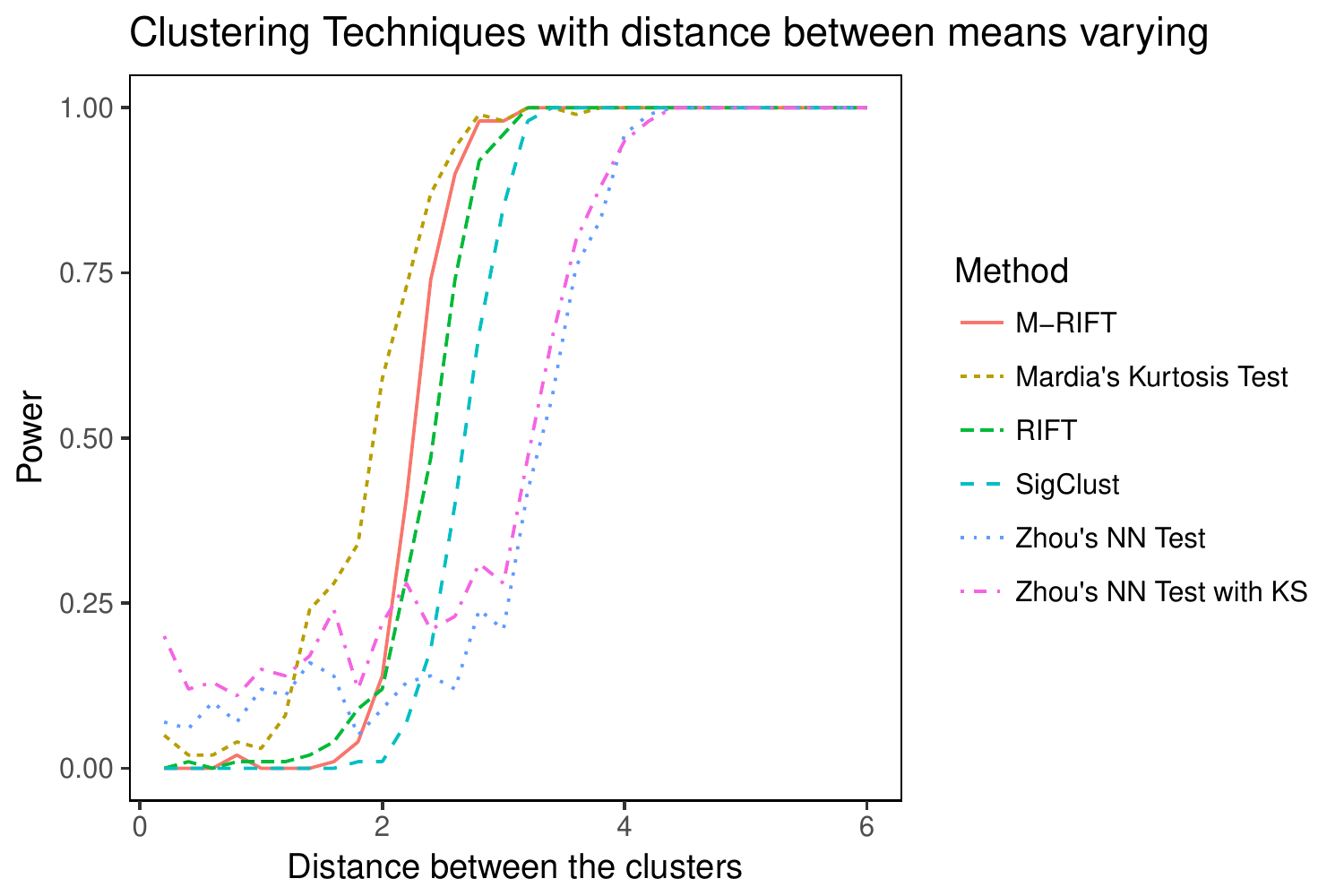}
\caption{Comparing the power of the tests 
with varying number of observations, $n$ and with increasing distance between the two mixture distributions (increasing $a$).}
\label{Compare Tests n}
\label{Compare Tests mean}
\end{figure}

First, we compare the effect of varying the number of
observations ($n$) for the different tests. We run $100$ simulations where we generate observations from a mixture of two 2D Gaussian distributions given by, 
$0.5 N(\mu, \mathbf{I}_d) + 0.5 N(-\mu, \mathbf{I}_d),$ with 
$d = 2$ and $\mu = (2, 0)$. Figure~\ref{Compare Tests n} gives the proportion of tests that reject the
null hypothesis that the underlying distribution has just one
cluster.  We see that \mriftspace and SigClust have comparable power,
and that they have higher power than the other tests. We also notice that Mardia's
Kurtosis test and \riftspace have comparable power, but they do not
perform as well as SigClust or \mriftnospace.

Next we vary the value of $a$ and see how increasing or decreasing
the distance between the two distributions changes the ability
of the tests to reject. Figure~\ref{Compare Tests mean} compares the
proportion of times the tests detect the two distributions as
we vary the distance between them. Notice that Mardia's Kurtosis
test and both the \riftspaces perform better than SigClust in this
case. In particular, they detect the two clusters for smaller values of $a$
when compared to SigClust. Also notice that SigClust does not detect the
presence of the two clusters at all when the distance between the two 
clusters is $\leq 1.5$. 
For the rest of our simulations, we consider comparisons between the \riftspaces and SigClust.

\subsubsection{Signal in one direction}

We compare the power of our test with SigClust while checking whether
the tests control the type-I error. We consider a mixture of two
normal distributions, $0.5N(0, \Sigma)+ 0.5N(\mu, \Sigma)$, where $\mu = (a, 0, \ldots, 0)$ 
with $a = 0, 10, 20$ and $\Sigma = \mathbf{I}_d$. The sample size is $n = 500$, we use $450$ points to
estimate the Gaussian mixture parameters and $50$ points to test the
hypothesis. The dimension is $d = 1000$. Notice that when $a = 0$, the
distribution reduces to a single Gaussian distribution and as we take
larger $a$, the signal strength grows. The empirical distributions of
p-values, after $30$ realizations of the experiment, for \riftnospace,
Median RIFT (\mriftnospace) and SigClust are shown in Figure~\ref{Signal
  1d}. We notice that the SigClust has very good power for both $a =
10$ and $a = 20$, whereas the \riftspaces catch up for $a = 20$.

\begin{figure}[h]
\centering
\includegraphics[width=0.45\linewidth]{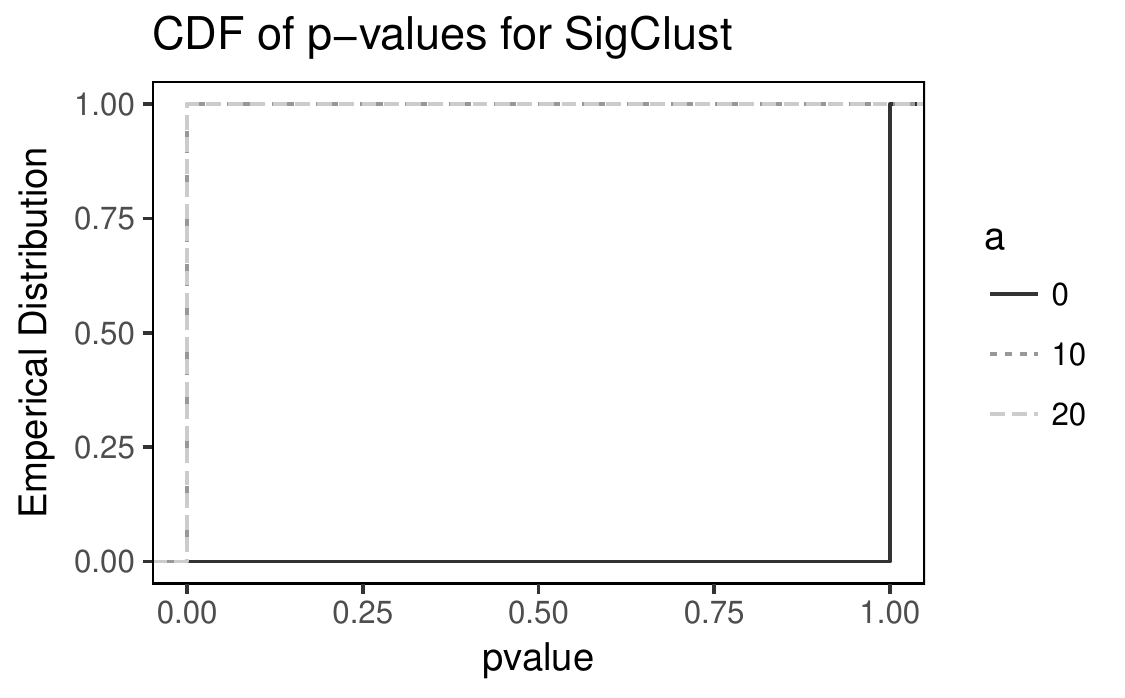}
\includegraphics[width=0.45\linewidth]{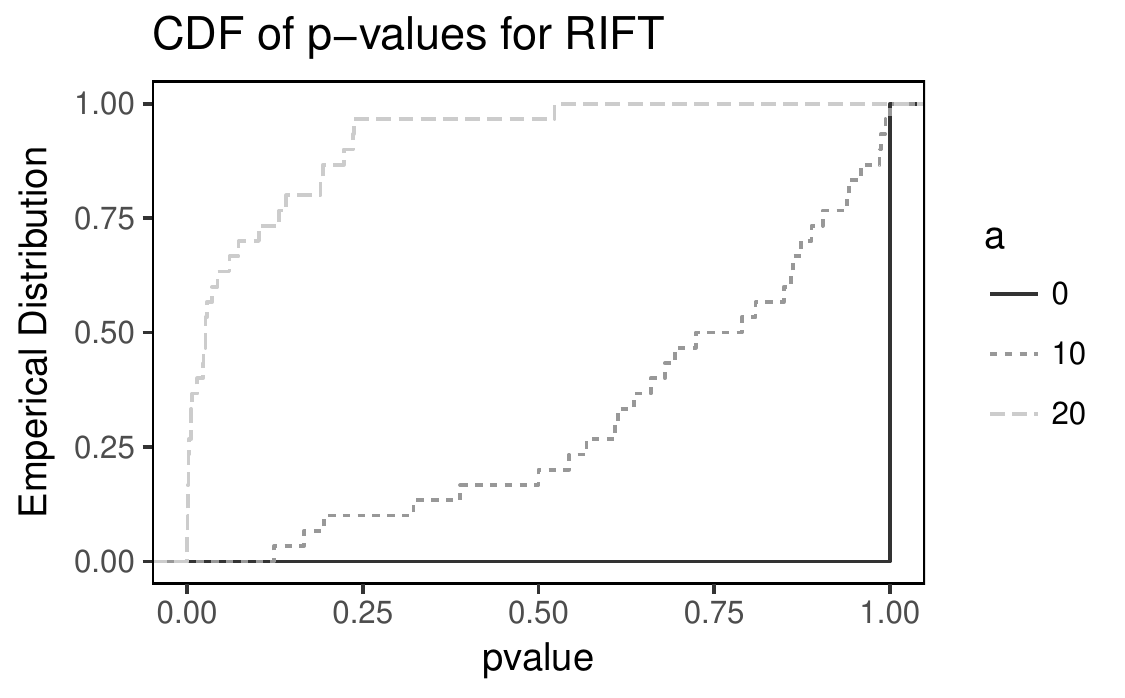}
\includegraphics[width=0.45\linewidth]{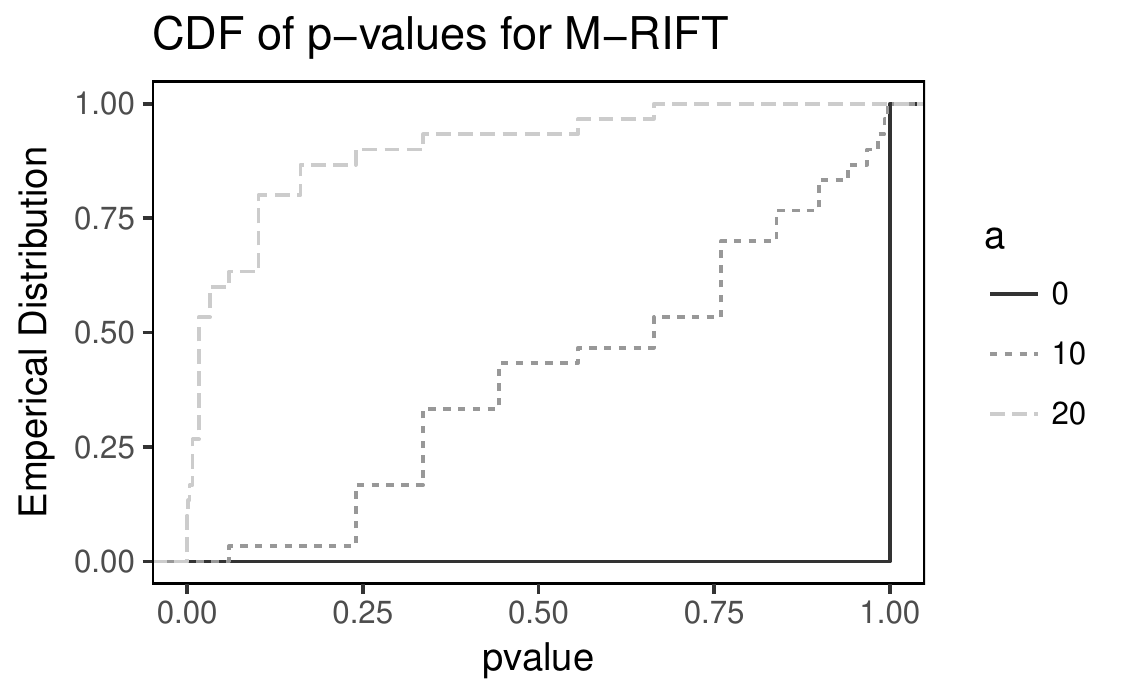}
\caption{Comparing the empirical distribution of the p-values when signal is exactly in one direction.}
\label{Signal 1d}
\end{figure}

\subsubsection{Signal in All Directions}
\begin{figure}[h]
\centering
\includegraphics[width=0.45\linewidth]{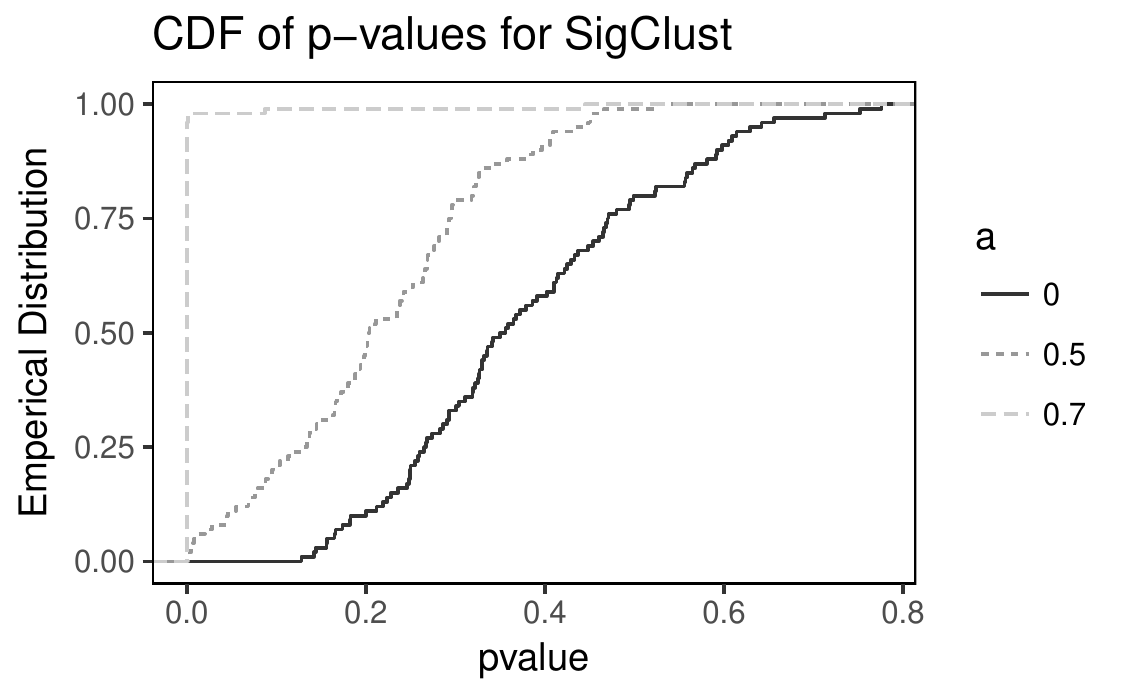}
\includegraphics[width=0.45\linewidth]{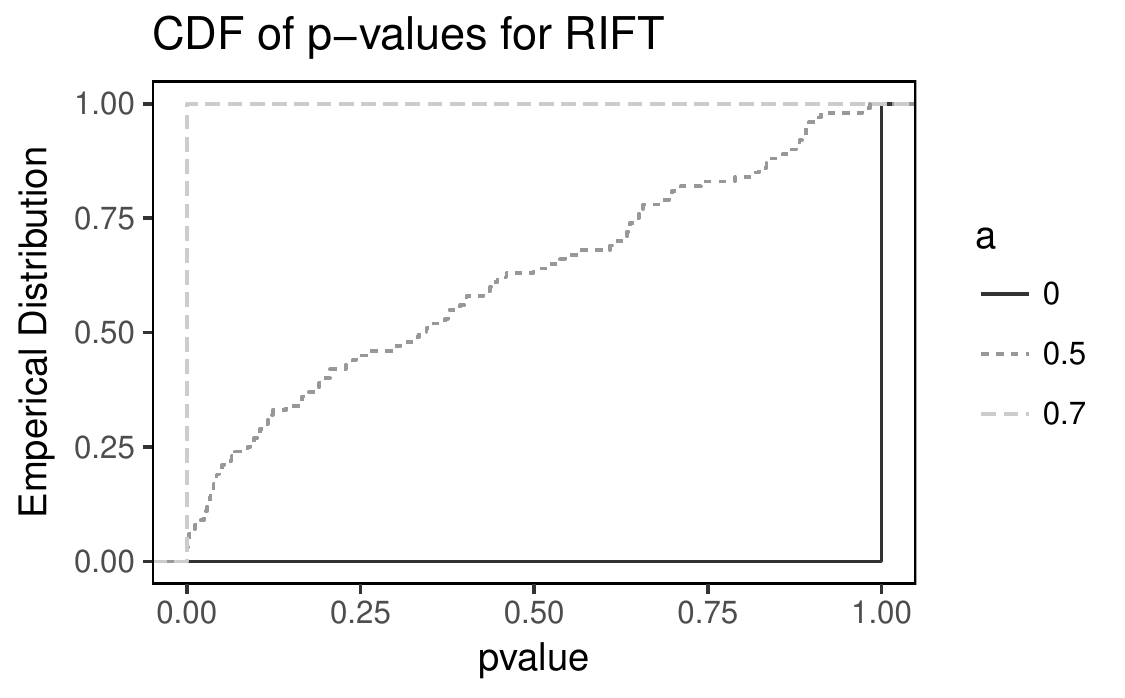}
\includegraphics[width=0.45\linewidth]{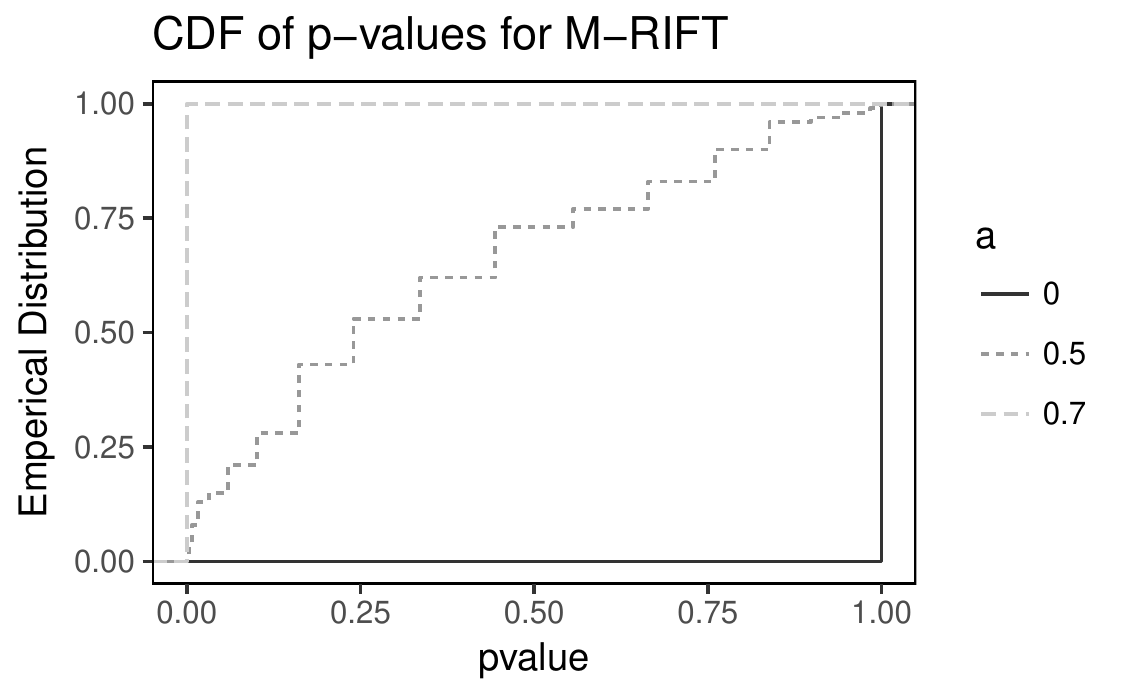}
\caption{Comparing the empirical distribution of the p-values when signal is in all directions.}
\label{Signal alld}
\end{figure}

Now we consider data with signal in all directions and
compare the tests.  We consider a mixture of two normal
distributions, $0.5N(0, \Sigma)+ 0.5N(\mu, \Sigma)$, where $\mu = (a,
a, \ldots , a)$ with $a = 0, 0.5, 0.7$ and $\Sigma$ is a diagonal
matrix with $\Sigma_{11} = 100$ and $\Sigma_{jj} = 1$ for $j = 2,
\ldots, d$. We consider a high-dimensional setting where 
the sample size is chosen to be $n = 100$ and the dimension is $d =
1000$. Figure~\ref{Signal alld} shows the p-values generated by each
of the tests. In this case, we see that all the tests perform
similarly well. SigClust performs only slightly better than the \riftnospaces.

\subsubsection{Example where SigClust Fails}

Finally, we compare the power of the \riftspaces with SigClust and Mardia's Kurtosis
test in detecting the signal in one direction if the variability in
another direction is very high. We consider a mixture of two normal
distributions, $0.5N(0, \Sigma)+ 0.5N(\mu, \Sigma)$, where $\mu = (a,
0, \ldots , 0)$ with $a = 0, 10, 20$ and $\Sigma$ is a diagonal matrix
with $\Sigma_{jj} = 400$ for $j = 2$ and $\Sigma_{jj} = 1$ for $j \neq
2$. That is, we are trying to detect the signal in the first dimension
while the variability in the second dimension is very high. The sample
size is $n = 100$ and dimension is $d = 5$.

\begin{figure}[h]
\centering
\includegraphics[width=0.45\linewidth]{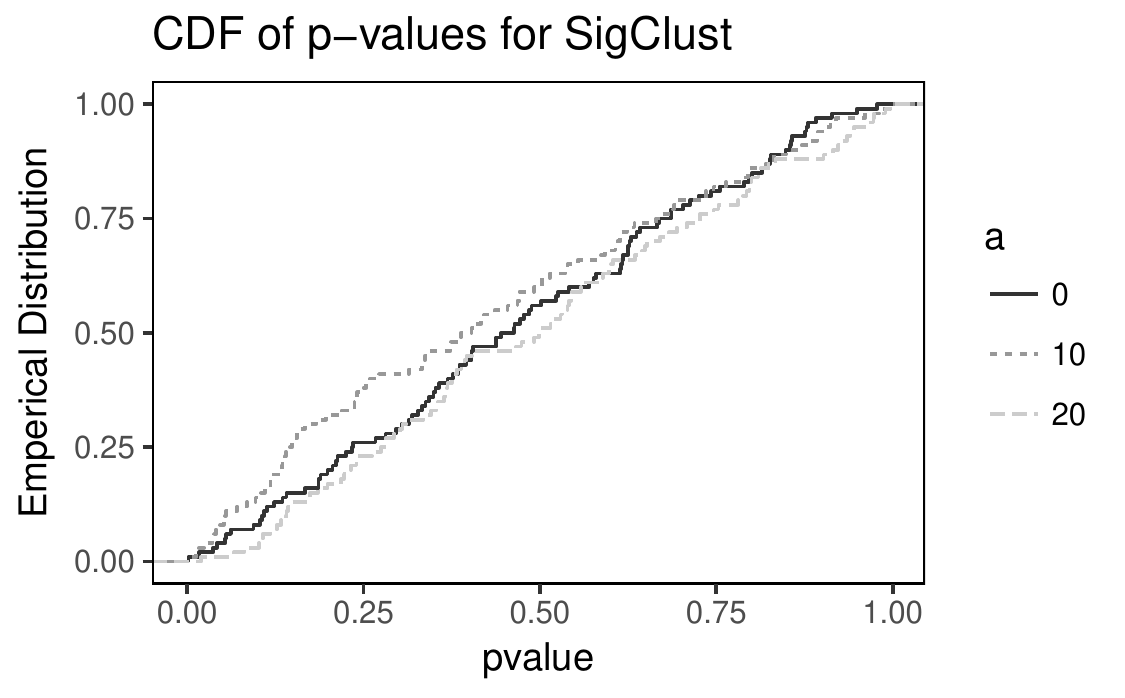}
\includegraphics[width=0.45\linewidth]{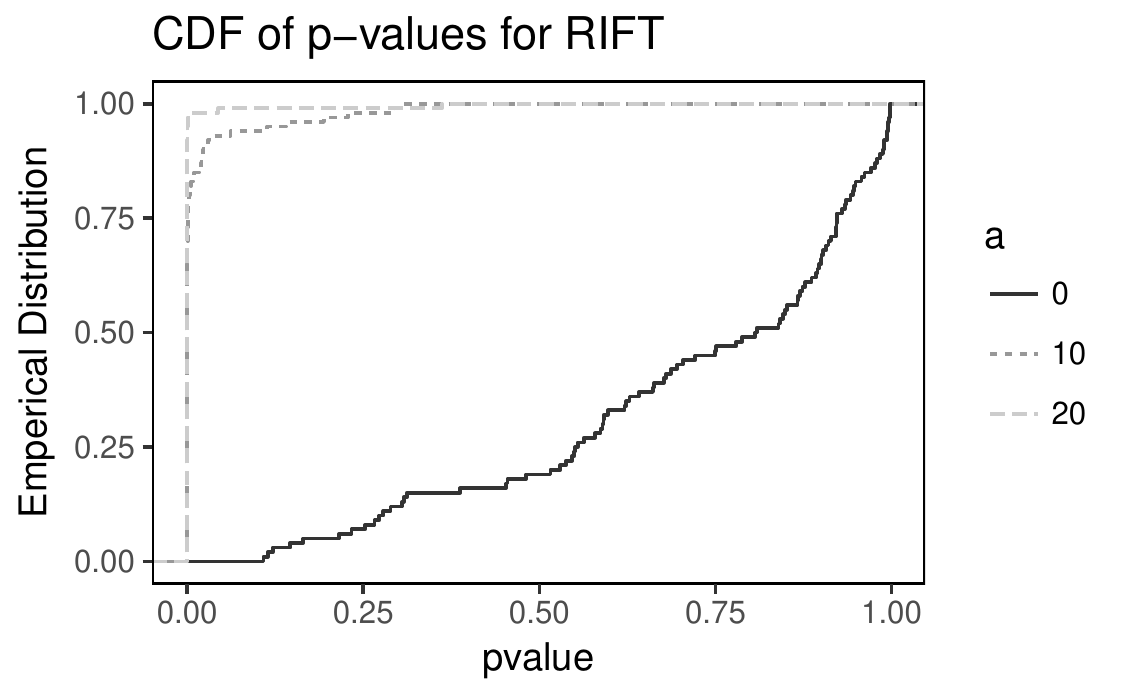}
\includegraphics[width=0.45\linewidth]{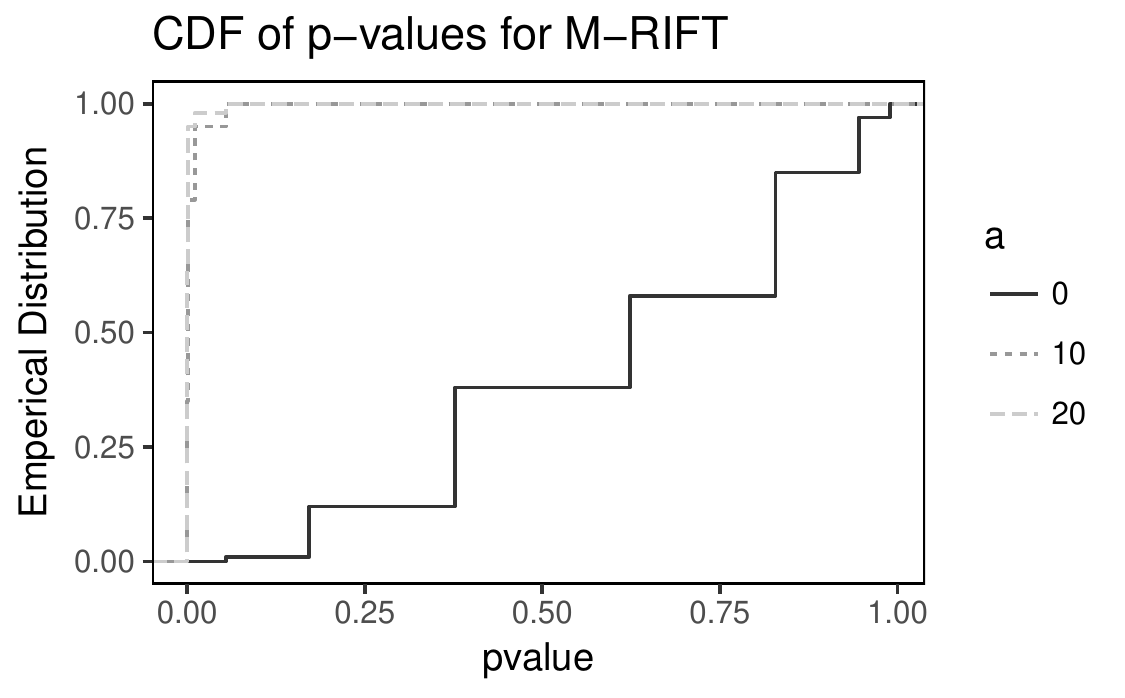}
\includegraphics[width=0.45\linewidth]{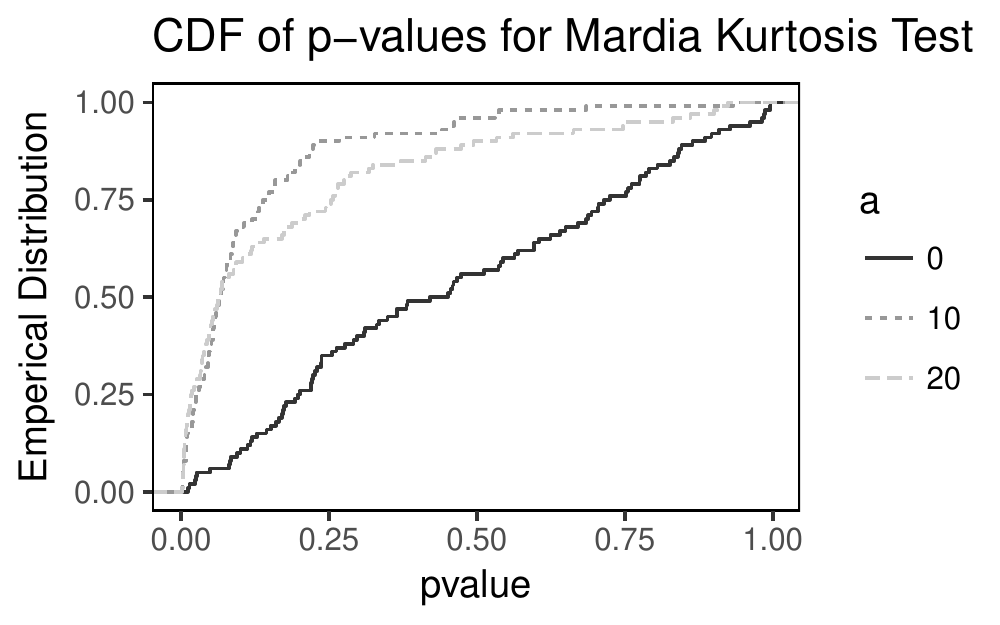}
\caption{Comparing the power of the tests with higher signal in one direction and high variability in another.}
\label{Highsignal 1d highvar 1d}
\end{figure}

The empirical distributions of the p-values are shown in Figure
\ref{Highsignal 1d highvar 1d}. We notice that SigClust has almost
no power in detecting the signal in one direction when there is high
variability in any other direction, whereas both the \riftspaces have high
power while controlling the type-I error. Mardia's Kurtosis test also has
higher power than SigClust but has lower power than the \riftnospaces.

\subsection{Hierarchical Clustering example: Four Cluster Setting (K = 4)}
\label{sec:simhierarchical}
In this section, we compare the tests in a hierarchical setting. We compare the \riftnospaces, SigClust and truncated SigClust, where for the SigClusts the
clustering is performed using k-means clustering with $k = 2$ and
mixture of Gaussians is used in the case of the \riftnospaces. We consider the
alternative setting in which observations are drawn from a mixture of
four clusters, each of which is a Gaussian distribution with
covariance matrix $\Sigma = {I}_d$. Our motive is to study how the different tests behave at each split in a hierarchical setting.

We compare the methods for two arrangements of the four Gaussian
components. In the first setting, the four components are placed at
the vertices of a square with side length $\delta$ and in the second
setting, the four components are placed at the vertices of a regular
tetrahedron with side length $\delta$. $50$ samples were drawn from
each of the Gaussian components for $100$ simulations.

\begin{table}[]
\caption{Comparing the different algorithms for hierarchical clustering when samples are generated from a mixture of four Gaussian distributions. The table gives the number of simulations that identify the particular number of significant clusters over 100 replications.}
\label{Four_cluster_example}
\centering
\begin{tabular}{ccccccccccccc}
\toprule
\toprule
 \multirow{2}{*}{Method} &  \multirow{2}{*}{Algorithm type} & & \multicolumn{3}{c}{Parameters} &  & \multicolumn{6}{c}{Number of clusters} \\  \cline{4-6} \cline{8-13}
 &  & & $d$ & $\delta$ & arr. &  & 1 & 2 & 3 & {\bf 4} & 5 & $\geq 6$ \\ \midrule
 \riftnospace & \multirow{4}{*}{Top-down} &  &  \multirow{4}{*}{2} & \multirow{4}{*}{4} & \multirow{4}{*}{square} & & 31 & 33 & 33 & {\bf 3} & 0 & 0 \\
\mriftnospace &  &  &  &  & & & 5 & 9 & 43 & {\bf 43} & 0 & 0 \\
SigClust &  &  &  &  & & & 63 & 0 &3  & {\bf 15} & 5 &14  \\
Trunc. SigClust &  &  &  &  & & & 63 & 0 & 8 & {\bf 8} & 7 & 14 \\ \midrule
\riftnospace & \multirow{4}{*}{Bottom-up} &  &  \multirow{4}{*}{2} & \multirow{4}{*}{4} & \multirow{4}{*}{square} & & 70 & 22 & 7 & {\bf 1} & 0 & 0 \\
\mriftnospace &  &  &  &  & & & 16 & 36 & 41 & {\bf 7} & 0 & 0 \\
SigClust &  &  &  &  & & & 65 & 0 &29  & {\bf 5} &1  & 0 \\
Trunc. SigClust &  &  &  &  & & & 57 & 0 & 30 & {\bf 5} & 8 & 0 \\ \midrule
 \riftnospace & \multirow{4}{*}{Top-down} &  &  \multirow{4}{*}{2} & \multirow{4}{*}{6} & \multirow{4}{*}{square} & & 1 & 3 & 22 & {\bf 74} & 0 & 0 \\
\mriftnospace &  &  &  &  & & & 0 & 0 & 4 & {\bf 96} & 0 & 0 \\
SigClust &  &  &  &  & & & 16 & 0 & 0  & {\bf 43} &13  & 28 \\
Trunc. SigClust &  &  &  &  & & & 16 & 0 & 0 & {\bf 46} & 14 & 24 \\ \midrule
\riftnospace & \multirow{4}{*}{Bottom-up} &  &  \multirow{4}{*}{2} & \multirow{4}{*}{6} & \multirow{4}{*}{square} & & 10 & 8 & 29 & {\bf 53} & 0 & 0 \\
\mriftnospace &  &  &  &  & & & 0 & 0 & 10 & {\bf 90} & 0 & 0 \\
SigClust &  &  &  &  & & & 77 & 0 &  0& {\bf 20} & 3 & 0 \\
Trunc. SigClust &  &  &  &  & & & 57 & 0 & 1 & {\bf 30} & 12 & 0 \\ \midrule
 \riftnospace & \multirow{4}{*}{Top-down} &  &  \multirow{4}{*}{3} & \multirow{4}{*}{4} & \multirow{4}{*}{tetrahedral} & & 1 & 5 & 27 & {\bf 67} & 0 & 0 \\
\mriftnospace &  &  &  &  & & & 0 & 0 & 5 & {\bf 95} & 0 & 0 \\
SigClust &  &  &  &  & & &  86& 0 & 1 & {\bf 5}  & 1 & 7 \\
Trunc. SigClust &  &  &  &  & & & 82 & 2 & 0 & {\bf 7} & 2 & 7 \\ \midrule
\riftnospace & \multirow{4}{*}{Bottom-up} &  &  \multirow{4}{*}{3} & \multirow{4}{*}{4} & \multirow{4}{*}{tetrahedral} & & 9 & 13 & 40 & {\bf 38} & 0 & 0 \\
\mriftnospace &  &  &  &  & & & 0 & 1 & 24 & {\bf 75} & 0 & 0 \\
SigClust &  &  &  &  & & & 58 & 0  & 24 & {\bf 8} &  10 & 0 \\
Trunc. SigClust &  &  &  &  & & & 50 & 0 & 23 & {\bf 12} & 15 & 0 \\ \midrule
 \riftnospace & \multirow{4}{*}{Top-down} &  &  \multirow{4}{*}{3} & \multirow{4}{*}{5} & \multirow{4}{*}{tetrahedral} & & 0 & 0 & 9 & {\bf 91} & 0 & 0 \\
\mriftnospace &  &  &  &  & & & 0 & 0 & 0 & {\bf 100} & 0 & 0 \\
SigClust &  &  &  &  & & & 71 & 0 & 0 & {\bf 7} & 4  & 18 \\
Trunc. SigClust &  &  &  &  & & & 72 & 2 & 0 & {\bf 8} & 5 & 13 \\ \midrule
\riftnospace & \multirow{4}{*}{Bottom-up} &  &  \multirow{4}{*}{3} & \multirow{4}{*}{5} & \multirow{4}{*}{tetrahedral} & & 0 & 0 & 27 & {\bf 73} & 0 & 0 \\
\mriftnospace &  &  &  &  & & & 0 & 0 & 1 & {\bf 99} & 0 & 0 \\
SigClust &  &  &  &  & & & 54 & 0 & 29 &{\bf 7}  & 10 & 0 \\
Trunc. SigClust &  &  &  &  & & & 48 & 0 & 26 & {\bf 11} & 15 & 0 \\ \midrule
\end{tabular}
\end{table}

For each of the simulations, we use the four tests \riftnospace, \mriftnospace, SigClust and truncated SigClust in the hierarchical setting and record the number of clusters given by each. Table \ref{Four_cluster_example}   gives the simulation results for some
values of $d$ and $\delta$. We notice that \mriftspace performs better than the
other tests in all the experiments. We also notice that the top-down
hierarchical algorithms tend to give more clusters than the bottom-up
hierarchical algorithms. In the case of \riftspace and \mriftnospace, we notice
that the top-down algorithms identify four as the correct number of
clusters more often than the bottom-up algorithms. In general, \mriftspace
and \riftspace identify four as the number of significant clusters present,
more often than SigClust or truncated SigClust.

\subsection{Sequential \riftnospace}
\label{sec:simsequential}

Now we compare the proposed sequential model selection approach (Sequential \riftspace
or S-\riftnospace) to AIC and BIC. We use two versions of the model
selection approach - one using the Kullback-Leibler distance and one using the $\ell_2$ distance between the estimated and the true densities. Using two simulated
experiments, we compare these methods to using AIC and BIC.

First, we reconsider the four cluster example used in the hierarchical
clustering setting where the four components are placed at the
vertices of a regular tetrahedron with side length $\delta$. $100$
samples are drawn from each of the Gaussian components for $100$
simulations. For each simulation, we use S-\riftspace with the two different distances - Kullback-Leibler distance and $\ell_2$ distance and record the number of clusters given by them. We also record the number of clusters that give the minimum AIC and BIC for each simulation. Table \ref{Model_selection_four_cluster_example} gives the results of the simulations. We
notice that S-\riftspace using Kullback-Leibler distance out-performs all the other methods. AIC performs very similar to it for $d = 10$ and $\delta = 10$, but we notice that for $d = 20$ and $\delta = 80$, S-\riftspace using Kullback-Leibler distance is the only one that detects the four clusters for some simulations.

\begin{table}
\caption{Comparing the different algorithms for selecting the ideal
  number of clusters when samples are generated from a mixture of four
  Gaussian distributions. The table gives the number of simulations
  that identify the particular number of significant clusters over 100
  replications.}
\label{Model_selection_four_cluster_example}
\centering
\begin{tabular}{cccccccccccc}
\toprule
\toprule
 \multirow{2}{*}{Method}  & & \multicolumn{3}{c}{Parameters} &  & \multicolumn{6}{c}{Number of clusters} \\  \cline{3-5} \cline{7-12}
 &  & $d$ & $\delta$ & arr. &  & 1 & 2 & 3 & {\bf 4} & 5 & $\geq 6$ \\ \midrule
 S-\riftspace (KL)  &  &  \multirow{4}{*}{10} & \multirow{4}{*}{6} & \multirow{4}{*}{Tetrahedral} & & 0 & 0 & 32 & {\bf 68} & 0 & 0 \\
S-\riftspace ($\ell_2$) &    &  &  & & & 60 & 22 & 11 & {\bf 7} & 0 & 0 \\
AIC &    &  &  & & & 0 & 0 & 46  & {\bf 54} & 0 &0  \\
BIC   &  &  &  & & & 1 & 41 & 58 & {\bf 0} & 0 & 0 \\ \midrule
S-\riftspace (KL)  &  &  \multirow{4}{*}{10} & \multirow{4}{*}{10} & \multirow{4}{*}{Tetrahedral} & & 0 & 0 & 5 & {\bf 93} & 2 & 0 \\
S-\riftspace ($\ell_2$) &    &  &  & & & 55 & 16 & 25 & {\bf 4} & 0 & 0 \\
AIC &    &  &  & & & 0 & 0 &7  & {\bf 93} & 0 &0  \\
BIC   &  &  &  & & & 0 & 0 & 99 & {\bf 1} & 0 & 0 \\ \midrule
S-\riftspace (KL)  &  &  \multirow{4}{*}{20} & \multirow{4}{*}{80} & \multirow{4}{*}{Tetrahedral} & & 0 & 4 & 86 & {\bf 10} & 0 & 0 \\
S-\riftspace ($\ell_2$) &    &  &  & & & 94 & 5 & 1 & {\bf 0} & 0 & 0 \\
AIC &    &  &  & & & 0 & 7 & 93  & {\bf 0} & 0 &0  \\
BIC   &  &  &  & & & 1 & 99 & 0 & {\bf 0} & 0 & 0 \\ \midrule
\end{tabular}
\end{table}

To further explore the properties of Sequential \riftnospace, we also study
a simulation with $10$ clusters. We generate $n$ data points from $10$ Gaussian components with means given by:
\begin{align*}
\mu_1 = (\veca , {\bf 0}, {\bf 0}, {\bf 0}, {\bf 0}), \hspace{1in} & \mu_6 = (-\veca , {\bf 0}, {\bf 0}, {\bf 0}, {\bf 0}),\\
\mu_2 = ({\bf 0}, \veca , {\bf 0}, {\bf 0}, {\bf 0}), \hspace{1in}& \mu_7 = ({\bf 0}, - \veca , {\bf 0}, {\bf 0}, {\bf 0}),\\
\mu_3 = ( {\bf 0}, {\bf 0}, \veca ,{\bf 0}, {\bf 0}), \hspace{1in}& \mu_8 = ( {\bf 0}, {\bf 0}, -\veca , {\bf 0}, {\bf 0}),\\
\mu_4 = ({\bf 0}, {\bf 0}, {\bf 0}, \veca , {\bf 0}), \hspace{1in}& \mu_9 = ({\bf 0}, {\bf 0}, {\bf 0}, -\veca , {\bf 0}),\\
\mu_5 = ({\bf 0}, {\bf 0}, {\bf 0}, {\bf 0}, \veca ), \hspace{1in}& \mu_6 = ({\bf 0}, {\bf 0}, {\bf 0}, {\bf 0}, -\veca ),
\end{align*}
where $\veca  = (a, a, \ldots, a)$ and ${\bf 0} = (0, 0, \ldots, 0)$
are vectors of length $p = d/5$. Each Gaussian component has mean
$0$ and variance $\sigma^2$. We generate $n/10$ data points from each
of the Gaussians.

We consider dimensions $d = 30$, so $p = 6$ and consider two values of
$n = 1000, 1500$. We vary the distance between the means by
considering two values of $a = 200, 500$ and consider three variances
$\sigma^2 = 0.001, 0.04, 0.16$. For each of the three variances, we
simulate $100$ samples and record the number of clusters given by
S-\riftnospace, AIC and BIC.

\begin{table}[]
\caption{Comparing the different algorithms for selecting the ideal number of clusters when samples are generated from a mixture of $10$ Gaussian distributions. The entries of the table give the numbers of simulations (out of a total of 100) for which a certain estimate of the number of clusters is obtained.}
\label{Model_selection_10_cluster_example}
\centering
\begin{tabular}{cccccccccccc}
\toprule
\toprule
 \multirow{2}{*}{Method}  & & \multicolumn{3}{c}{Parameters} &  & \multicolumn{6}{c}{Number of clusters} \\  \cline{3-5} \cline{7-12}
 &  & $n$ & $a$ & $\sigma^2$ &  & $\leq 5$ & 6 & 7 & 8 & 9 & {\bf 10} \\ \midrule
 S-\riftspace (KL)  &  &  \multirow{4}{*}{1000} & \multirow{4}{*}{200} & \multirow{4}{*}{0.001} & & 0 & 1 & 0 & 45 & 54 & {\bf 0} \\
S-\riftspace ($\ell_2$) &    &  &  & & & 100 & 0 & 0 & 0 & 0 & {\bf 0} \\
AIC &    &  &  & & & 0 & 1 & 1  & 52 & 46 & {\bf 0}  \\
BIC   &  &  &  & & & 13 & 3 & 46 & 38 & 0 & {\bf 0} \\ \midrule
S-\riftspace (KL)  &  &  \multirow{4}{*}{1000} & \multirow{4}{*}{200} & \multirow{4}{*}{0.04} & & 0 & 2 & 7 & 71 & 20 & {\bf 0} \\
S-\riftspace ($\ell_2$) &    &  &  & & & 100 & 0 & 0 & 0 & 0 & {\bf 0} \\
AIC &    &  &  & & & 2 & 0 & 7  & 84 & 7 & {\bf 0}  \\
BIC   &  &  &  & & & 100 & 0 & 0 & 0 & 0 & {\bf 0} \\ \midrule
S-\riftspace (KL)  &  &  \multirow{4}{*}{1000} & \multirow{4}{*}{200} & \multirow{4}{*}{0.16} & & 1 & 2 & 21 & 65 & 11 & {\bf 0} \\
S-\riftspace ($\ell_2$) &    &  &  & & & 100 & 0 & 0 & 0 & 0 & {\bf 0} \\
AIC &    &  &  & & & 3 & 0 & 22  & 75 & 0 & {\bf 0}  \\
BIC   &  &  &  & & & 100 & 0 & 0 & 0 & 0 & {\bf 0} \\ \midrule
S-\riftspace (KL)  &  &  \multirow{4}{*}{1500} & \multirow{4}{*}{200} & \multirow{4}{*}{0.001} & & 0 & 0 & 0 & 0 & 22 & {\bf 78} \\
S-\riftspace ($\ell_2$) &    &  &  & & & 96 & 3 & 1 & 0 & 0 & {\bf 0} \\
AIC &    &  &  & & & 0 & 0 & 0  & 0 & 67 & {\bf 33}  \\
BIC   &  &  &  & & & 0 & 0 & 0 & 0 & 100 & {\bf 0} \\ \midrule
S-\riftspace (KL)  &  &  \multirow{4}{*}{1500} & \multirow{4}{*}{200} & \multirow{4}{*}{0.04} & & 0 & 0 & 0 & 0 & 72 & {\bf 28} \\
S-\riftspace ($\ell_2$) &    &  &  & & & 93 & 2 & 0 & 5 & 0 & {\bf 0} \\
AIC &    &  &  & & & 0 & 0 & 0  & 0 & 100 & {\bf 0}  \\
BIC   &  &  &  & & & 0 & 0 & 0 & 77 & 23 & {\bf 0} \\ \midrule
S-\riftspace (KL)  &  &  \multirow{4}{*}{1500} & \multirow{4}{*}{200} & \multirow{4}{*}{0.16} & & 0 & 0 & 0 & 0 & 92 & {\bf 8} \\
S-\riftspace ($\ell_2$) &    &  &  & & & 95 & 2 & 3 & 0 & 0 & {\bf 0} \\
AIC &    &  &  & & & 0 & 0 & 0  & 0 & 100 & {\bf 0}  \\
BIC   &  &  &  & & & 0 & 0 & 0 & 100 & 0 & {\bf 0} \\ \midrule
S-\riftspace (KL)  &  &  \multirow{4}{*}{1500} & \multirow{4}{*}{500} & \multirow{4}{*}{0.001} & & 0 & 0 & 0 & 0 & 8 & {\bf 92} \\
S-\riftspace ($\ell_2$) &    &  &  & & & 96 & 3 & 1 & 0 & 0 & {\bf 0} \\
AIC &    &  &  & & & 0 & 0 & 0  & 0 & 31 & {\bf 69}  \\
BIC   &  &  &  & & & 0 & 0 & 0 & 0 & 100 & {\bf 0} \\ \midrule
S-\riftspace (KL)  &  &  \multirow{4}{*}{1500} & \multirow{4}{*}{500} & \multirow{4}{*}{0.04} & & 0 & 0 & 0 & 0 & 45 & {\bf 55} \\
S-\riftspace ($\ell_2$) &    &  &  & & & 94 & 1 & 0 & 5 & 0 & {\bf 0} \\
AIC &    &  &  & & & 0 & 0 & 0  & 0 & 95 & {\bf 5}  \\
BIC   &  &  &  & & & 0 & 0 & 0 & 4 & 96 & {\bf 0} \\ \midrule
S-\riftspace (KL)  &  &  \multirow{4}{*}{1500} & \multirow{4}{*}{500} & \multirow{4}{*}{0.16} & & 0 & 0 & 0 & 0 & 73 & {\bf 27} \\
S-\riftspace ($\ell_2$) &    &  &  & & & 95 & 2 & 3 & 0 & 0 & {\bf 0} \\
AIC &    &  &  & & & 0 & 0 & 0  & 0 & 97 & {\bf 3}  \\
BIC   &  &  &  & & & 0 & 0 & 0 & 55 & 45 & {\bf 0} \\ \midrule
\end{tabular}
\end{table}

The estimates of the number of clusters given by S-\riftnospace, AIC and BIC
are recorded in Table \ref{Model_selection_10_cluster_example}. We
notice that in every case S-\riftspace using Kullback-Leibler distance outperforms
all the other methods. AIC performs the next best. We notice that both
S-\riftspace using $\ell_2$ loss and BIC tend to under estimate the number of
clusters.

\subsection{Summary of the Simulations}

For two clusters which are separated in just one of the dimensions, if
the variance in the other dimensions isn't too large,
SigClust out-performs all the other methods for small sample sizes. 
\riftspace and Mardia's Kurtosis Test show comparable results. But when the
distance between the clusters is small, or when the variance in some
other dimension is much larger than the separation, SigClust
loses power completely and \riftspace and Mardia's Kurtosis Test
out-perform SigClust.
We also observe that as the dimension increases, 
\riftspace has lower power than the SigClust.

For the simulated examples that have more than two clusters, 
hierarchical clustering using
\riftspace detects the true number of clusters much better than
hierarchical clustering using SigClust. 
Finally, we notice that using S-\riftspace to detect the correct number of
clusters is better than minimizing the AIC or BIC. We also see that
the version using the Kullback-Leibler distance out-performs
the one using $\ell_2$ distance, which tends to under-estimate the number
of clusters.

\section{Application to Gene Expression Data}
\label{sec:gene}
To further compare the power of the \riftspaces to the power of the
SigClusts in the hierarchical setting, we apply the approach to a
cancer gene expression dataset. We consider a dataset consisting 
of three different
cancer types - head and neck squamous cell carcinoma (HNSC), lung
squamous cell carcinoma (LUSC) and lung adenocarcinoma (LUAD). Since
we have samples from three distinctively different cancers, we expect
the methods to be able to detect the presence of three different
clusters. We compare the clusterings given by hierarchical \riftspace and \mriftspace with
hierarchical SigClust at level $\alpha = 0.05$.

We combine data on $100$ tumor samples from each of HNSC, LUSC and
LUAD to create a data set of $300$ samples, similar to
\cite{kimes2017statistical}. The data is obtained from The Cancer
Genome Atlas (TCGA) project \citep{cancer2012comprehensive,cancer2014comprehensive}
whose RNA sequence data v2 is
available at \url{https://wiki.nci.nih.gov/display/TCGA/RNASeq+
  Version+2}. We used the R package TCGA2STAT
\citep{wan2015tcga2stat} to download the TCGA data into a format that
can be directly used for our statistical analysis.

There are a total of 20,501 genes of which we use the $500$ genes that
have the highest median absolute deviation (MAD) about the median. To
scale the data appropriately, we consider a log-transformation of the
data. In order to do so, first we replace all expression values that
are zero with the smallest non-zero expression value for all genes
over the data and then take a log-transformation.

\begin{table}[]
\caption{Clusterings given by RIFT and SigClust for the multi-cancer gene expression dataset.}
\label{Multi-cancer}
\centering
\begin{tabular}{cccc}
\toprule
\toprule
\multirow{2}{*}{True Classes} & \multicolumn{3}{c}{\riftspaces Classes} \\ \cmidrule(l){2-4} 
 & HNSC & LUSC & LUAD \\ \midrule
HNSC & 79 & 21 & 0 \\
LUSC & 7 & 70 & 23 \\
LUAD & 0 & 1 & 99 \\ \bottomrule
\end{tabular}
\qquad
\begin{tabular}{cccc}
\toprule
\toprule
\multirow{2}{*}{True Classes} & \multicolumn{3}{c}{SigClust Classes} \\ \cmidrule(l){2-4} 
 & HNSC & LUSC & LUAD \\ \midrule
HNSC & 90 & 10 & 0 \\
LUSC & 4 & 74 & 22 \\
LUAD & 0 & 1 & 99 \\ \bottomrule
\end{tabular}
\end{table}

SigClust was implemented with $1000$ simulations at every node. 
The top-down and the bottom-up versions of both \riftspace and \mriftspace
correctly give $3$ clusters. The top-down version of SigClust gives
$9$ clusters and the bottom-up version gives $5$ clusters. All the
algorithms first create a split between LUAD and the other two cancers
and then the next split separates HNSC and LUSC. Table
\ref{Multi-cancer} gives the clusterings given by the first two splits
for the \riftspaces and SigClust. Note that even though SigClust gives
better clusters, it splits all the clusters further into smaller clusters.

Hence, similar to the simulations with multiple clusters in Section~\ref{sec:simhierarchical}, 
in this case also hierarchical clustering using
\riftspace detects the true number of clusters much better than
hierarchical clustering using SigClust.

\section{Conclusion}
\label{section::conclusion}
We presented an analysis of the SigClust procedure of \citet{liu2012statistical}
in certain examples when the dimension $d$ was held fixed. 
On the other hand, increasing dimension was considered in the work of
\citet{liu2012statistical}, but only under restrictive conditions.
A more thorough understanding of the power of hypothesis testing based
approaches when $d$ increases is warranted.

We subsequently presented a different
hypothesis testing based approach for
clustering with mixtures of Normals based on \emph{relative fit}.
By testing the relative fit of different mixtures based on data splitting we get
a simple test statistic with a Normal limiting distribution.
As with any method, there are cases where the method works well
but there are also cases where it fails.
The main advantage of our approach
is that it uses a test with a simple limiting distribution
and the test does not rely on the assumption that the model is correct.

\bibliographystyle{apalike}
\bibliography{paper}

\begin{thebibliography}{}

\bibitem[Balakrishnan et~al., 2017]{balakrishnan2017statistical}
Balakrishnan, S., Wainwright, M.~J., Yu, B., et~al. (2017).
\newblock Statistical guarantees for the em algorithm: From population to
  sample-based analysis.
\newblock {\em The Annals of Statistics}, 45(1):77--120.

\bibitem[Bickel and Breiman, 1983]{bickel1983sums}
Bickel, P.~J. and Breiman, L. (1983).
\newblock Sums of functions of nearest neighbor distances, moment bounds, limit
  theorems and a goodness of fit test.
\newblock {\em The Annals of Probability}, pages 185--214.

\bibitem[Bock, 1985]{Bock1985}
Bock, H.~H. (1985).
\newblock On some significance tests in cluster analysis.
\newblock {\em Journal of Classification}, 2(1):77--108.

\bibitem[Boyd and Vandenberghe, 2004]{boyd2004convex}
Boyd, S. and Vandenberghe, L. (2004).
\newblock {\em Convex optimization}.
\newblock Cambridge university press.

\bibitem[Charnigo and Sun, 2004]{charnigo2004testing}
Charnigo, R. and Sun, J. (2004).
\newblock Testing homogeneity in a mixture distribution via the l 2 distance
  between competing models.
\newblock {\em Journal of the American Statistical Association},
  99(466):488--498.

\bibitem[Chen, 2017]{chen2017finite}
Chen, J. (2017).
\newblock On finite mixture models.
\newblock {\em Statistical Theory and Related Fields}, 1(1):15--27.

\bibitem[Chen et~al., 2009]{chen2009hypothesis}
Chen, J., Li, P., et~al. (2009).
\newblock Hypothesis test for normal mixture models: The em approach.
\newblock {\em The Annals of Statistics}, 37(5A):2523--2542.

\bibitem[Chen et~al., 2012]{chen2012inference}
Chen, J., Li, P., and Fu, Y. (2012).
\newblock Inference on the order of a normal mixture.
\newblock {\em Journal of the American Statistical Association},
  107(499):1096--1105.

\bibitem[Dacunha-Castelle et~al., 1999]{dacunha1999testing}
Dacunha-Castelle, D., Gassiat, E., et~al. (1999).
\newblock Testing the order of a model using locally conic parametrization:
  population mixtures and stationary arma processes.
\newblock {\em The Annals of Statistics}, 27(4):1178--1209.

\bibitem[Engelman and Hartigan, 1969]{engelman1969percentage}
Engelman, L. and Hartigan, J.~A. (1969).
\newblock Percentage points of a test for clusters.
\newblock {\em Journal of the American Statistical Association},
  64(328):1647--1648.

\bibitem[Fraley and Raftery, 2002]{fraley2002model}
Fraley, C. and Raftery, A.~E. (2002).
\newblock Model-based clustering, discriminant analysis, and density
  estimation.
\newblock {\em Journal of the American statistical Association},
  97(458):611--631.

\bibitem[Garcia-Escudero et~al., 2009]{garcia2009robust}
Garcia-Escudero, L., Gordaliza, A., Matran, C., and Mayo-Iscar, A. (2009).
\newblock A robust maximal f-ratio statistic to detect clusters structure.
\newblock {\em Communications in Statistics-Theory and Methods},
  38(5):682--694.

\bibitem[Gassiat, 2002]{gassiat2002likelihood}
Gassiat, E. (2002).
\newblock Likelihood ratio inequalities with applications to various mixtures.
\newblock {\em Annales de l'Institut Henri Poincare (B) Probability and
  Statistics}, 38:897--906.

\bibitem[Ghosh and Sen, 1984]{ghosh1984asymptotic}
Ghosh, J.~K. and Sen, P.~K. (1984).
\newblock On the asymptotic performance of the log likelihood ratio statistic
  for the mixture model and related results.
\newblock {\em Berkeley Conference In Honor of Jerzy Neyman and Jack Kiefer}.

\bibitem[Gu et~al., 2017]{gu2017testing}
Gu, J., Koenker, R., and Volgushev, S. (2017).
\newblock Testing for homogeneity in mixture models.
\newblock {\em Econometric Theory}, pages 1--46.

\bibitem[Hartigan, 1978]{hartigan1978asymptotic}
Hartigan, J. (1978).
\newblock Asymptotic distributions for clustering criteria.
\newblock {\em The Annals of Statistics}, pages 117--131.

\bibitem[Hartigan, 1985]{hartigan1985failure}
Hartigan, J. (1985).
\newblock A failure of likelihood asymptotics for normal mixtures.
\newblock In {\em Proc. Barkeley Conference in Honor of J. Neyman and J.
  Kiefer}, volume~2, pages 807--810.

\bibitem[Hartigan, 1975]{hartigan1975clustering}
Hartigan, J.~A. (1975).
\newblock {\em Clustering algorithms}.
\newblock Wiley.

\bibitem[Huang et~al., 2015]{huang2015statistical}
Huang, H., Liu, Y., Yuan, M., and Marron, J. (2015).
\newblock Statistical significance of clustering using soft thresholding.
\newblock {\em Journal of Computational and Graphical Statistics},
  24(4):975--993.

\bibitem[Kimes et~al., 2017]{kimes2017statistical}
Kimes, P.~K., Liu, Y., Neil~Hayes, D., and Marron, J.~S. (2017).
\newblock Statistical significance for hierarchical clustering.
\newblock {\em Biometrics}, 73(3):811--821.

\bibitem[Lee, 1979]{lee1979multivariate}
Lee, K.~L. (1979).
\newblock Multivariate tests for clusters.
\newblock {\em Journal of the American Statistical Association},
  74(367):708--714.

\bibitem[Li and Chen, 2010]{li2010testing}
Li, P. and Chen, J. (2010).
\newblock Testing the order of a finite mixture.
\newblock {\em Journal of the American Statistical Association},
  105(491):1084--1092.

\bibitem[Liu and Shao, 2004]{liu2004asymptotics}
Liu, X. and Shao, Y. (2004).
\newblock Asymptotics for the likelihood ratio test in a two-component normal
  mixture model.
\newblock {\em Journal of Statistical Planning and Inference}, 123(1):61--81.

\bibitem[Liu et~al., 2008]{liu2008statistical}
Liu, Y., Hayes, D.~N., Nobel, A., and Marron, J. (2008).
\newblock Statistical significance of clustering for high-dimension,
  low--sample size data.
\newblock {\em Journal of the American Statistical Association},
  103(483):1281--1293.

\bibitem[Liu et~al., 2012]{liu2012statistical}
Liu, Y., Hayes, D.~N., Nobel, A., and Marron, J.~S. (2012).
\newblock Statistical significance of clustering for high-dimension,
  low--sample size data.
\newblock {\em Journal of the American Statistical Association}.

\bibitem[Maitra et~al., 2012]{maitra2012bootstrapping}
Maitra, R., Melnykov, V., and Lahiri, S.~N. (2012).
\newblock Bootstrapping for significance of compact clusters in
  multidimensional datasets.
\newblock {\em Journal of the American Statistical Association},
  107(497):378--392.

\bibitem[Mardia, 1970]{mardia1970measures}
Mardia, K.~V. (1970).
\newblock Measures of multivariate skewness and kurtosis with applications.
\newblock {\em Biometrika}, 57(3):519--530.

\bibitem[Mardia, 1974]{mardia1974applications}
Mardia, K.~V. (1974).
\newblock Applications of some measures of multivariate skewness and kurtosis
  in testing normality and robustness studies.
\newblock {\em Sankhy{\=a}: The Indian Journal of Statistics, Series B}, pages
  115--128.

\bibitem[McLachlan and Peel, 2004]{mclachlan2004finite}
McLachlan, G. and Peel, D. (2004).
\newblock {\em Finite mixture models}.
\newblock John Wiley \& Sons.

\bibitem[McLachlan and Rathnayake, 2014]{mclachlan2014number}
McLachlan, G.~J. and Rathnayake, S. (2014).
\newblock On the number of components in a gaussian mixture model.
\newblock {\em Wiley Interdisciplinary Reviews: Data Mining and Knowledge
  Discovery}, 4(5):341--355.

\bibitem[McShane et~al., 2002]{mcshane2002methods}
McShane, L.~M., Radmacher, M.~D., Freidlin, B., Yu, R., Li, M.-C., and Simon,
  R. (2002).
\newblock Methods for assessing reproducibility of clustering patterns observed
  in analyses of microarray data.
\newblock {\em Bioinformatics}, 18(11):1462--1469.

\bibitem[Milligan and Cooper, 1985]{milligan1985examination}
Milligan, G.~W. and Cooper, M.~C. (1985).
\newblock An examination of procedures for determining the number of clusters
  in a data set.
\newblock {\em Psychometrika}, 50(2):159--179.

\bibitem[Network et~al., 2012]{cancer2012comprehensive}
Network, C. G. A.~R. et~al. (2012).
\newblock Comprehensive genomic characterization of squamous cell lung cancers.
\newblock {\em Nature}, 489(7417):519.

\bibitem[Network et~al., 2014]{cancer2014comprehensive}
Network, C. G. A.~R. et~al. (2014).
\newblock Comprehensive molecular profiling of lung adenocarcinoma.
\newblock {\em Nature}, 511(7511):543.

\bibitem[Pollard, 1982]{Pollard1982}
Pollard, D. (1982).
\newblock A central limit theorem for k-means clustering.
\newblock {\em The Annals of Probability}, 10(4):919--926.

\bibitem[Qiu, 2010]{QIU20101701}
Qiu, D. (2010).
\newblock A comparative study of the k-means algorithm and the normal mixture
  model for clustering: Bivariate homoscedastic case.
\newblock {\em Journal of Statistical Planning and Inference}, 140(7):1701 --
  1711.

\bibitem[Rosenbaum, 1961]{rosenbaum1961moments}
Rosenbaum, S. (1961).
\newblock Moments of a truncated bivariate normal distribution.
\newblock {\em Journal of the Royal Statistical Society: Series B
  (Methodological)}, 23(2):405--408.

\bibitem[Rousseeuw, 1987]{rousseeuw1987silhouettes}
Rousseeuw, P.~J. (1987).
\newblock Silhouettes: a graphical aid to the interpretation and validation of
  cluster analysis.
\newblock {\em Journal of computational and applied mathematics}, 20:53--65.

\bibitem[Suzuki and Shimodaira, 2006]{suzuki2006pvclust}
Suzuki, R. and Shimodaira, H. (2006).
\newblock Pvclust: an r package for assessing the uncertainty in hierarchical
  clustering.
\newblock {\em Bioinformatics}, 22(12):1540--1542.

\bibitem[Tibshirani and Walther, 2005]{tibshirani2005cluster}
Tibshirani, R. and Walther, G. (2005).
\newblock Cluster validation by prediction strength.
\newblock {\em Journal of Computational and Graphical Statistics},
  14(3):511--528.

\bibitem[Tibshirani et~al., 2001]{tibshirani2001estimating}
Tibshirani, R., Walther, G., and Hastie, T. (2001).
\newblock Estimating the number of clusters in a data set via the gap
  statistic.
\newblock {\em Journal of the Royal Statistical Society: Series B (Statistical
  Methodology)}, 63(2):411--423.

\bibitem[Vogt and Schmid, 2017]{vogt2017clustering}
Vogt, M. and Schmid, M. (2017).
\newblock Clustering with statistical error control.
\newblock {\em arXiv preprint arXiv:1702.02643}.

\bibitem[Wan et~al., 2015]{wan2015tcga2stat}
Wan, Y.-W., Allen, G.~I., and Liu, Z. (2015).
\newblock Tcga2stat: simple tcga data access for integrated statistical
  analysis in r.
\newblock {\em Bioinformatics}, 32(6):952--954.

\bibitem[Zhou and Jammalamadaka, 1993]{zhou1993goodness}
Zhou, S. and Jammalamadaka, S.~R. (1993).
\newblock Goodness of fit in multidimensions based on nearest neighbour
  distances.
\newblock {\em Journal of Nonparametric Statistics}, 2(3):271--284.

\end{thebibliography}

\newpage

\begin{appendices}

\section{Hierarchical Clustering Algorithm Descriptions}

\begin{figure}[h]
\centering
\includegraphics[width=0.9\linewidth]{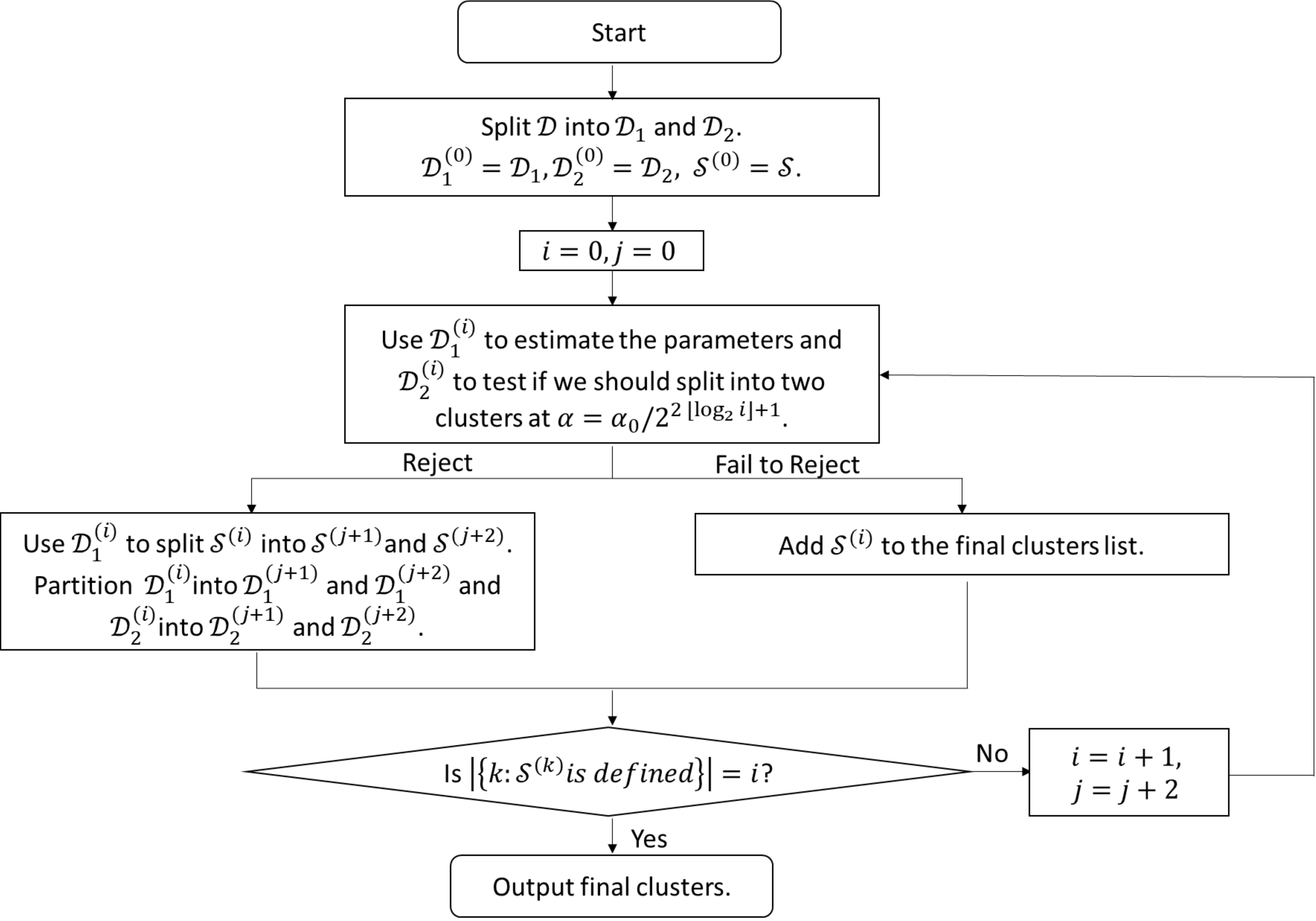}
\caption{Top-down cluster tree algorithm.}
\label{TopDownFlowchart}
\end{figure}

\begin{figure}[h]
\centering
\includegraphics[width=0.9\linewidth]{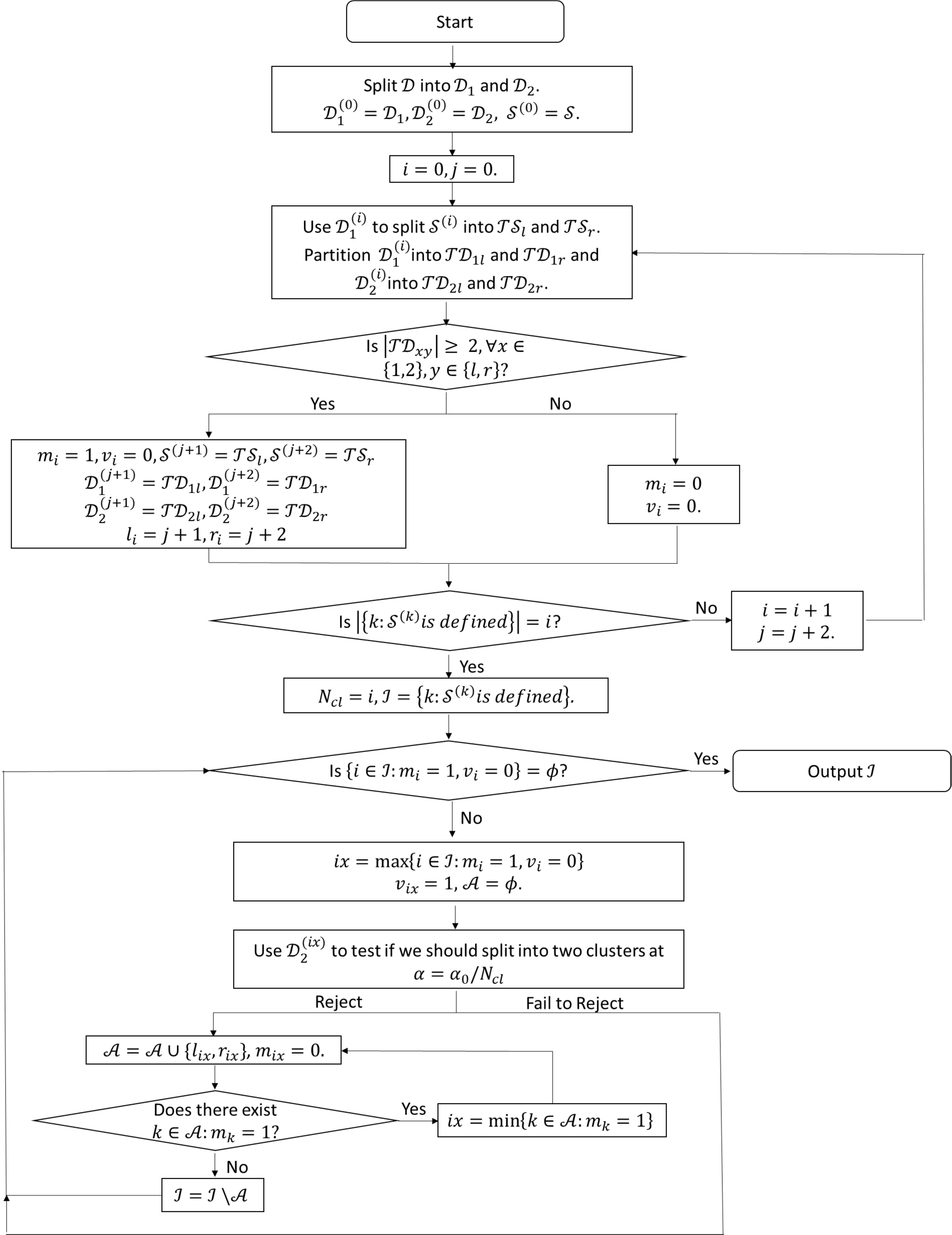}
\caption{Bottom-up cluster tree algorithm.}
\label{BottomUpFlowchart}
\end{figure}

In this Appendix,
we summarize the algorithms.
The summaries are in
Figures
\ref{fig::top-down}
\ref{TopDownFlowchart}, 
\ref{fig::bottom-up} and
\ref{BottomUpFlowchart}.

\begin{figure}[h]
\fbox{\parbox{6in}{
{\bf Top-Down Algorithm:}
\begin{enumerate}
\item Split the data into two halves $\mathcal{D}_1$ and $\mathcal{D}_2$. We estimate the parameters using $\mathcal{D}_1$ and create the tree using $\mathcal{D}_2$.
\item Set $\mathcal{D}_1^{T^{(0)}} = \mathcal{D}_1$, $\mathcal{D}_2^{T^{(0)}} = \mathcal{D}_2$. Use $\mathcal{D}_1^{T^{(0)}}$ to estimate the parameters required and $\mathcal{D}_2^{T^{(0)}}$ to test the null hypothesis that the data comes from a single cluster using the algorithms mentioned before. 
\item If you accept the test at level $\alpha/2$, stop. If you reject the test at level $\alpha/2$, partition the sample space into two pieces $T_1^{(1)}$ and $T_2^{(1)}$. The partition also partitions $\mathcal{D}_1$ and $\mathcal{D}_2$, say into $\mathcal{D}_1^{T_1^{(1)}}$ and $\mathcal{D}_1^{T_1^{(1)}}$ and $\mathcal{D}_2^{T_2^{(1)}}$ and $\mathcal{D}_2^{T_2^{(1)}}$ respectively. Set $\text{Depth} = 1$.
\item If $\left\lbrace i: T_i^{(\text{Depth})} \text{ is defined} \right\rbrace$ is not an empty set, then for every $i \in \left\lbrace i: T_i^{(\text{Depth})} \text{ is defined} \right\rbrace$:
\begin{itemize}
\item[(a)] Set $T = T_i^{(\text{Depth})}$.
\item[(b)] Use the corresponding sample sets $\mathcal{D}_1^{T}$ to estimate the parameters required and $\mathcal{D}_2^{T}$ to test the null hypothesis that the data comes from a single cluster using the algorithms mentioned before.
\item[(c)] If you reject the test at level $\alpha/2^{2 \text{Depth} + 1}$, partition the sample space $T$
into two pieces $T_{2i - 1}^{(\text{Depth} + 1)}$ and $T_{2i}^{(\text{Depth} + 1)}$. Also partition $\mathcal{D}_1^{T}$ and $\mathcal{D}_2^{T}$, say into $\mathcal{D}_1^{T_{2i - 1}^{(\text{Depth} + 1)}}$ and $\mathcal{D}_1^{T_{2i}^{(\text{Depth} + 1)}}$ and $\mathcal{D}_2^{T_{2i - 1}^{(\text{Depth} + 1)}}$ and $\mathcal{D}_2^{T_{2i}^{(\text{Depth} + 1)}}$ respectively.
\end{itemize}
  
\item Set $\text{Depth} = \text{Depth} + 1$ and repeat step 4 till $\left\lbrace i: T_i^{(\text{Depth})} \text{ is defined} \right\rbrace$ is an empty set.
\end{enumerate}
}}
\caption{\em \riftspace applied in a top-down manner.}
\label{fig::top-down}
\end{figure}

\begin{figure}
\fbox{\parbox{6in}{
{\bf Bottom-Up Algorithm:}
\begin{enumerate}
\item Split the data into two halves $\mathcal{D}_1$ and $\mathcal{D}_2$. We estimate the parameters using $\mathcal{D}_1$ and create the tree. Then using $\mathcal{D}_2$ we prune the tree by testing the significance of every split.
\item Let the entire sample space be denoted by $T_1^{(0)}$. Set $\mathcal{D}_1^{T_1^{(0)}} = \mathcal{D}_1$, $\mathcal{D}_2^{T_1^{(0)}} = \mathcal{D}_2$. Set $\text{Depth} = 0$.
\item If $\left\lbrace i: T_i^{(\text{Depth})} \text{ is defined} \right\rbrace$ is not an empty set, then for every $i \in \left\lbrace i: T_i^{(\text{Depth})} \text{ is defined} \right\rbrace$:
\begin{itemize}
\item[(a)] Set $T = T_i^{(\text{Depth})}$.
\item[(b)] Use the corresponding sample sets $\mathcal{D}_1^{T}$ to estimate a mixture of two Gaussians and use that to partition the sample space $T$
into two pieces $T_{2i - 1}^{(\text{Depth} + 1)}$ and $T_{2i}^{(\text{Depth} + 1)}$. Also partition $\mathcal{D}_1^{T}$ and $\mathcal{D}_2^{T}$, say into $\mathcal{D}_1^{T_{2i - 1}^{(\text{Depth} + 1)}}$ and $\mathcal{D}_1^{T_{2i}^{(\text{Depth} + 1)}}$ and $\mathcal{D}_2^{T_{2i - 1}^{(\text{Depth} + 1)}}$ and $\mathcal{D}_2^{T_{2i}^{(\text{Depth} + 1)}}$ respectively. 
\end{itemize}
\item Set $\text{Depth} = \text{Depth} + 1$ and repeat step 3 till $\left\lbrace i: T_i^{(\text{Depth})} \text{ is defined} \right\rbrace$ is an empty set.
\item If the total number of nodes in the tree is given by $N_{nodes}$, then set $\text{Depth} = \text{Depth} - 1$ and for every $i$ such that $\{T_{2i - 1}^{(\text{Depth})}, T_{2i}^{(\text{Depth})}\}$ is not empty:
\begin{itemize}
\item[(a)] Set $T = T_i^{(\text{Depth} - 1)}$.
\item[(b)] Use the sample set $\mathcal{D}_2^{T}$ to test the null hypothesis that the data comes from a single cluster using the algorithms mentioned before. If you fail to reject the test at level $\alpha/N_{nodes}$, then delete $T_{2i - 1}^{(\text{Depth})}$ and $T_{2i}^{(\text{Depth})}$. Also delete their sub-trees.
\end{itemize}
 \item Set $\text{Depth} = \text{Depth} - 1$ and repeat step 5 till $\text{Depth} = 0$.
\end{enumerate}
}}
\caption{\em \riftspace applied in a bottom-up manner.}
\label{fig::bottom-up}
\end{figure}

\clearpage

\section{Proof of results presented under the null hypothesis of SigClust}

In this Appendix, we prove Theorem~\ref{thm:null} and all the results required to prove it. We first note that the regularity conditions ((ii), (iii) and (iv)) of \cite{Pollard1982} and hence of Corollary 6.5 in \cite{Bock1985} are satisfied by a $N(0,\Sigma)$ distribution. Furthermore, the $2$-means solution is unique, under the conditions on $\Sigma$. Additionally, Lemma \ref{lemma::verify} in Appendix~\ref{app:GammaNull} shows that (v) holds. Thus it follows from Pollard's result that
\begin{align*}
\sqrt{n}(\vecbn - \vecmu) \rightsquigarrow N(0, G_0^{-1} V G_0^{-1}),
\end{align*}
where $\vecmu$ is the vector that minimizes the population within
cluster sum of squares for the 2-means clustering, $V$ is the $kd \times kd$ diagonal matrix with
\begin{equation}
V_i = 4 \mathbb{E}\left[ (X - \mu_i)(X - \mu_i)^T \mathbb{I}_{A_i} \right]
\end{equation}
as its $i$th diagonal block and $G_0$ is analogously defined to the $G$ as defined in equation (\ref{eqn::Gamma}) for the alternative. So, $G_0$ is a matrix made up of $d \times
d$ matrixes of the form,
\begin{equation}
\label{eqn::GammaNull}
(G_{0})_{ij} = \left\lbrace \begin{array}{c c}
2\mathbb{P}(A_i) \mathbf{I}_d - 2  r_{i j}^{-1} \int_{M_{i j}} f(x) (x - \mu_i) (x - \mu_i)^T \ d\sigma(x) & \mbox{ for } \ \ i = j\\
-2 r_{ij}^{-1}  \int_{M_{i j}} f(x) (x - \mu_i) (x - \mu_j)^T \ d\sigma(x) & \mbox{ for } \ \ i \neq j,
\end{array} \right.
\end{equation}
for $i, j \in \{ 1, 2 \}$ where $r_{ij} = \| \mu_i - \mu_j\|$, $f(\cdot)$ is the corresponding density function and $\sigma(\cdot)$ is the $(d - 1)$ dimensional Lebesgue measure. Here $A_i$ denotes the set of points in $\mathbb{R}^d$ closer to $\mu_i$ than to any other $\mu_j$, $M_{ij}$ denotes the face common to $A_i$ and $A_j$ and $\mathbf{I}_d$ denotes the $d \times d$ identity matrix. We show that $G_0$ is positive definite in Lemma~\ref{lemma::verify}.

\noindent Now using Corollary 6.5 in \cite{Bock1985}, Lemma~\ref{lem:bock} follows immediately. In the next section, Appendix~\ref{app:claims}, we prove Claims~\eqref{eqn:muopt} and~\eqref{eqn:tausq} which together with Lemma~\ref{lem:bock} give Theorem~\ref{thm:null}.

\subsection{Proof of Claims~\eqref{eqn:muopt} and~\eqref{eqn:tausq}}
\label{app:claims}

\vspace{0.2cm}

\noindent {\bf Proof of Claim~\eqref{eqn:muopt}: } The vector $\vecmu$ that minimizes the population within
cluster sum of squares for the 2-means clustering has components given by 
\begin{align*}
\mu_1 &= \left(-\sigma_1 \sqrt{\frac{2}{\pi}}, 0, \ldots, 0\right)^T,~~~~\text{and} \\
\mu_2 &= \left(\sigma_1 \sqrt{\frac{2}{\pi}}, 0, \ldots, 0\right)^T.
\end{align*}
The corresponding (optimal) population clusters are 
\begin{align*}
A_1 &= \{x = (x_1, \ldots, x_d) \in
\mathbb{R}^d : x_1 \leq 0 \}~~~~\text{and}, \\
A_2 &= \{x = (x_1, \ldots, x_d) \in
\mathbb{R}^d : x_1 \geq 0 \}. 
\end{align*}
Thus, it follows that,
\begin{align*}
W(\vecmu) &= \mathbb{E}\left[\|X - \mu_1\|^2 \mathbb{I}_{\{X_1 < 0 \}}\right] + \mathbb{E}\left[\|X - \mu_2\|^2 \mathbb{I}_{\{X_1 > 0 \}}\right] \\
&= 2 \mathbb{E}\left[\|X - \mu_1\|^2 \mathbb{I}_{\{X_1 < 0 \}}\right]\\
&= 2 \left( \mathbb{E}\left[(X_1 - \mu_{11})^2 \mathbb{I}_{\{X_1 < 0 \}}\right] + \sum_{i = 2}^d \mathbb{E}\left[X_i^2 \mathbb{I}_{\{X_1 < 0 \}}\right] \right)\\
&= 2 \left( \frac{\sigma_1^2 }{2} \left(1 - \frac{2}{\pi} \right) + \sum_{i = 2}^d \frac{\sigma_i^2}{2} \right)\\
&= \sum_{i = 1}^d \sigma_i^2 - \frac{2 \sigma_1^2}{\pi},
\end{align*}
which yields Claim~\eqref{eqn:muopt}.

\vspace{0.2cm}

\noindent {\bf Proof of Claim~\eqref{eqn:tausq}: } In a similar fashion we can compute $\tau^2$. Observe that,
\begin{align*}
\tau^2 + \left[W(\vecmu) \right]^2 &= \frac{1}{2} \left\lbrace \mathbb{E}\left[\left. \|X - \mu_1\|^4 \right\vert X_1 < 0 \right] + \mathbb{E}\left[\left. \|X - \mu_2\|^4 \right\vert X_1 > 0 \right] \right\rbrace \\
&= \mathbb{E}\left[\left. \|X - \mu_1\|^4 \right\vert X_1 < 0 \right]\\
&= \mathbb{E}\left[\left. \left((X_1 - \mu_{11})^2 + \sum_{i = 2}^d X_i^2 \right)^2 \right\vert X_1 < 0 \right]\\
&= \mathbb{E}\left[\left. (X_1 - \mu_{11})^4 \right\vert X_1 < 0 \right] + 2 \sum_{i = 2}^d \mathbb{E}\left[\left. (X_1 - \mu_{11})^2 \right\vert X_1 < 0 \right] \mathbb{E}\left[X_i^2 \right] \\
\end{align*}
\begin{align*}
& \ \ \ \ \ +  2 \sum_{i = 2}^d \sum_{j = 2, j \neq i}^d  \mathbb{E}\left[X_i^2 \right] \mathbb{E}\left[X_j^2 \right] + \sum_{i = 2}^d \mathbb{E}\left[X_i^4 \right]\\
&= \sigma_1^4\left( 3 - \frac{4}{\pi} - \frac{12}{\pi^2} \right) + 2 \sum_{i = 2}^d \sigma_1^2 \sigma_i^2 \left(1 - \frac{2}{\pi} \right) + 2 \sum_{i = 2}^d \sum_{j = 2, j \neq i}^d  \sigma_i^2 \sigma_j^2 +  3 \sum_{i = 2}^d \sigma_i^4.
\end{align*}
Plugging in the value of $W(\vecmu)$ we have,
\begin{align*}
\tau^2 &=  \sigma_1^4\left( 2 - \frac{16}{\pi^2} \right) + 2 \sum_{i = 2}^d \sigma_i^4\\
&= 2 \sum_{i = 1}^d \sigma_i^4 - \frac{16 \sigma_1^4}{\pi^2},
\end{align*}
which is precisely Claim~\eqref{eqn:tausq}.

$\Box$

\subsection{Proof of $G_0$ being positive definite}
\label{app:GammaNull}

\noindent In order to use the result in \citet{Pollard1982} to prove Lemma~\ref{lem:bock} we need to verify that condition (v) holds. The vector $\vecmu$ that minimizes the population within
cluster sum of squares for the 2-means clustering is given above along with the two optimum population clusters. We additionally have that $M_{12} = \{ x = (x_1,
\ldots, x_d) \in \mathbb{R}^d : x_1 = 0 \}$, $\mathbb{P}(A_1) = \mathbb{P}(A_2) = 0.5$
and $r_{12} = 2 \sigma_1 \sqrt{\frac{2}{\pi}}$. The form of $V$ and $G_0$ can then be given by:

\begin{lemma}
\label{lemma::verify}
If $X = (X_1, \ldots, X_d) \in \mathbb{R}^d$ follows $N(0, \Sigma)$, 
that is, $\P$ is the distribution of $N(0, \Sigma)$, 
where $\Sigma$ has diagonal elements 
$\sigma_1^2 > \sigma_2^2 \geq \sigma_3^2 \geq \ldots \geq \sigma_d^2 > 0$, then for $i, j \in \{1, 2 \}$, $i \neq j$,
\[V_i =  \left( \begin{array}{c c c c}
\frac{\sigma_1^2 }{2} \left(1 - \frac{2}{\pi} \right) & 0 & \ldots & 0\\
0 & \frac{\sigma^2_2}{2} & \ldots & 0\\
\vdots & \vdots & \ddots & \vdots\\
0 & 0 & \ldots & \frac{\sigma^2_d}{2}
\end{array} \right)\]
and the matrix $G_0$ as defined in equation (\ref{eqn::GammaNull}) is positive definite.
\end{lemma}

\noindent {\bf Proof of Lemma \ref{lemma::verify}.}
 The different blocks of the variance matrix are given by,
\begin{align*}
V_1 = V_2 &= \mathbb{E}\left[(X - \mu_1)(X - \mu_1)^T \mathbb{I}_{\{X_1 < 0 \}} \right] \\
&= \left( \begin{array}{c c c c}
\mathbb{E}\left[ (X_1 - \mu_{11})^2 \mathbb{I}_{\{X_1 < 0 \}} \right] & \mathbb{E}\left[ (X_1 - \mu_{11}) X_2 \mathbb{I}_{\{X_1 < 0 \}} \right] & \ldots & \mathbb{E}\left[ (X_1 - \mu_{11}) X_d \mathbb{I}_{\{X_1 < 0 \}} \right]\\
\mathbb{E}\left[ (X_1 - \mu_{11}) X_2 \mathbb{I}_{\{X_1 < 0 \}} \right] & \mathbb{E}\left[ X_2^2 \mathbb{I}_{\{X_1 < 0 \}} \right] & \ldots & \mathbb{E}\left[X_2 X_d \mathbb{I}_{\{X_1 < 0 \}} \right]\\
\vdots & \vdots & \ddots & \vdots\\
\mathbb{E}\left[ (X_1 - \mu_{11}) X_d \mathbb{I}_{\{X_1 < 0 \}} \right] & \mathbb{E}\left[ X_2 X_d \mathbb{I}_{\{X_1 < 0 \}} \right] & \ldots & \mathbb{E}\left[X_d^2 \mathbb{I}_{\{X_1 < 0 \}} \right]
\end{array} \right)
\end{align*}
\begin{align*}
\mathbb{E}\left[ (X_1 - \mu_{11})^2 \mathbb{I}_{\{X_1 < 0 \}} \right] &= \mathbb{E}\left[ (X_1^2 - 2 \mu_{11} X_1 + \mu_{11}^2 ) \mathbb{I}_{\{X_1 < 0 \}} \right] \\
&= \frac{1}{2} \mathbb{E}[X_1^2] - 2 \mu_{11} \left(\frac{1}{2} \mu_{11} \right) + \frac{1}{2} \mu_{11}^2  \\
&= \frac{1}{2} \sigma_1^2 - \frac{1}{2} \left(\frac{2}{\pi} \sigma_1^2 \right)  \\
&= \frac{\sigma_1^2 }{2} \left(1 - \frac{2}{\pi} \right)
\end{align*}
For $j \neq 1$,
\begin{align*}
\mathbb{E}\left[ (X_1 - \mu_{11}) X_j \mathbb{I}_{\{X_1 < 0 \}} \right] &= \mathbb{E}\left[ (X_1 - \mu_{11})\mathbb{I}_{\{X_1 < 0 \}} \right] \mathbb{E}[X_j] = 0.
\end{align*}
Let $i \neq j,$ and $ i,j \in \{ 2, \ldots, d\}$,
\[ \mathbb{E}\left[ X_i X_d \mathbb{I}_{\{X_1 < 0 \}} \right] = \mathbb{E}\left[ X_i \right] \mathbb{E}\left[X_d \right] \mathbb{E}\left[\mathbb{I}_{\{X_1 < 0 \}} \right] = 0.\]
For $j \neq 1$,
\begin{align*}
\mathbb{E}\left[ X_j^2 \mathbb{I}_{\{X_1 < 0 \}} \right] &= \frac{1}{2} \mathbb{E}\left[ X_j^2\right] = \frac{\sigma_j^2}{2}.
\end{align*}
Therefore,
\[V_1 = V_2  =  \left( \begin{array}{c c c c}
\frac{\sigma_1^2 }{2} \left(1 - \frac{2}{\pi} \right) & 0 & \ldots & 0\\
0 & \frac{\sigma^2_2}{2} & \ldots & 0\\
\vdots & \vdots & \ddots & \vdots\\
0 & 0 & \ldots & \frac{\sigma^2_d}{2}
\end{array} \right).\]

Now for $x \in M_{12}$,
\begin{align*}
(x - \mu_1) (x - \mu_1)^T &= \left(\begin{array}{c c c c}
(x_1 - \mu_{11})^2 & (x_1 - \mu_{11})(x_2 - \mu_{12}) & \hdots &  (x_1 - \mu_{11})(x_d - \mu_{1d})\\
(x_1 - \mu_{11})(x_2 - \mu_{12}) & (x_2 - \mu_{12})^2 & \hdots &  (x_2 - \mu_{12})(x_d - \mu_{1d})\\
\vdots & \vdots & \ddots & \vdots \\
(x_1 - \mu_{11})(x_d - \mu_{1d}) & (x_2 - \mu_{12})(x_d - \mu_{1d}) & \hdots &  (x_d - \mu_{1d})^2
\end{array} \right) \\
&= \left(\begin{array}{c c c c}
\mu_{11}^2 &  - \mu_{11} x_2 & \hdots &  - \mu_{11} x_d\\
- \mu_{11} x_2 & x_2^2 & \hdots &  x_2 x_d\\
\vdots & \vdots & \ddots & \vdots \\
- \mu_{11} x_d &  x_2 x_d & \hdots &  x_d^2
\end{array} \right),
\end{align*}
\begin{align*}
(x - \mu_1) (x - \mu_2)^T &=  \left(\begin{array}{c c c c}
(x_1 - \mu_{11})(x_1 - \mu_{21}) & (x_1 - \mu_{11})(x_2 - \mu_{22}) & \hdots & (x_1 - \mu_{11})(x_d - \mu_{2d})\\
(x_2 - \mu_{12})(x_1 - \mu_{21}) & (x_2 - \mu_{12})(x_2 - \mu_{22}) & \hdots & (x_2 - \mu_{12})(x_d - \mu_{2d})\\
\vdots & \vdots & \ddots & \vdots \\
(x_d - \mu_{1d})(x_1 - \mu_{21}) & (x_d - \mu_{1d})(x_2 - \mu_{22}) & \hdots & (x_d - \mu_{1d})(x_d - \mu_{2d})
\end{array} \right) \\
&= \left(\begin{array}{c c c c}
- \mu_{11}^2 &  - \mu_{11} x_2 & \hdots &  - \mu_{11} x_d\\
- \mu_{21} x_2 & x_2^2 & \hdots & x_2 x_d\\
\vdots & \vdots & \ddots & \vdots \\
- \mu_{21} x_d & x_2 x_d & \hdots & x_d^2
\end{array} \right).
\end{align*}
Also, note that $\mathbb{I}_{M_{12}} = \mathbb{I}_{\{X \in M_{12}\}} = \mathbb{I}_{\{X_1 = 0\}}$.
Therefore,
\[
 \mu_{11}^2 \int_{M_{12}} f(x) \ d\sigma(x) = \frac{2 \sigma_1^2}{\pi} \frac{1}{\sqrt{2 \pi } \sigma_1} = \sqrt{\frac{2}{\pi^3}} \sigma_1.
\]
 For $2 \leq j \leq d$,
\begin{align*}
 \mu_{11} \ \int_{M_{12}} x_j \ f(x) \ d\sigma(x)  &= \mu_{11} \ \mathbb{E}[X_j] \ \frac{1}{\sqrt{2 \pi } \sigma_1} = 0,\\
 \mu_{21} \ \int_{M_{12}} x_j \ f(x) \ d\sigma(x) &= \mu_{21} \ \mathbb{E}[X_j] \ \frac{1}{\sqrt{2 \pi } \sigma_1} = 0,\\
\int_{M_{12}} x_j^2 \ f(x) \ d\sigma(x) &=  \mathbb{E}[X_j^2] \ \frac{1}{\sqrt{2 \pi } \sigma_1} = \frac{\sigma_j^2}{\sqrt{2 \pi } \sigma_1}.
\end{align*}
Let $i \neq j,$ and $ i,j \in \{ 2, \ldots, d\}$,
\[\int_{M_{12}} x_i \ x_j \ f(x) \ d\sigma(x) =  \mathbb{E}[X_i] \ \mathbb{E}[X_j] \ \frac{1}{\sqrt{2 \pi } \sigma_1}  = 0.\] 
Then the matrix $G_0$ can be derived as,
\begin{align*}
(G_0)_{22} = (G_0)_{11} &= \mathbf{I}_d - \frac{1}{\sigma_1} \sqrt{\frac{\pi}{2}} \left(\begin{array}{c c c c}
\sqrt{\frac{2}{\pi^3}} \sigma_1 & 0 & \hdots & 0 \\
0 & \frac{\sigma_2^2}{\sqrt{2 \pi} \sigma_1} & \hdots & 0 \\
\vdots & \vdots & \ddots & \vdots \\
0 & 0 & \hdots & \frac{\sigma_d^2}{\sqrt{2 \pi} \sigma_1}
\end{array} \right)
= \left(\begin{array}{c c c c}
1 - {\frac{1}{\pi}}  & 0 & \hdots & 0\\
0 & 1 - \frac{\sigma_2^2}{ 2 \sigma_1^2 } & \hdots & 0 \\
\vdots & \vdots & \ddots & \vdots \\
0 & 0 & \hdots & 1 - \frac{\sigma_d^2}{ 2 \sigma_1^2 }
\end{array} \right),\\
(G_0)_{21} = (G_0)_{12} &= - \frac{1}{\sigma_1} \sqrt{\frac{\pi}{2}} \left(\begin{array}{c c c c}
- \sqrt{\frac{2}{\pi^3}} \sigma_1 & 0 & \hdots & 0\\
0 & \frac{\sigma_2^2}{\sqrt{2 \pi} \sigma_1} & \hdots & 0\\
\vdots & \vdots & \ddots & \vdots \\
0 & 0 & \hdots & \frac{\sigma_d^2}{\sqrt{2 \pi} \sigma_1}
\end{array} \right) = \left(\begin{array}{c c c c}
{\frac{1}{\pi}}  & 0 & \hdots & 0\\
0 & - \frac{\sigma_2^2}{ 2 \sigma_1^2} & \hdots & 0\\
\vdots & \vdots & \ddots & \vdots \\
0 & 0 & \hdots & - \frac{\sigma_d^2}{ 2 \sigma_1^2 }
\end{array} \right). 
\end{align*}

Using the result from \citet{boyd2004convex}, we have that the symmetric matrix $G_0$ is positive definite if and only if $(G_0)_{11}$ and $G_0 / (G_0)_{11}$ (the Schur complement of $(G_0)_{11}$ in $G_0$) are both positive definite. $(G_0)_{11}$ is a diagonal matrix with strictly positive entries on its diagonal since $\sigma_1^2 > \sigma_j^2$ for $j \neq 1$. Therefore,  $(G_0)_{11}$ is trivially a positive definite matrix. To show $G_0 / (G_0)_{11}$ is also positive definite first we simplify it.

\begin{align*}
G_0 / (G_0)_{11} &= (G_0)_{22} - (G_0)_{21} \left[ (G_0)_{11} \right]^{-1} (G_0)_{12}\\
&= \left(\begin{array}{c c c c}
1 - {\frac{1}{\pi}}  & 0 & \hdots & 0\\
0 & 1 - \frac{\sigma_2^2}{ 2 \sigma_1^2 } & \hdots & 0 \\
\vdots & \vdots & \ddots & \vdots \\
0 & 0 & \hdots & 1 - \frac{\sigma_d^2}{ 2 \sigma_1^2 }
\end{array} \right) - 
\left(\begin{array}{c c c c}
{\frac{1}{\pi^2}} \ \frac{\pi}{\pi - 1} & 0 & \hdots & 0\\
0 &  \frac{\sigma_2^4}{ 4 \sigma_1^4} \ \frac{2 \sigma_1^2}{2 \sigma_1^2 - \sigma_2^2} & \hdots & 0\\
\vdots & \vdots & \ddots & \vdots \\
0 & 0 & \hdots &  \frac{\sigma_d^4}{ 4 \sigma_1^4 } \ \frac{2 \sigma_1^2}{2 \sigma_1^2 - \sigma_d^2}
\end{array} \right)\\
&= \left(\begin{array}{c c c c}
1 - \frac{1}{\pi - 1} & 0 & \hdots & 0\\
0 & \frac{2(\sigma_1^2 - \sigma_2^2)}{ 2 \sigma_1^2 - \sigma_2^2} & \hdots & 0 \\
\vdots & \vdots & \ddots & \vdots \\
0 & 0 & \hdots & \frac{2(\sigma_1^2 - \sigma_d^2)}{ 2 \sigma_1^2 - \sigma_d^2}
\end{array} \right),
\end{align*}
which is again a diagonal matrix with strictly positive entries on its diagonal since $\sigma_1^2 > \sigma_j^2$ for $j \neq 1$. Therefore,  $G_0 / (G_0)_{11}$ is also a positive definite matrix, which implies $G_0$ itself is a positive definite matrix. 

$\Box$

\subsection{Limiting distribution under the null when $\sigma_1^2 = \sigma_2^2$}
\label{sec:spherical}

We will generally focus on the non-spherical case
since it yields tractable limiting distributions.
But here we briefly mention what happens when the null distribution is spherical.
In this case,
the limiting distribution is quite complicated,
For simplicity, we only consider the special case $d=2$.
To find the distribution of the test
statistic, we first find the distribution of the between-cluster sum
of squares, where the between-cluster sum of squares for a partition
given by centers $\veca  = (a_1, \ldots, a_k)$ and the set of
corresponding convex polyhedrons $A_1, \ldots, A_k$ is defined as,
\[ 
B_n(a) = 
\frac{1}{n} \sum_{j = 1}^k n_j \|a_j - \overline{X}\|^2, \ \ \ n_j = \sum_{i = 1}^n \mathbb{I}_{\{ X_i \in A_j \}}. 
\]

When $2$-means clustering is applied to two dimensional data, the
two partitions can also be uniquely identified using the separating
line dividing them. The line containing the optimal centers is
perpendicular to this line. 
Consider the line
joining the centers and the point where it meets the separating line,
say $p$. This line can uniquely be identified by the angle the line
makes with the $x$-axis, ${\bf \beta}$, and its distance from the
origin, $c$. Therefore, instead of defining between-cluster sum of
squares as a function of the centers of the partition, we can also
define it as a function of ${\bf \beta}, c$ and $p$ denoted by $B_n
({\bf \beta}, c, p)$. Therefore corresponding to the two centers of the optimal
partition $\vecbn = (b_{n1}, b_{n2})$, we can also find the optimal hyperplane for the
data, denoted by $({ \beta_n}, c_n, p_n)$.

We perform a 2-means
clustering on the data which finds the optimal partition of the data
in order to minimize the within-cluster sum of squares $W_n (\vecbn)$ 
and maximize the between-cluster sum of squares denoted by
$B_n(\beta_n, c_n, p_n)$. Then,
\[ B_n (\beta_n , c_n, p_n) =  \max_{\beta} \max_{c} \max_p B_n(\beta, c, p).\]
We also define $B_n(\beta) = \max_{c} \max_p B_n(\beta, c, p)$.

\begin{theorem}
\label{thm::spherical}
If $X_1, \ldots, X_n \sim N(0, \Sigma), \ X_i \in \mathbb{R}^2$, where
$\Sigma$ has diagonal elements $\sigma_1^2 = \sigma_2^2 = 1$, then
$\sqrt{n}(B_n(\beta_n, c_n, p_n) - 2/\pi)$ is asymptotically
distributed as the maximum of a Gaussian process $Z(\beta)$ on the
circle $0 \leq \beta < 2 \pi$, where $Z(\beta)$ has mean $0$ and the
covariance between $Z(\beta)$ and $Z(\phi)$ is given by
\[ 
\left\lbrace \frac{8}{\pi ^2} 
\left(\sin \alpha + (\pi - \alpha) \cos \alpha - 2 \right)\right\rbrace, \ \ \alpha = |\beta - \phi| \leq \pi.
\]
\end{theorem} 

Note that $B_n(\beta_n, c_n, p_n) = \max_{\beta} \max_{c} \max_p
B_n(\beta, c, p) = \max_{\beta} B_n (\beta)$. Let $Z(\beta) = \sqrt{n}
\left( B_n(\beta) - \frac{2}{\pi} \right)$, then $\sqrt{n}\left(
B_n(\beta_n, c_n) - \frac{2}{\pi} \right) = \max_{\beta}
Z(\beta)$. Then the proof of the theorem follows directly from Lemmas
\ref{lemma::H1}
and \ref{lemma::H2} stated and proved below. $\Box$

\begin{lemma}
\label{lemma::H1}
If $X_1, \ldots, X_n \sim N(0, \Sigma), \ X_i \in \mathbb{R}^2$, where
$\Sigma$ has diagonal elements $\sigma_1^2 = \sigma_2^2 = 1$, then
$\forall \ 0 \leq \beta < 2 \pi$,
\[
\sqrt{n} \left( B_n(\beta) - \frac{2}{\pi} \right) \rightsquigarrow 
N\left( 0, \frac{8}{\pi}\left( 1 - \frac{2}{\pi}\right)\right) \ \text{ as } \ n \to \infty.\]
\end{lemma}

\noindent {\bf Proof of Lemma \ref{lemma::H1}.}  As the bivariate circular normal
is invariant to the angle $\beta$, without loss of generality we can assume $\beta =
0$. Then the optimal centers of the partition $b_{n1}, b_{n2}$ lie on
a line parallel to the x-axis. Now if we condition on $c_n$, 
then the
line containing the centers is deterministic and hence the
between-cluster sum of squares after performing $2$-means clustering
on the data is same as the between-cluster sum of squares after
projecting the data onto the line joining the centers. So $B_n(\beta)$
is the same as between-cluster sum of squares for $Y_1, \ldots, Y_n$
where $Y_i$ has the same distribution as $X_{i1}$. Now $Y_i's$ are
univariate with $Y_i \sim N(0,1)$.

\citet{hartigan1978asymptotic} showed that for univariate normal data,
$Y_1, \ldots Y_n \sim N(0, 1),$ on performing $2$-means clustering the
asymptotic distribution of between-cluster sum of squares, 
$B_n({\bf   b_n})$ can be given as
\[ 
\sqrt{n} \left( B_n(\vecbn) - \frac{2}{\pi} \right) \rightsquigarrow 
N\left( 0, \frac{8}{\pi}\left( 1 - \frac{2}{\pi}\right)\right) \ \text{ as } \ n \to \infty,
\]
where $\vecbn$ is the vector of cluster centers for the optimal partition.
Therefore,
\[ 
\left.\sqrt{n} \left( B_n(\beta) - \frac{2}{\pi} \right) \right\vert
c_n \rightsquigarrow N\left( 0, \frac{8}{\pi}\left( 1 -     \frac{2}{\pi}\right)\right) \ 
\text{ as } \ n \to \infty,
\]
which does not depend on $c_n$. Hence,
\[
\sqrt{n} \left( B_n(\beta) - \frac{2}{\pi} \right) \rightsquigarrow 
N\left( 0, \frac{8}{\pi}\left( 1 - \frac{2}{\pi}\right)\right) \ \text{ as } \ n \to \infty.\ \Box
\]

\begin{lemma}
\label{lemma::H2}
The asymptotic covariance between $\sqrt{n} B_n(\beta)$ and $\sqrt{n}
B_n(\phi)$ is given by,
\[ 
\lim_{n \to \infty} n \ \text{Cov}(B_n(\beta), B_n(\phi)) = 
\frac{16}{\pi ^2} \left(\sin \alpha + \left(\frac{\pi}{2} - \alpha\right) \cos \alpha - 1 \right), \ \ 
\alpha = |\beta - \phi| \leq \pi.\]
\end{lemma}

\noindent {\bf Proof of Lemma \ref{lemma::H2}.}  \citet{hartigan1978asymptotic}
provides a Taylor's expansion of $B_n(\beta)$ about the population
between-cluster sum of squares $B_n (\mu) = \frac{2}{\pi}$ as,
\[ 
B_n(\beta) = \frac{2}{\pi} + \frac{1}{2} \left(\|b_{n1}(\beta) - b_{n2}(\beta)\|_2 - 
\|\mu_1(\beta) - \mu_2 (\beta)\|_2\right) \| \mu_1(\beta) - 
\mu_2 (\beta) \|_2 + o_p(n^{-1}),
\]
where $(b_{n1}(\beta), b_{n2}(\beta))$ are the centers for the optimal
partition of the data corresponding to $B_n(\beta)$ and $(\mu_1,
\mu_2)$ are the centers for the optimal partition of the entire
population. For $\beta = 0$, the optimal centers are $\mu_1(0) =
(-\sqrt{2/\pi}, 0)$ and $\mu_2(0) = (\sqrt{2/\pi}, 0)$ and hence 
$\|\mu_1(0) - \mu_2 (0)\|_2 = 2 \sqrt{\frac{2}{\pi}}$. Also as the
density of $N(0, \mathbf{I}_d)$ is rotationally invariant, $ \|\mu_1(\beta) -
\mu_2 (\beta)\|_2 = 2 \sqrt{\frac{2}{\pi}}$ for any
$\beta$. Therefore,
\[ 
B_n(\beta) = \frac{2}{\pi} +  
\left(\|b_{n1}(\beta) - b_{n2}(\beta)\|_2 - 2 \sqrt{\frac{2}{\pi}} \right) \sqrt{\frac{2}{\pi}} + o_p(n^{-1}).
\]
 
Due to the rotational invariance of the bivariate circular normal,
$\text{Cov}(B_n(\beta), B_n(\phi)) = \text{Cov}(B_n(0), B_n(\alpha))$, where $\alpha
= |\beta - \phi| \leq \pi$. Therefore it is enough to consider
$\text{Cov}(B_n(0), B_n(\alpha))$ where,
\[
\lim_{n \to \infty} n \ \text{Cov}(B_n(0), B_n(\alpha)) = 
\lim_{n \to \infty} \frac{2}{\pi} \ \text{Cov} \left(\sqrt{n} \ \|b_{n1}(0) - b_{n2}(0)\|_2, 
\sqrt{n} \ \|b_{n1}(\alpha) - b_{n2}(\alpha)\|_2 \right).
\]
To find $\|b_{n1}(\alpha) - b_{n2}(\alpha)\|_2$, if we rotate the axes
by an angle of $-\alpha$, any point $(X_{i1}, X_{i2})$ is now given by
\[
(X_{i1} \cos \alpha + X_{i2} \sin \alpha, X_{i2} \cos \alpha - X_{i1} \sin \alpha).
\] 
For easier notation let us define $Z_i := X_{i1} \cos \alpha + X_{i2}
\sin \alpha$ for $i = 1, 2, \ldots, n$. Let us also define $p_n :=
\frac{1}{n} \sum_{i = 1}^n \mathbb{I}_{\{X_{i1} > 0 \}}$ and $p_n' :=
\frac{1}{n} \sum_{i = 1}^n \mathbb{I}_{\{Z_i > 0 \}}$. Then,
\[ 
\|b_{n1}(0) - b_{n2}(0)\|_2 = 
\frac{\sum_{i = 1}^n X_{i1} \mathbb{I}_{\{X_{i1} > 0 \}}}{n p_n} - \frac{\sum_{i = 1}^n X_{i1} \mathbb{I}_{\{X_{i1} < 0 \}}}
{n (1 - p_n)},
\]
\[
\text{ and } \ \ \|b_{n1}(\alpha) - b_{n2}(\alpha)\|_2 = 
\frac{\sum_{i = 1}^n Z_i \mathbb{I}_{\{Z_i > 0 \}}}{n p_n'} - \frac{\sum_{i = 1}^n Z_i \mathbb{I}_{\{Z_i < 0 \}}}{n (1 - p_n')}. 
\]
Using the Law of Total Covariance,
\begin{align*}
\text{Cov} &\left(\|b_{n1}(0) - b_{n2}(0)\|_2, \|b_{n1}(\alpha) - b_{n2}(\alpha)\|_2 \right) \\
&= \mathbb{E} \left[ \text{Cov} \left( \left.\|b_{n1}(0) - b_{n2}(0)\|_2, 
\|b_{n1}(\alpha) - b_{n2}(\alpha)\|_2 \right\vert p_n, p_n' \right)\right] \\
& \ \ \ \ \ \ +  \text{Cov} \left( \mathbb{E} \left[\left.\|b_{n1}(0) - b_{n2}(0)\|_2 \right\vert p_n \right], 
\mathbb{E} \left[\left. \|b_{n1}(\alpha) - b_{n2}(\alpha)\|_2  \right\vert  p_n' \right]\right)\\
&= \RNum{1} + \RNum{2} \ \ \text{(say)}.
\end{align*} 
The second term ($\RNum{2}$) can be easily simplified as
\begin{align*}
\text{Cov} &\left( \mathbb{E} \left[\left.\frac{\sum_{i = 1}^n X_{i1} \mathbb{I}_{\{X_{i1} > 0 \}}}{n p_n} - 
\frac{\sum_{i = 1}^n X_{i1} \mathbb{I}_{\{X_{i1} < 0 \}}}{n (1 - p_n)} \right\vert p_n \right], 
\mathbb{E} \left[\left.\frac{\sum_{i = 1}^n Z_i \mathbb{I}_{\{Z_i > 0 \}}}{n p_n'} - 
\frac{\sum_{i = 1}^n Z_i \mathbb{I}_{\{Z_i < 0 \}}}{n (1 - p_n')} \right\vert  p_n' \right]\right)\\
&= \text{Cov} \left( {\mathbb{E}[X_{i1} \mathbb{I}_{\{X_{i1} > 0 \}}]} - {\mathbb{E}[ X_{i1} \mathbb{I}_{\{X_{i1} < 0 \}}]}, 
{\mathbb{E}[ Z_i \mathbb{I}_{\{Z_i > 0 \}}]} - { \mathbb{E}[ Z_i \mathbb{I}_{\{Z_i < 0 \}}]} \right)\\
&=  \text{Cov} \left( 2 \frac{1}{\sqrt{2 \pi}}, 2 \frac{1}{\sqrt{2 \pi}} \right) = 0.
\end{align*}
The first term ($\RNum{1}$) becomes,
\begin{align*}
\RNum{1} &= \mathbb{E}\left[ \text{Cov} \left( \left.\frac{\sum_{i = 1}^n X_{i1} \mathbb{I}_{\{X_{i1} > 0 \}}}
{n p_n} - \frac{\sum_{i = 1}^n X_{i1} \mathbb{I}_{\{X_{i1} < 0 \}}}{n (1 - p_n)}, 
\frac{\sum_{i = 1}^n Z_i \mathbb{I}_{\{Z_i > 0 \}}}{n p_n'} - \frac{\sum_{i = 1}^n 
Z_i \mathbb{I}_{\{Z_i < 0 \}}}{n (1 - p_n')} \right\vert p_n, p_n' \right)\right]\\
  &= \mathbb{E}\left[ \text{Cov}\left( X_{11}, Z_1 \vert X_{11} > 0, Z_1 > 0 \right) 
\frac{\sum_{i = 1}^n  \mathbb{I}_{\{X_{i1} > 0, Z_{i} > 0 \}}}{n^2 p_n p_n'} \right] \\
  & \hspace{3cm} - \mathbb{E}\left[ \text{Cov}\left( X_{11}, Z_1 \vert X_{11} > 0, Z_1 < 0 \right) 
\frac{\sum_{i = 1}^n  \mathbb{I}_{\{X_{i1} > 0, Z_{i} < 0 \}}}{n^2 p_n (1 - p_n')} \right]\\
  & \hspace{3cm} - \mathbb{E}\left[ \text{Cov}\left( X_{11}, Z_1 \vert X_{11} < 0, Z_1 > 0 \right) 
\frac{\sum_{i = 1}^n  \mathbb{I}_{\{X_{i1} < 0, Z_{i} > 0 \}}}{n^2 (1 - p_n) p_n'} \right]\\
  & \hspace{3cm} + \mathbb{E}\left[ \text{Cov}\left( X_{11}, Z_1 \vert X_{11} > 0, Z_1
      > 0 \right) \frac{\sum_{i = 1}^n \mathbb{I}_{\{X_{i1} < 0, Z_{i} < 0
        \}}}{n^2 (1 - p_n) (1 - p_n')} \right].
\end{align*}
Due to the symmetry of the Gaussian distribution about the origin, 
\[
\text{Cov}\left( X_{11}, Z_1 \vert X_{11} > 0, Z_1 > 0 \right) =  
\text{Cov}\left( X_{11}, Z_1 \vert X_{11} > 0, Z_1 > 0 \right),\]
\[ \text{and } \ \ 
\text{Cov}\left( X_{11}, Z_1 \vert X_{11} > 0, Z_1 < 0 \right) = 
\text{Cov}\left( X_{11}, Z_1 \vert X_{11} < 0, Z_1 > 0 \right).
\]
Hence the first term becomes,
\begin{align*}
\RNum{1} &= \text{Cov} \left( X_{11}, Z_1 \vert X_{11} > 0, Z_1 > 0 \right) \mathbb{E}\left[ \frac{\sum_{i = 1}^n  \mathbb{I}_{\{X_{i1} > 0, Z_{i} > 0 \}}}{n^2 p_n p_n'} + \frac{\sum_{i = 1}^n  \mathbb{I}_{\{X_{i1} < 0, Z_{i} < 0 \}}}{n^2 (1 - p_n) (1 - p_n')} \right] \\
& \hspace{2cm} - \text{Cov}\left( X_{11}, Z_1 \vert X_{11} > 0, Z_1 < 0 \right) \mathbb{E}\left[  \frac{\sum_{i = 1}^n  \mathbb{I}_{\{X_{i1} > 0, Z_{i} < 0 \}}}{n^2 p_n (1 - p_n')} + \frac{\sum_{i = 1}^n  \mathbb{I}_{\{X_{i1} < 0, Z_{i} > 0 \}}}{n^2 (1 - p_n) p_n'}  \right]
\end{align*}
As $n p_n \sim \text{Bin}(n, 0.5)$ and $n p_n' \sim \text{Bin}(n, 0.5)$, ${1}/{p_n} \overset{p}{\to} 2$ and ${1}/{p_n'} \overset{p}{\to} 2$. Hence by Slutsky's theorem and weak law of large numbers,
\begin{align*}
\lim_{n \to \infty} n \ \mathbb{E}\left[ \frac{\sum_{i = 1}^n  \mathbb{I}_{\{X_{i1} > 0, Z_{i} > 0 \}}}{n^2 p_n p_n'} + \frac{\sum_{i = 1}^n  \mathbb{I}_{\{X_{i1} < 0, Z_{i} < 0 \}}}{n^2 (1 - p_n) (1 - p_n')} \right] &= 4 \left( \mathbb{P}(X_{11} > 0, Z_{1} > 0) + \mathbb{P}(X_{11} < 0, Z_{1} < 0) \right)\\
&= 8 \mathbb{P}(X_{11} > 0, Z_{1} > 0).
\end{align*}
Similarly using Slutsky's theorem and weak law of large numbers,
\[\lim_{n \to \infty} n \ \mathbb{E}\left[  \frac{\sum_{i = 1}^n  \mathbb{I}_{\{X_{i1} > 0, Z_{i} < 0 \}}}{n^2 p_n (1 - p_n')} + \frac{\sum_{i = 1}^n  \mathbb{I}_{\{X_{i1} < 0, Z_{i} > 0 \}}}{n^2 (1 - p_n) p_n'}  \right] = 8 \mathbb{P}(X_{11} > 0, Z_{1} < 0).\]
On the other hand, we can find the covariances as,
\begin{align*}
\text{Cov} \left( X_{11}, Z_1 \vert X_{11} > 0, Z_1 > 0 \right) &= \mathbb{E}\left[ X_{11} Z_1 \vert X_{11} > 0, Z_1 > 0 \right] - \mathbb{E}[X_{11}| X_{11} > 0] \mathbb{E}[Z_1 | Z_1 > 0]\\
&= \frac{\mathbb{E}\left[ X_{11} Z_1 \mathbb{I}_{\{X_{11} > 0, Z_1 > 0\}} \right]}{\mathbb{P}(X_{11} > 0, Z_1 > 0)} - \mathbb{E}[X_{11}| X_{11} > 0] \mathbb{E}[Z_1 | Z_1 > 0],\\
\text{Cov} \left( X_{11}, Z_1 \vert X_{11} > 0, Z_1 < 0 \right) &=  \frac{\mathbb{E}\left[ X_{11} Z_1 \mathbb{I}_{\{X_{11} > 0, Z_1 < 0\}} \right]}{\mathbb{P}(X_{11} > 0, Z_1 < 0)} - \mathbb{E}[X_{11}| X_{11} > 0] \mathbb{E}[Z_1 | Z_1 < 0].
\end{align*}
As $X_{11}, X_{12} \sim N(0, 1),$ it implies $Z_1 = X_{11} \cos \alpha + X_{12} \sin \alpha \sim N(0, 1)$. Hence,
\[ \mathbb{E}[X_{11}| X_{11} > 0] = \mathbb{E}[Z_1 | Z_1 > 0] = \sqrt{\frac{2}{\pi}},\]
\[ \text{and} \ \ \  \mathbb{E}[Z_1 | Z_1 < 0] = - \sqrt{\frac{2}{\pi}}. \]
To find the first expectation, we define $R = \sqrt{X_{11}^2 + X_{12}^2}$ and $\beta = \tan^{-1}\left({X_{12}}/{X_{11}} \right)$ such that $X_{11} = R \cos \beta$ and $X_{12} = R \sin \beta$. Then the Jacobian, ${\partial(x_1, x_2)}/{\partial(r, \beta)}$ can be given by,
\[\frac{\partial(x_1, x_2)}{\partial(r, \beta)} = \left\vert \begin{array}{c c}
\frac{\partial x_1}{\partial r} & \frac{\partial x_1}{\partial \beta}\\
\frac{\partial x_2}{\partial r} & \frac{\partial x_2}{\partial \beta}
\end{array} \right\vert = \left\vert \begin{array}{c c}
\cos \beta & - r \sin \beta\\
\sin \beta & r \cos \beta
\end{array} \right\vert = r.\]
Also $x_{1} > 0$ can be written as $\beta \in \left[- {\pi}/{2}, {\pi}/{2}  \right]$ and assuming $0 < \alpha < {\pi}/{2}$, $x_{1} \cos \alpha + x_{2} \sin \alpha = r \cos (\beta - \alpha)$ and $x_{1} \cos \alpha + x_{2} \sin \alpha > 0$ can be written as $\beta - \alpha \in \left[- {\pi}/{2}, {\pi}/{2}  \right]$ or $\beta \in \left[\alpha - ({\pi}/{2}), \alpha + ({\pi}/{2})  \right]$. $x_{1} \cos \alpha + x_{2} \sin \alpha < 0$ can be written as $\beta - \alpha \in \left[- {3\pi}/{2}, - {\pi}/{2}  \right]$ or $\beta \in \left[\alpha - {3 \pi}/{2}, \alpha - {\pi}/{2}  \right]$. Therefore,
\begin{align*}
\mathbb{E}\left[ X_{11} Z_1 \mathbb{I}_{\{X_{11} > 0, Z_1 > 0\}} \right] &=
\mathbb{E}\left[ X_{11} (X_{11} \cos \alpha + X_{12} \sin \alpha) \mathbb{I}_{\{X_{11} > 0 \}}  \mathbb{I}_{\{X_{11} \cos \alpha + X_{12} \sin \alpha > 0 \}}\right] \\
&= \int_{- \infty}^\infty \int_{- \infty}^\infty x_{1} (x_{1} \cos \alpha + x_{2} \sin \alpha) \mathbb{I}_{\{x_{1} > 0 \}}  \mathbb{I}_{\{x_{1} \cos \alpha + x_{2} \sin \alpha > 0 \}} \ \frac{1}{2 \pi} e^{- \left( \frac{x_1^2}{2} + \frac{x_2^2}{2} \right)} dx_1 \ dx_2 \\
&= \int_{0}^\infty \int_{\alpha - \frac{\pi}{2}}^{\frac{\pi}{2}} r \cos \beta \ r \cos (\beta - \alpha) \ \frac{1}{2 \pi} e^{- \frac{r^2}{2}} r \ d\beta \ dr \\
&= \frac{1}{\sqrt{2 \pi}} \ \int_{0}^\infty r^3 \ \frac{1}{\sqrt{2 \pi}} \ e^{- \frac{r^2}{2}} \ dr \  \int_{\alpha - \frac{\pi}{2}}^{\frac{\pi}{2}} \frac{1}{2} \left( \cos (2 \beta - \alpha) + \cos \alpha\right) \ d\beta  \\
&= \frac{1}{\sqrt{2 \pi}} \ \sqrt{\frac{2}{\pi}} \  \int_{\alpha - \frac{\pi}{2}}^{\frac{\pi}{2}} \frac{1}{2} \left( \cos (2 \beta - \alpha) + \cos \alpha\right) \ d\beta  \\
&= \frac{1}{2 \pi}\left( \left. \frac{\sin(2 \beta - \alpha)}{2}  + \beta  \cos \alpha  \right) \right\vert_{\alpha - \frac{\pi}{2}}^{\frac{\pi}{2}}\\
&= \frac{1}{2 \pi}\left( \frac{\sin(\pi - \alpha) - \sin(\alpha - \pi)}{2}  + (\pi - \alpha) \cos \alpha \right)\\
&= \frac{1}{2 \pi}\left( \sin \alpha  + (\pi - \alpha)\cos \alpha \right)
\end{align*}
Similarly,
\begin{align*}
\mathbb{E}\left[ X_{11} Z_1 \mathbb{I}_{\{X_{11} > 0, Z_1 < 0\}} \right] &=
\mathbb{E}\left[ X_{11} (X_{11} \cos \alpha + X_{12} \sin \alpha) \mathbb{I}_{\{X_{11} > 0 \}}  \mathbb{I}_{\{X_{11} \cos \alpha + X_{12} \sin \alpha < 0 \}}\right] \\
&= \frac{1}{2 \pi} \ \int_{- \frac{\pi}{2}}^{\alpha - \frac{\pi}{2}}  \cos (2 \beta - \alpha) + \cos \alpha \ d\beta  \\
&= \frac{1}{2 \pi}\left( \left. \frac{\sin(2 \beta - \alpha)}{2}  + \beta  \cos \alpha  \right) \right\vert_{ - \frac{\pi}{2}}^{\alpha - \frac{\pi}{2}}\\
&= \frac{1}{2 \pi}\left( \frac{\sin(\alpha - \pi) - \sin(- \alpha - \pi)}{2}  +  \alpha \cos \alpha \right)\\
&= \frac{1}{2 \pi}\left( \frac{\sin(\pi + \alpha) - \sin(\pi - \alpha) }{2}  +  \alpha \cos \alpha \right)\\
&= \frac{1}{2 \pi}\left(  \alpha \cos \alpha - \sin \alpha \right)
\end{align*}
Plugging in all the derivations we get,
\begin{align*}
\lim_{n \to \infty} n \RNum{1} &= \left(\frac{\frac{1}{2 \pi}\left( \sin \alpha  + (\pi - \alpha)\cos \alpha \right)}{\mathbb{P}(X_{11} > 0, Z_1 > 0)} - \frac{2}{\pi} \right)  8 \mathbb{P}(X_{11} > 0, Z_1 > 0) \\
& \hspace{2cm} - \left(\frac{\frac{1}{2 \pi}\left(  \alpha \cos \alpha - \sin \alpha \right)}{\mathbb{P}(X_{11} > 0, Z_1 < 0)} + \frac{2}{\pi} \right) 8 \mathbb{P}(X_{11} > 0, Z_1 < 0)\\
&= \frac{4}{\pi} (2 \sin \alpha + (\pi - 2 \alpha) \cos \alpha) - \frac{16}{\pi} \left(\mathbb{P}(X_{11} > 0, Z_1 > 0) + \mathbb{P}(X_{11} > 0, Z_1 < 0) \right)\\
&= \frac{8}{\pi} \left(\sin \alpha + \left(\frac{\pi}{2} - \alpha\right) \cos \alpha\right) - \frac{8}{\pi}.
\end{align*}
Therefore,
\begin{align*}
\lim_{n \to \infty} n \ \text{Cov} \left(B_n(0), B_n(\alpha)\right)
&= \frac{2}{\pi} \left( \frac{8}{\pi} \left(\sin \alpha + \left(\frac{\pi}{2} - \alpha\right) \cos \alpha\right) - \frac{8}{\pi} \right)\\
&= \frac{16}{\pi^2} \left(\sin \alpha + \left(\frac{\pi}{2} - \alpha\right) \cos \alpha - 1\right). \ \Box
\end{align*}
We also note that setting $\alpha = 0$, gives us $\lim_{n \to \infty} n~\text{V}(B_n
(\beta)) = \frac{8}{\pi}\left(1 - \frac{2}{\pi}\right)$.
$\Box$
\clearpage

\section{Proof of results presented under the alternate hypothesis of SigClust}

In Section~\ref{sec::alternate}, we study the geometry of $k-$means under the alternative. Recall, under the alternative of SigClust we suppose that, we observe $n$ samples from:
\begin{align}
X \sim \frac{1}{2} N(-\theta_1, D) +
\frac{1}{2} N(\theta_1, D)
\end{align}
where $\theta_1 = (a/2, 0, \ldots, 0) \in \mathbb{R}^d$ and $a > 0$. Furthermore, $D$ is a diagonal
matrix with elements $\Sigma_{jj} = \sigma_j^2$, such that $\sigma_1^2, \sigma_2^2 > \sigma_3^2 \geq \ldots \geq \sigma_d^2$. 
Throughout this Appendix and the next, we denote,
\begin{align*} 
u &:= \frac{a}{2\sigma_1}, \\
\kappa &:= \left[ \frac{a}{2} \mathbb{P}(|Z| \leq u) + \sqrt{\frac{2}{\pi}} \sigma_1 \exp(-u^2/2) \right], \\
\widetilde{\sigma}^2 &= \left[\sum_{i=1}^d \sigma_i^2\right] + \frac{a^2}{4}.
\end{align*}

In this Appendix, in Section~\ref{sec:NewPollard} we first prove Theorem~\ref{thm::newPollard}, which is a result analogous to Theorem 6.4 (b) of \citet[pp.~101]{Bock1985} for symmetric $2$-means clustering, that gives the limiting distribution of the within sum of squares under the alternative. This Theorem assumes two things: first, the existence of a unique minimizer of the within sum of squares and second, the positive definiteness of the matrix $G$ defined in equation (\ref{eqn::Gamma}).

We prove Lemma~\ref{lemma::Gamma alternate} that shows the positive definiteness of $G$ in Appendix~\ref{sec:GammaAlt}. In Section~\ref{sec:AlternativeSplit}, we prove Theorem~\ref{thm::optimalsplit} that gives the optimal population split which results in the minimum within sum of squares under the alternative. The idea behind the proof is that when condition (\ref{eqn::firstdimsplit}) is true,  the population-level optimal $2$-means solution is unique and is given by:
\begin{align}
\label{eqn:optsplit1}
\vecmu^* = \left( \left[\begin{matrix}
\kappa \\
0 \\
\vdots \\
0
\end{matrix}\right],\left[\begin{matrix}
- \kappa \\
0 \\
\vdots \\
0
\end{matrix}\right] \right),
\end{align}
and when condition (\ref{eqn::seconddimsplit}) is true then the population-level optimal $2$-means solution is unique and is given by:
\begin{align}
\label{eqn:optsplit2}
\vecmu^* = \left(  \left[\begin{matrix}
0 \\
\sqrt{\frac{2}{\pi}} \sigma_2 \\
\vdots \\
0
\end{matrix}\right],\left[\begin{matrix}
0 \\
-\sqrt{\frac{2}{\pi}} \sigma_2  \\
\vdots \\
0
\end{matrix}\right], \right). 
\end{align}
Now the reason that we have $\sqrt{\frac{2}{\pi}} \sigma_2$ in equation (\ref{eqn:optsplit2}) is because $E[X_2 | X_2 > 0] = \sqrt{\frac{2}{\pi}} \sigma_2$ and similarly we have $\kappa$ in equation (\ref{eqn:optsplit1}) because $E[X_1 | X_1 > 0] = \kappa$, which is given by the following lemma:

\begin{lemma}
\label{lemma::4.1}
Suppose that 
\begin{align*}
Y \sim \frac{1}{2} N(-a/2, \sigma^2) + \frac{1}{2} N(a/2, \sigma^2),
\end{align*}
and that $Z \sim N(0,1)$, then we have that,
\begin{align*}
\mathbb{E}[Y | Y > 0] =  \frac{a}{2} \mathbb{P}(|Z| \leq u) + \sqrt{\frac{2}{\pi}} \sigma_1 \exp(-u^2/2)  =  \kappa.
\end{align*}
\end{lemma}

Additionally, in order to find the within sum of squares for a particular split, we first introduce two lemmas that give the resulting within sum of squares $W(b)$ corresponding to particular forms of seperating hyperplanes. More specifically, the following lemmas give the within sum of squares $W(b)$ corresponding to any separating hyperplane $\mathcal{H}(b)$ where $b$ is of the form $\{b \in \mathbb{R}^d : b_1 \geq 0, \ \sum_{i = 1}^d b_i^2 = 1 \}$. Recollect that,
\[W(b) =  E\left[\|X- E[X| b^{T} X > 0]\|^2 | b^{T} X > 0\right]. \]

\begin{lemma}
For a separating hyperplane $\mathcal{H}(b) = \{y \in \mathbb{R}^d: b^Ty = 0\}$ when
\[b \in \{b \in \mathbb{R}^d: b_1 \geq 0, \ b_1^2 + b_2^2 = 1, \ b_j = 0 \ \forall \ j \geq 3 \},\]
 the corresponding within sum of squares $W(b)$ is given by:
 \[W(b) = \sum_{j = 1}^d \sigma_j^2 + \frac{a^2}{4} - \left[ \left( 2  \Phi\left( \frac{ab_1}{2 \sqrt{b^TDb}}\right)  - 1 \right) \frac{a}{2} + \frac{2 b_1 \sigma_1^2}{\sqrt{b^T D b}}  \phi \left(\frac{ a b_1}{2 \sqrt{b^T D b}} \right)\right]^2 - \frac{4 b_2^2 \sigma_2^4}{b^T D b}  \phi^2 \left(\frac{ a b_1}{2 \sqrt{b^T D b}} \right) , \]
 where $\phi(\cdot)$  and $\Phi(\cdot)$ are respectively the density function and the distribution function of standard normal distribution.
\label{lemma::WSSb}
\end{lemma}


\begin{lemma}
For any fixed $i \geq 2$ and a separating hyperplane $\mathcal{H}(b) = \{y \in \mathbb{R}^d: b^Ty = 0\},$ when 
\[b \in \{b \in \mathbb{R}^d: b_i = 1 \ \text{ and } \ b_j = 0 \ \forall \ j  \neq i \},\]
the corresponding within sum of squares $W(b)$ is given by:
 \[W(b) = \sum_{j = 1}^d \sigma_j^2 +  \frac{a^2}{4}  - \frac{2}{\pi} \sigma_i^2. \]
\label{lemma::WSSbi1}
\end{lemma}

Further more, to prove Theorem~\ref{thm::optimalsplit} we extend projection arguments made in \citet{QIU20101701} to the $d-$dimensional scenario. So we provide a lemma that gives the projection of a cluster center onto the separating hyperplane.
\begin{lemma}
If $Y \sim N(\theta_1, D)$, where $\theta_1 = (a/2, 0, \ldots, 0)
\in \mathbb{R}^d$ and $D$ is a diagonal matrix. Then the $i^{th}$ coordinate of the projection of $E[Y|b^T Y > 0]$ onto the separating hyperplane $\mathcal{H}(b)$ when $\sum_{i = 1}^d b_i^2 = 1$, is given by:
\[\mathcal{P}_i = \frac{a}{2} \mathbb{I}\{i = 1\} - \frac{a b_1 b_i}{2} + \frac{b_i Var(Y_i) - b_i \left(b^T D b\right)}{b^T D b} \ \left(E\left[\left. b^T Y \right\vert b^T Y > 0 \right] - \frac{a b_1}{2}\right).\]
\label{lemma::1Nprojection}
\end{lemma}

Finally, in order to compare the resulting within sum of squares from different seperating hyperplanes we need the following lower bound on $\kappa^2$:
\begin{lemma}
\begin{align*}
\kappa^2 - \frac{2}{\pi} \left( \sigma_1^2 + \frac{a^2}{4} \right) \geq \begin{cases}
\frac{a^4}{240 \sigma^2 \pi}~~~\text{for}~~0 \leq a \leq 4 \sigma_1\\
\frac{a^2}{40}~~~\text{for}~~a \geq 4\sigma_1.
\end{cases}
\end{align*}
\label{lemma::lowerboundW}
\end{lemma}
\noindent We include the proofs of all of these additional lemmas that help us prove Theorem~\ref{thm::optimalsplit} in Appendix~\ref{sec:AdditionalProofs}.

\subsection{Proof of Theorem \ref{thm::newPollard}}
\label{sec:NewPollard}
In order to prove this Theorem, we trace the steps followed by \citet{Pollard1982} to prove their main theorem. First we define for every vector $\mathbf{t} = [t_1, t_2] \in \mathbb{R}^{2d}$ and every $x \in \mathbb{R}^{d}$,
\begin{equation}
{\phi}(x, \mathbf{t}) = \min \{ \|x - t_1\|^2, \|x - t_2\|^2\},
\end{equation}
as defined by \citet{Pollard1982}. We additionally define a symmetric version of the function for $t^* \in \mathbb{R}^{d}$ and every $x \in \mathbb{R}^{d}$,
\begin{equation}
\tilde{\phi}(x, t^*) = {\phi}(x, (t^*, - t^*)) = \min \{ \|x - t^*\|^2, \|x + t^*\|^2\}.
\end{equation}
Let us also define a map $T$ from $\mathbb{R}^d$ to $\mathbb{R}^{2d}$ as $T(t^*) = (t^*, -t^*)$. Then,
\[\tilde{\phi}(x, t^*) = {\phi}(x, T(t^*)). \]
Now note that an analogous version of Lemma A in  \citet{Pollard1982} also holds for the map $t^* \rightarrow \tilde{\phi}(\cdot, t^*)$ as the composite function of two differentiables functions is also differentiable. Now Lemma B  in  \citet{Pollard1982} also holds for $\tilde{\phi}(\cdot, t^*)$ since the class of functions that are to be considered now is a subset of the class of functions ($\mathcal{G}$) considered for  ${\phi}(x, \mathbf{t})$, as we only consider $\mathbf{t}$ of the form $\mathbf{t} = (t^*, -t^*)$ for some $t^* \in \mathbb{R}^d$. Since a subset of a Donsker class is also a Donsker class, an analogous Lemma B holds for $\tilde{\phi}(\cdot, t^*)$.

We can also derive an analogous version of Lemma C and Lemma D by just using the chain rule for finding the second derivative of a composite function and using the results as obtained in Lemma A and B. Now putting them all together we can derive an analogous version of the main theorem. 

Note that the assumptions (i) and (v) of the main theorem in \citet{Pollard1982} are the same as the ones assumed here. The assumptions (ii) - (iv) are met by the mixtures of two Normals assumed in the statement. Therefore, following the arguments presented in the proof of the main theorem in \citet{Pollard1982}  and in the proof of Theorem 6.4 (b) on page 101 of \citet{Bock1985}, we arrive at the result that:
\[\sqrt{n}( W_n^{(0)}(\mathbf{b_n}^{(0)}) - W(\mu^*)) \rightsquigarrow N(0, \tau^{*2}).\] 
$\Box$


\subsection{Proof of Theorem \ref{thm::optimalsplit}}
\label{sec:AlternativeSplit}

To prove this lemma, we extend projection arguments made in \citet{QIU20101701} to the $d-$dimensional scenario. As shown earlier, for any separating hyperplane, $\mathcal{H}(b) = \{y \in \mathbb{R}^d: b^Ty = 0\}$ when $b_1 \geq 0$ and $\sum_{i = 1}^d b_i^2 = 1$, the corresponding within sum of squares is given by:
\begin{align*}
W(b) &= P(b^T X > 0) E[\|X - E[X| b^T X > 0]\|^2 | b^T X > 0] \\
& \ \ \ \ \ \ \ \ +  P(b^T X < 0) E[\|X- E[X| b^T X < 0]\|^2 | b^T X < 0]\\
&= E\left[\left. \|X- E[X| b^T X > 0]\|^2 \right\vert b^T X > 0 \right], \ \ \ \ \ (\text{Since, } - X \overset{d}{=} X).
\end{align*}
The corresponding cluster centers are given by $C_1$ and $C_2$ where $C_1 = E[ X | b^TX > 0]$ and $C_2 = E[X| b^TX < 0] = - C_1$, due to symmetry. 

\textbf{Step 1: Finding the projection of the $2 -$means clustering centers onto the separating hyperplane.}

As in the proof of Lemma \ref{lemma::WSSb}, we define $f$ to be the pdf of $N(-\theta_1, D)$ and $g$ to be the pdf of $N(\theta_1, D)$. We also define a latent variable $Q \sim \mbox{Ber}(0.5)$ and $Y \sim f$ if $Q = 0$ and $Y \sim g$ if $Q = 1$. Then $ X \overset{d}{=} Y$ and similar to the proof of  Lemma \ref{lemma::WSSb},
\begin{align*}
C_1 =  E[X| b^TX > 0] = \alpha \ E_f[ Y | b^TY > 0] + (1 - \alpha) \ E_g[ Y | b^TY > 0],
\end{align*}
where $\alpha = P\left(Q = 0| b^TY > 0 \right) = 1 - \Phi\left({ab_1}/{2 \sqrt{b^TDb}} \right)$ as shown in equation \ref{eqn::splitalpha}, $E_f$ is the expectation when the distribution of $Y$ has a pdf $f$ and $E_g$ is the expectation when the pdf is $g$.

To find the projection of $C_1$ and $C_2$ onto the separating hyperplane $\mathcal{H}(b)$, we use lemma \ref{lemma::1Nprojection}. Since $C_1$ is the weighted mean of $E_f[ Y | b^TY > 0]$ and $E_g[ Y | b^TY > 0]$, the projection of $C_1$ is the weighted mean of their projections. Using Lemma \ref{lemma::1Nprojection}, we get that the  $i^{th}$ coordinate of the projection of $C_1$ onto the separating hyperplane $\mathcal{H}(b)$ is given by:
\begin{align*}
 \mathcal{P}(C_1)_i  &=  \alpha \left[-\frac{a}{2} \mathbb{I}\{i = 1\} + \frac{a b_1 b_i}{2} + \frac{b_i \sigma_i^2 - b_i \left(b^T D b\right)}{b^T D b} \ \left(E_f\left[\left. b^T Y \right\vert b^T Y > 0 \right] + \frac{a b_1}{2}\right) \right]\\
 & \ \ \ \ \ \ \ \ + (1 - \alpha) \left[ \frac{a}{2} \mathbb{I}\{i = 1\} - \frac{a b_1 b_i}{2} + \frac{b_i \sigma_i^2 - b_i \left(b^T D b\right)}{b^T D b} \ \left(E_g\left[\left. b^T Y \right\vert b^T Y > 0 \right] - \frac{a b_1}{2}\right) \right]\\
 &= (1 - 2 \alpha) \left[\frac{a}{2} \mathbb{I}\{i = 1\} - \frac{a b_1 b_i}{2} - \frac{b_i \sigma_i^2 - b_i \left(b^T D b\right)}{b^T D b} \  \frac{a b_1}{2} \right] \\
 & \ \ \ \ \ \ \ \ +  \frac{b_i \sigma_i^2 - b_i \left(b^T D b\right)}{b^T D b}\left( \alpha E_f\left[\left. b^T Y \right\vert b^T Y > 0 \right] + (1 - \alpha)  E_g\left[\left. b^T Y \right\vert b^T Y > 0 \right]\right)\\
  &= (1 - 2 \alpha) \left[\frac{a}{2} \mathbb{I}\{i = 1\}  - \frac{a b_1 b_i \sigma_i^2 }{2 b^T D b} \right] \\
 & \ \ \ \ \ \ \ \ +  \frac{b_i \left(\sigma_i^2 -  b^T D b\right)}{b^T D b}\left( \alpha E_f\left[\left. b^T Y \right\vert b^T Y > 0 \right] + (1 - \alpha)  E_g\left[\left. b^T Y \right\vert b^T Y > 0 \right]\right),
\end{align*}
where $E_f$ is the expectation when the distribution of $Y$ has a pdf $f$ and $E_g$ is the expectation when the pdf is $g$.

\textbf{Step 2: The projection of the optimum centers has to be the origin.}

Since $\mathcal{H}(b)$ is a separating hyperplane for $2$-means clustering, $b^Ty > 0$ for some $y \in \mathbb{R}^d$ if and only if $\| y - C_1\| > \| y - C_2\|$. This gives that the line joining the centers $C_1$ and $C_2$ is perpendicular to the separating hyperplane and the separating hyperplane bisects the line segment joining the centers. Since the midpoint of the centers $({C_1 + C_2})/{2} = 0$, the projection of $C_1$ and $C_2$ onto the separating hyperplane is the origin. Therefore, $\mathcal{P}(C_1)_i = 0$ for every $i$. This implies that for the optimal separating hyperplane:

\begin{enumerate}
\item For $i = 1$, 
\begin{equation}
 (1 - 2 \alpha) \left[\frac{a}{2}  - \frac{a b_1^2 \sigma_1^2 }{2 b^T D b} \right]  +  \frac{b_1 \left(\sigma_1^2 -  b^T D b\right)}{b^T D b} \mathbf{A} = 0,
 \label{eqn::spliti1}
\end{equation}
where $\mathbf{A} = \alpha E_f\left[\left. b^T Y \right\vert b^T Y > 0 \right] + (1 - \alpha)  E_g\left[\left. b^T Y \right\vert b^T Y > 0 \right] > 0.$
\item For $i \geq 2$, 
\begin{equation}
- (1 - 2 \alpha) \left[ \frac{a b_1 b_i \sigma_i^2 }{2 b^T D b} \right] +  \frac{b_i \left(\sigma_i^2 -  b^T D b\right)}{b^T D b} \mathbf{A} = 0.
 \label{eqn::spliti2}
\end{equation}
\end{enumerate}

\textbf{Step 3: Finding values of ${b} \in \mathbb{R}^d$ that satisfy the above equations.}

Since we assume  $b_1 \geq 0$,  $\alpha = 1 - \Phi\left({ab_1}/({2 \sqrt{b^TDb})} \right) \leq 0.5$, where equality occurs if and only if $b_1 = 0$. Hence $1 - 2 \alpha \geq 0$. Now let us consider two cases. One when $\sigma_1^2 \geq \sigma_2^2$ and the other when $\sigma_2^2 > \sigma_1^2$. 

\begin{enumerate}
\item \textbf{Case 1: $\mathbf{\sigma_1^2 \geq \sigma_2^2}$}\\
Note that $b^T D b \leq \max_i \sigma_i^2 = \sigma_1^2$ and therefore the second expression in equation (\ref{eqn::spliti1}) is greater than or equal to $0$. Also $b_1^2 \sigma_1^2 \leq b^T D b$, therefore the first expression in equation (\ref{eqn::spliti1}) is also greater than or equal to $0$. Now for equation (\ref{eqn::spliti1}) to hold, we need both the expressions to be zero. Now the first expression is zero if either $b_1 = 0$ or $b_1^2 \sigma_1^2 = b^T D b$. Note that $b_1^2 \sigma_1^2 = b^T D b \iff b_1 = 1$ and $b^T D b = \sigma_1^2$. The second expression is also zero if either $b_1 = 0$ or $\{b_1 = 1$ and $b^T D b = \sigma_1^2\}$. So for equation (\ref{eqn::spliti1}) to hold, we need 
\[  \{b_1 = 0\} \ \text{ OR } \{ b_1 = 1 \ \text{ and } \ b^T D b = \sigma_1^2\}.\]

If $b_1 = 1$ and $b^T D b = \sigma_1^2$, then $b_i = 0$ for all $i \geq 2$, so equation (\ref{eqn::spliti2}) holds for all $i \geq 2$. But if $b_1 = 0$, equation (\ref{eqn::spliti2}) simplifies to $b_i \left(\sigma_i^2 -  b^T D b\right) = 0$ which holds if and only if $b_i = 0 \ \text{ or } \ b^T D b = \sigma_i^2$. Therefore equations (\ref{eqn::spliti1}) and (\ref{eqn::spliti2}) hold iff 
\[\left\lbrace b_1 = 1 \ \text{ and } \ b_i = 0, \ i \geq 2 \right\rbrace   \ \text{ OR } \  \left\lbrace b_1 = 0  \ \text{ and } \  \left( b_i = 0  \ \text{ or } \ b^T D b = \sigma_i^2, \ i \geq 2\right)\right\rbrace. \]

\item \textbf{Case 2: $\mathbf{\sigma_1^2 < \sigma_2^2}$}\\
First, we consider equation  (\ref{eqn::spliti2}) for $i = d$. $b^T D b \geq \min_i \sigma_i^2 = \sigma_d^2$, therefore the second expression in equation (\ref{eqn::spliti2}) has the same sign as $b_d$. The first expression in equation (\ref{eqn::spliti2}) also has the same sign as $b_d$. Therefore, for their sum to be zero, i.e., for equation (\ref{eqn::spliti2}) to hold for $i = d$, we need both the expressions to be zero. The first expression is zero if either $b_1 = 0$ or $b_d = 0$. The second expression is zero if either $b_i = 0$ or $b^TDb = \sigma_d^2$. So for $i = d$, equation (\ref{eqn::spliti2}) holds iff $ \left\lbrace b_1 = 0  \ \text{ and } \  \left( b_d= 0  \ \text{ or } \ b^T D b = \sigma_d^2, \ i \geq 2\right)\right\rbrace  \ \text{ OR } \  \left\lbrace b_1 \neq 0  \ \text{ and } \  b_d = 0\right\rbrace. $

If $b_1 = 0$ for $2 \leq i < d$,  equation (\ref{eqn::spliti2}) simplifies to $b_i \left(\sigma_i^2 -  b^T D b\right) = 0$ which holds if and only if $b_i = 0 \ \text{ or } \ b^T D b = \sigma_i^2$. Now we concentrate on what happens when $b_1 \neq 0$, i.e. when $b_1 > 0$. We know so far that $b_d = 0$.

Now we consider equation  (\ref{eqn::spliti2}) for $i = d - 1$. Since $b_d = 0$, $b^T D b \geq \min_{1 \leq i \leq d-1} \sigma_i^2 = \sigma_{d-1}^2$, therefore the second expression in equation (\ref{eqn::spliti2}) has the same sign as $b_{d - 1}$. The first expression in equation (\ref{eqn::spliti2}) also has the same sign as $b_{d - 1}$. Therefore, for equation (\ref{eqn::spliti2}) to hold for $i = d - 1$, we need both the expressions to be zero. The first expression is zero iff $b_{d - 1} = 0$ since $b_1 > 0$. Similar arguments can be made by considering  equation (\ref{eqn::spliti2}) for $i = d - 2, \ldots, 3$ sequentially. We can show that if $b_1 > 0$, $b_i = 0$ for $i \geq 3$. Therefore the separating hyperplane is of the form 
\[\mathcal{H}(b) = \{y \in \mathbb{R}^d: b^Ty = 0\} \ \text{where} \ b_1 > 0, \ b_1^2 + b_2^2 = 1, \ b_j = 0 \ \forall \ j \geq 3. \]

We now consider equation (\ref{eqn::spliti1}) and equation (\ref{eqn::spliti2}) for $i = 2$. Note that $b_1 = 1$ and $b_2 = 0$ is a feasible solution. 

Let us now consider $b_1 > 0$ and $0 < b_2^2 < 1$. In this case to study the feasibility of equation (\ref{eqn::spliti1}) and equation (\ref{eqn::spliti2}) for $i = 2$ we need to look at $\mathbf{A}$. Now recall that if $Y \sim f$, then $b^T Y \sim N(-ab_1/2, b^T D b)$ and if $Y \sim g$, then $b^T Y \sim N(ab_1/2, b^T D b)$. Therefore,
\begin{align*}
 \mathbf{A} &= \alpha E_f\left[\left. b^T Y \right\vert b^T Y > 0 \right] + (1 - \alpha)  E_g\left[\left. b^T Y \right\vert b^T Y > 0 \right] \\
  &= \alpha \left( - \frac{ab_1}{2} + \sqrt{\frac{2 b^T D b}{\pi}} \ \frac{e^{- \frac{a^2 b_1^2}{8 b^T D b}}}{2 \Phi \left( - \frac{ab_1}{2 \sqrt{b^T D b}} \right)}\right) + (1 - \alpha)  \left(  \frac{ab_1}{2} + \sqrt{\frac{2 b^T D b}{\pi}} \ \frac{e^{- \frac{a^2 b_1^2}{8 b^T D b}}}{2 \Phi \left( \frac{ab_1}{2 \sqrt{b^T D b}} \right)}\right).
\end{align*}

Recall that $\alpha = 1 - \Phi\left(\frac{ab_1}{2 \sqrt{b^TDb}} \right) =  \Phi\left(- \frac{ab_1}{2 \sqrt{b^TDb}} \right).$ 

Therefore,
\begin{align*}
 \mathbf{A} &= \alpha \left( - \frac{ab_1}{2} + \sqrt{\frac{2 b^T D b}{\pi}} \ \frac{e^{- \frac{a^2 b_1^2}{8 b^T D b}}}{2 \alpha}\right) + (1 - \alpha)  \left(  \frac{ab_1}{2} + \sqrt{\frac{2 b^T D b}{\pi}} \ \frac{e^{- \frac{a^2 b_1^2}{8 b^T D b}}}{2 (1 - \alpha)}\right)\\
 &= (1 - 2 \alpha) \frac{ab_1}{2} + \sqrt{\frac{2 b^T D b}{\pi}} \ e^{- \frac{a^2 b_1^2}{8 b^T D b}}\\
 &= (1 - 2 \alpha) \frac{ab_1}{2} + 2 \sqrt{b^T D b} \ \phi\left(\frac{ab_1}{2 \sqrt{b^TDb}} \right).
\end{align*}

Plugging this into the L.H.S. of equation (\ref{eqn::spliti1}) we get:
\begin{align*}
&  (1 - 2 \alpha) \left[\frac{a}{2}  - \frac{a b_1^2 \sigma_1^2 }{2 b^T D b} \right]  +  \frac{b_1 \left(\sigma_1^2 -  b^T D b\right)}{b^T D b} \left( (1 - 2 \alpha) \frac{ab_1}{2} + 2 \sqrt{b^T D b} \ \phi\left(\frac{ab_1}{2 \sqrt{b^TDb}} \right)\right)\\
= \ & (1 - 2 \alpha) \frac{a}{2}  \left[ 1 + \frac{ - b_1^2 \sigma_1^2 + b_1^2 \left(\sigma_1^2 -  b^T D b\right)}{b^T D b} \right]  +    \frac{2 b_1 \left(\sigma_1^2 (b_1^2 + b_2^2) - (b_1^2 \sigma_1^2 + b_2^2 \sigma_2^2) \right)}{\sqrt{b^T D b}}  \phi\left(\frac{ab_1}{2 \sqrt{b^TDb}} \right)\\
= \ & (1 - 2 \alpha) \frac{a}{2}  \left[ 1 - b_1^2 \right]  -    \frac{2 b_1 b_2^2 \left(\sigma_2^2 - \sigma_1^2 \right)}{\sqrt{b^T D b}}  \phi\left(\frac{ab_1}{2 \sqrt{b^TDb}} \right)\\
= \ & (1 - 2 \alpha) \frac{a}{2} \ b_2^2  -    \frac{2 b_1 b_2^2 \left(\sigma_2^2 - \sigma_1^2 \right)}{\sqrt{b^T D b}}  \phi\left(\frac{ab_1}{2 \sqrt{b^TDb}} \right)\\
= \ & b_2^2 \left[ (1 - 2 \alpha) \frac{a}{2}  -    \frac{2 b_1 \left(\sigma_2^2 - \sigma_1^2 \right)}{\sqrt{b^T D b}}  \phi\left(\frac{ab_1}{2 \sqrt{b^TDb}} \right) \right].
\end{align*}

Similarly simplifying L.H.S. of equation (\ref{eqn::spliti2}) for $i = 2$ we get:
\begin{align*}
&  - (1 - 2 \alpha) \left[ \frac{a b_1 b_2 \sigma_2^2 }{2 b^T D b} \right] +  \frac{b_2\left(\sigma_2^2 -  b^T D b\right)}{b^T D b} \left( (1 - 2 \alpha) \frac{ab_1}{2} + 2 \sqrt{b^T D b} \ \phi\left(\frac{ab_1}{2 \sqrt{b^TDb}} \right)\right)\\
= \ &   (1 - 2 \alpha)  \frac{a b_1 b_2}{2}  \left[ \frac{- \sigma_2^2 + \sigma_2^2 -  b^T D b }{ b^T D b} \right] +  \frac{2 b_2\left(\sigma_2^2 (b_1^2 + b_2^2) - (b_1^2 \sigma_1^2 + b_2^2 \sigma_2^2)\right)}{\sqrt{b^T D b}}  \phi\left(\frac{ab_1}{2 \sqrt{b^TDb}} \right)\\
= \ &  - (1 - 2 \alpha)  \frac{a b_1 b_2}{2} +  \frac{2 b_2 b_1^2 \left(\sigma_2^2 - \sigma_1^2 \right)}{\sqrt{b^T D b}}  \phi\left(\frac{ab_1}{2 \sqrt{b^TDb}} \right)\\
= \ & - b_1 b_2 \left[ (1 - 2 \alpha) \frac{a}{2}  -    \frac{2 b_1 \left(\sigma_2^2 - \sigma_1^2 \right)}{\sqrt{b^T D b}}  \phi\left(\frac{ab_1}{2 \sqrt{b^TDb}} \right) \right].
\end{align*}

Since $b_1 > 0$, $0 < b_2^2 < 1$ and $\alpha = 1 - \Phi\left(\frac{ab_1}{2 \sqrt{b^TDb}} \right)$, equation (\ref{eqn::spliti1})  and equation (\ref{eqn::spliti2}) for $i = 2$ hold if and only if

\begin{equation}
 \left(2 \ \Phi\left(\frac{ab_1}{2 \sqrt{b^TDb}} \right)  - 1 \right) \frac{a}{2}  -    \frac{2 b_1 \left(\sigma_2^2 - \sigma_1^2 \right)}{\sqrt{b^T D b}}  \phi\left(\frac{ab_1}{2 \sqrt{b^TDb}} \right) = 0
 \label{eqn::spliti4}
\end{equation}

Now we know that for $x > 0$, $2 \Phi(x) - 1 > 2 x \phi(x)$. So since $b_1 > 0$, 
\[\left( 2 \ \Phi\left(\frac{ab_1}{2 \sqrt{b^TDb}} \right)  - 1 \right)  > 2 \ \frac{ab_1}{2 \sqrt{b^TDb}}  \ \phi\left(\frac{ab_1}{2 \sqrt{b^TDb}} \right).\]
Therefore, if $\sigma_2^2 \leq \sigma_1^2 + \frac{a^2}{4}$, the L.H.S. of equation (\ref{eqn::spliti4}) becomes
\begin{align*}
&  \left(2 \ \Phi\left(\frac{ab_1}{2 \sqrt{b^TDb}} \right)  - 1 \right) \frac{a}{2}  -    \frac{2 b_1 \left(\sigma_2^2 - \sigma_1^2 \right)}{\sqrt{b^T D b}}  \phi\left(\frac{ab_1}{2 \sqrt{b^TDb}} \right) \\
& > 2 \ \frac{a^2 b_1}{4 \sqrt{b^TDb}}  \ \phi\left(\frac{ab_1}{2 \sqrt{b^TDb}} \right) - 2 \ \frac{a^2 b_1}{4 \sqrt{b^TDb}}  \ \phi\left(\frac{ab_1}{2 \sqrt{b^TDb}} \right) \\
&= 0.
\end{align*}

Therefore if  $\sigma_2^2 \leq \sigma_1^2 + \frac{a^2}{4}$, then equation (\ref{eqn::spliti4}) can not hold and hence either $b_1 = 0$ or $b_2 = 0$. Putting everything together, this implies equations (\ref{eqn::spliti1})  and (\ref{eqn::spliti2}) hold iff
\[\left\lbrace b_1 = 1 \ \text{ and } \ b_i = 0, \ i \geq 2 \right\rbrace   \ \text{ OR } \  \left\lbrace b_1 = 0  \ \text{ and } \  \left( b_i = 0  \ \text{ or } \ b^T D b = \sigma_i^2, \ i \geq 2\right)\right\rbrace. \]

For  $\sigma_2^2 > \sigma_1^2 + \frac{a^2}{4}$, we have shown that if $b_1 = 0,$ we require $b_i = 0$ or $b^T D b = \sigma_i^2$ for all $i \geq 2$ and if $b_1 > 0$, we require $b_i = 0$ for all $i \geq 3$. Therefore  equations (\ref{eqn::spliti1})  and (\ref{eqn::spliti2}) hold iff
\[\left\lbrace b_1 =1 \ \text{ and } \ b_i = 0, \ i \geq 2 \right\rbrace   \ \text{ OR } \  \left\lbrace b_1 = 0  \ \text{ and } \  \left( b_i = 0  \ \text{ or } \ b^T D b = \sigma_i^2, \ i \geq 2\right)\right\rbrace \]
\[  \ \text{ OR } \ \left\lbrace 0 < b_1 < 1, \  b_1^2 + b_2^2 = 1, \ b_i = 0, \ i \geq 3,  \ \text{ and Eq \ref{eqn::spliti4} holds}  \  \right\rbrace.\]
\end{enumerate}

From cases 1 and 2 we finally come to the conclusion that for equations (\ref{eqn::spliti1}) and (\ref{eqn::spliti2}) to hold for $\sigma_2^2 \leq \sigma_1^2 + \frac{a^2}{4}$, we need
\begin{align}\label{eqn::sigma2small}
\left\lbrace b_1 = 1 \ \text{ and } \ b_i = 0, \ i \geq 2 \right\rbrace   \ \text{ OR } \  \left\lbrace b_1 = 0  \ \text{ and } \  \left( b_i = 0  \ \text{ or } \ b^T D b = \sigma_i^2, \ i \geq 2\right)\right\rbrace.
\end{align}

For $\sigma_2^2 > \sigma_1^2 + \frac{a^2}{4}$, we need
\begin{align}\label{eqn::sigma2big}
\left\lbrace b_1 =1 \ \text{ and } \ b_i = 0, \ i \geq 2 \right\rbrace   \ \text{ OR } \  \left\lbrace b_1 = 0  \ \text{ and } \  \left( b_i = 0  \ \text{ or } \ b^T D b = \sigma_i^2, \ i \geq 2\right)\right\rbrace \nonumber \\
 \ \text{ OR } \ \left\lbrace 0 < b_1 < 1, \  b_1^2 + b_2^2 = 1, \ b_i = 0, \ i \geq 3,  \ \text{ and Eq \ref{eqn::spliti4} holds}  \  \right\rbrace.
\end{align}

\textbf{Step 4: Among the possible values of ${b}$, finding ${b^*}$ that gives the minimum within sum of squares.}

\begin{enumerate}
\item \textbf{Case 1:} $\mathbf{\sigma_2^2 < \sigma_1^2 + \frac{a^2}{4}}$

\textbf{If every $\mathbf{\sigma_i^2}$ is distinct,} then for (\ref{eqn::sigma2small}) to hold, for some unique $i = i_0$, 
\[b^T D b = \sigma^2_{i_0}, \ b_{i_0} = 1 \ \text{ and } \  b_j = 0 \text{ for } j \neq i_0.\]
Notice that in this case, the optimal separating hyperplane is $\mathcal{H}(b) = \{ y \in \mathbb{R}^d : y_{i_0} = 0\}$ for some $i_0$. Now if $b_1 = 1$ and $b_i = 0$ for $i \geq 2$, we use Lemma \ref{eqn::WSSb} to find the corresponding within sum of squares as
\begin{equation}\label{eqn::W1}
W_1^* := W(b) = \sum_{j = 1}^d \sigma_j^2 +  \frac{a^2}{4}  - \left( \sqrt{\frac{2}{\pi}} \ \sigma_1 \ e^{-\frac{a^2}{8 \sigma_1^2}} + \frac{a}{2} \ P \left(|Z| < \frac{a}{2 \sigma_1}\right) \right)^{2}.
\end{equation}
For $i_0 = 2$, that is, when $b_2 = 1$ and $b_i = 0$ for $i \neq 2$ we can similarly use Lemma \ref{eqn::WSSb} to find the corresponding within sum of squares as
\begin{equation}\label{eqn::W2}
W_2^* := W(b) = \sum_{j = 1}^d \sigma_j^2 +  \frac{a^2}{4}  - \frac{2}{\pi} \sigma_2^2.
\end{equation}
From Lemma \ref{lemma::WSSbi1} we get that the corresponding within sum of squares when $b_{i_0} = 1$ and $b_j = 0$ for $j \neq i_0$, corresponding to any $i_0 \geq 2$ is
\begin{equation}\label{eqn::Wi0}
W_{i_0}^* := W(b) = \sum_{j = 1}^d \sigma_j^2 +  \frac{a^2}{4}  - \frac{2}{\pi} \sigma_{i_0}^2. 
\end{equation}

Now using the lower bound given by Lemma \ref{lemma::lowerboundW} we see that,
\[\left[ \frac{a}{2} \mathbb{P}(|Z| \leq u) + \sqrt{\frac{2}{\pi}} \sigma_1 \exp(-u^2/2) \right]^2 \geq  \frac{2}{\pi} \left( \sigma_1^2 + \frac{a^2}{4} \right) > \frac{2}{\pi}  \sigma_2^2 > \frac{2}{\pi}  \sigma_2^2, \ \text{ for } j \geq 3, \]
since $\sigma_2^2 > \sigma_j^2$  for $j \geq 3$. Therefore, the minimum within sum of squares is achieved if $W(b^*) = W_1^*$, that is, if $b_1^* = 1$ and $b_i^* = 0$ for every $i \neq 1$. Therefore, the unique optimal separating hyperlane is given by $\mathcal{H}(b^*) = \{y \in \mathbb{R}^d: y_1 = 0 \}$.\\~\\

\textbf{If $\mathbf{\sigma^2_i}$ are not distinct.} Since, $\sigma_1^2, \sigma_2^2 > \sigma_3^2 \geq \ldots \geq \sigma_d^2$, suppose for some $i_0 \geq 3$ and $i_0 + m \leq d$, $\sigma_{i_0}^2 = \sigma_{i_0 + 1}^2 = \ldots = \sigma_{i_0 + m}^2 = \sigma^2$. Then the optimal separating hyperplane can be given by either $\mathcal{H}(b) = \{ y \in \mathbb{R}^d : y_{i} = 0\}$, where $i \notin [i_0, i_0 + m]$, that is, $b_i = 1$ for some $i \notin [i_0, i_0 + m]$ and $b_j = 0$ for all $j \neq i$ or by $\mathcal{H}(b) = \{ y \in \mathbb{R}^d : \sum_{j = i_0}^{i_0 + m} b_j y_j = 0\}$, where $\sum_{j = i_0}^{i_0 + m} {b_j^2} = 1$. 

Suppose if the separating hyperplane is $\mathcal{H}(b) = \{ y \in \mathbb{R}^d : \sum_{j = i_0}^{i_0 + m} b_j y_j = 0\}$, where $\sum_{j = i_0}^{i_0 + m} {b_j^2} = 1$, then the corresponding within sum of squares is given by 
\[\tilde{W}_{i_0, m}^* := W(b) = \sum_{j \notin [i_0, i_0 + m]} \sigma_j^2 + \frac{a^2}{4} + \sum_{k = i_0}^{i_0 + m} E\left[\left. \left(X_k - E\left[X_k | b^TX > 0 \right] \right)^2 \right\vert b^T X > 0 \right]. \]
Since $b_i$ for $i \in [i_0, i_0 + m]$ are not all zero and $\sum_{i = i_0}^{i_0 + m} b_i^2 = 1$, we can construct a orthogonal matrix $\tilde{A}$ of dimension $(m + 1) \times (m+1)$ whose first row is given by $(b_{i_0}, \ldots, b_{i_0 + m})$. Now define a rotated space such that $v = (y_{i_0}, \ldots, y_{i_0 + m})$ is transformed to ${u} = \tilde{A} v$ and define $U = \tilde{A}V$, where $V = (X_{i_0}, \ldots, X_{i_0 + m})$. Then the separating hyperplane now becomes $\mathcal{H}(b) = \{u \in \mathbb{R}^d : {u}_{i_0} = 0 \}$ since ${u}_{i_0} = \sum_{j = i_0}^{i_0 + m} b_j y_j = 0$. As $V \sim N_{m + 1}(0, \sigma^2 I)$, we have that $U \sim N_{m + 1}(0, \sigma^2 I)$ and therefore, we can write the within sum of squares as:

\begin{align*}
\tilde{W}_{i_0, m}^* &= \sum_{j \notin [i_0, i_0 + m]} \sigma_j^2 + \frac{a^2}{4} + \sum_{k = 1}^{m} E\left[\left. \left(U_k - E\left[U_k | U_1 > 0 \right] \right)^2 \right\vert U_1 > 0 \right]\\
&= \sum_{j \notin [i_0, i_0 + m]} \sigma_j^2 + \frac{a^2}{4} + \sum_{k = i_0}^{i_0 + m} E\left[\left. \left(X_k - E\left[X_k | X_{i_0} > 0 \right] \right)^2 \right\vert X_{i_0} > 0 \right]\\
&= \sum_{j \neq i_0} \sigma_j^2 + \frac{a^2}{4} + E\left[\left. \left(X_{i_0} - E\left[X_{i_0} | X_{i_0} > 0 \right] \right)^2 \right\vert X_{i_0} > 0 \right].
\end{align*}

Similar to the calculations in Lemma \ref{lemma::WSSbi1}, we get 
\[\tilde{W}_{i_0, m}^* =  \sum_{j  = 1}^d \sigma_j^2 + \frac{a^2}{4} - \frac{2}{\pi} \sigma_{i_0}^2. \]

Now similar to the previous case, $\left( \sqrt{\frac{2}{\pi}} \ \sigma_1 \ e^{-\frac{a^2}{8 \sigma_1^2}} + \frac{a}{2} \ P \left(|Z| < \frac{a}{2 \sigma_1}\right) \right)^{2} \geq \frac{2}{\pi} \left(\sigma_1^2 + \frac{a^2}{4}\right) > \frac{2}{\pi} \sigma_j^2 $ for $j \geq 2$. Therefore, the minimum within sum of squares is achieved for $b^*$ if $b_1^* = 1$ and $b_i^* = 0$ for every $i \neq 1$ and the optimal separating hyperplane is given by $\mathcal{H}(b) = \{ y \in \mathbb{R}^d : y_1 = 0\}$.

\item \textbf{Case 2:} $\mathbf{\sigma_2^2 > \sigma_1^2 + \frac{a^2}{4}}$

Looking at the first possibility in Expression (\ref{eqn::sigma2big}), that is, if $b_1 = 1$ and $b_i = 0$ for every $i \neq 1$, the corresponding minimum within sum of squares, as seen in the previous case in Equation (\ref{eqn::W1}), is given by $W_1^*$.

Following similar reasonings as presented in Case 1 for the second possibility (\ref{eqn::sigma2big}), we argue that if every $\sigma_i^2$ is distinct, then $b_1 = 0$ and for a unique $i_0 \geq 2$, $b^T D b = \sigma^2_{i_0}, \ b_{i_0} = 1$  and $ b_j \to 0$  for $j \neq i_0.$ As seen in the previous case, if $i_0 = 2$ the corresponding minimum within sum of squares is $W_2^*$ (Equation (\ref{eqn::W2})) and for $i_0 \geq 2$, it is $W_{i_0}^*$ (Equation (\ref{eqn::Wi0})).

To study the third possibility in \ref{eqn::sigma2big}, we use Lemma \ref{lemma::WSSb} to get the within sum of squares for any $b$ such that $0 < b_1 < 1, b_1^2 + b_2^2 = 1$ and find the minimum possible $W(b)$ such that Equation (\ref{eqn::spliti4}) holds.
\begin{align*}
W(b) &= \sum_{i = 1}^d \sigma_i^2 + \frac{a^2}{4} - \left[ \left( 2  \Phi\left( \frac{ab_1}{2 \sqrt{b^TDb}}\right)  - 1 \right) \frac{a}{2} + \frac{2 b_1 \sigma_1^2}{\sqrt{b^T D b}}  \phi \left(\frac{ a b_1}{2 \sqrt{b^T D b}} \right)\right]^2\\
& \ \ \ \ \ \ \ \ \ \  - \frac{4 b_2^2 \sigma_2^4}{b^T D b}  \phi^2 \left(\frac{ a b_1}{2 \sqrt{b^T D b}} \right).
\end{align*}
In order to minimize $W(b)$ notice that we have to maximize 
\begin{align*}
 \left[ \left( 2  \Phi\left( \frac{ab_1}{2 \sqrt{b^TDb}}\right)  - 1 \right) \frac{a}{2} + \frac{2 b_1 \sigma_1^2}{\sqrt{b^T D b}}  \phi \left(\frac{ a b_1}{2 \sqrt{b^T D b}} \right)\right]^2 + \frac{4 b_2^2 \sigma_2^4}{b^T D b}  \phi^2 \left(\frac{ a b_1}{2 \sqrt{b^T D b}} \right),
\end{align*}
and for Equation (\ref{eqn::spliti4}) to hold, we have
\[\left(2 \Phi\left(\frac{ab_1}{2 \sqrt{b^TDb}} \right)  - 1 \right) \frac{a}{2}  =  \frac{2 b_1 \left(\sigma_2^2 - \sigma_1^2 \right)}{\sqrt{b^T D b}}  \phi\left(\frac{ab_1}{2 \sqrt{b^TDb}} \right). \]
\end{enumerate}
Therefore, we have to maximize the following given $b_1$ satisfies Equation (\ref{eqn::spliti4}).
\begin{align*}
 &\left[ \left( 2  \Phi\left( \frac{ab_1}{2 \sqrt{b^TDb}}\right)  - 1 \right) \frac{a}{2} + \frac{2 b_1 \sigma_1^2}{\sqrt{b^T D b}}  \phi \left(\frac{ a b_1}{2 \sqrt{b^T D b}} \right)\right]^2 + \frac{4 b_2^2 \sigma_2^4}{b^T D b}  \phi^2 \left(\frac{ a b_1}{2 \sqrt{b^T D b}} \right)\\
 &= \left[\frac{2 b_1 \left(\sigma_2^2 - \sigma_1^2 \right)}{\sqrt{b^T D b}}  \phi\left(\frac{ab_1}{2 \sqrt{b^TDb}} \right)+ \frac{2 b_1 \sigma_1^2}{\sqrt{b^T D b}}  \phi \left(\frac{ a b_1}{2 \sqrt{b^T D b}} \right)\right]^2 + \frac{4 b_2^2 \sigma_2^4}{b^T D b}  \phi^2 \left(\frac{ a b_1}{2 \sqrt{b^T D b}} \right)\\
&=\frac{4 b_1^2 \sigma_2^4}{b^T D b}  \phi^2 \left(\frac{ a b_1}{2 \sqrt{b^T D b}} \right) + \frac{4 b_2^2 \sigma_2^4}{b^T D b}  \phi^2 \left(\frac{ a b_1}{2 \sqrt{b^T D b}} \right)\\
&=\frac{4  \sigma_2^4}{b^T D b}  \phi^2 \left(\frac{ a b_1}{2 \sqrt{b^T D b}} \right).
\end{align*}

Again by using Equation (\ref{eqn::spliti4}) we get that 
\begin{align*}
\frac{4  \sigma_2^4}{b^T D b} \ \phi^2 \left(\frac{ a b_1}{2 \sqrt{b^T D b}} \right)  = 4 \sigma_2^4 \ \left(2 \Phi\left(\frac{ab_1}{2 \sqrt{b^TDb}} \right)  - 1 \right)^2 \frac{a^2}{4} \ \frac{1}{4 b_1^2 (\sigma_2^2 - \sigma_1^2)^2}.
\end{align*}

Since $2 \Phi\left(\frac{ab_1}{2 \sqrt{b^TDb}} \right)  - 1 \leq \frac{1}{\sqrt{2 \pi}} \ \left(\frac{ab_1}{\sqrt{b^TDb}} \right) $, if $b_1$ satisfies  Equation (\ref{eqn::spliti4}),
\begin{align*}
\frac{4  \sigma_2^4}{b^T D b}  \phi^2 \left(\frac{ a b_1}{2 \sqrt{b^T D b}} \right)  &\leq 4 \sigma_2^4 \ \frac{1}{2 \pi} \ \frac{a^2b_1^2}{b^TDb} \ \frac{a^2}{4} \ \frac{1}{4 b_1^2 (\sigma_2^2 - \sigma_1^2)^2}\\
&= \frac{2}{\pi} \sigma_2^2 \ \frac{\sigma_2^2}{b^TDb} \ \left(\frac{a^2/4}{\sigma_2^2 - \sigma_1^2} \right)^2\\
&\leq \frac{2}{\pi} \sigma_2^2 \ \frac{\sigma_2^2}{\sigma_1^2} \ \left(\frac{a^2/4}{\sigma_2^2 - \sigma_1^2} \right)^2,
\end{align*}
since $\sigma_1^2 < \sigma_2^2$ and therefore $\sigma_1^2 \leq b^TDb$. We now show that for large enough $\sigma_2^2$, $\frac{\sigma_2^2}{\sigma_1^2} \ \left(\frac{a^2/4}{\sigma_2^2 - \sigma_1^2} \right)^2 \leq 1$, that is, we want to show $\sigma_1^2 (\sigma_2^2 - \sigma_1^2)^2 \geq \sigma_2^2 \ \frac{a^4}{16}$. In order to show, we plug in $x$ instead of $\sigma_2^2$ and consider the equation:
\begin{equation}
\sigma_1^2 (x - \sigma_1^2)^2 - x \ \frac{a^4}{16} = 0 \iff \sigma_1^2 x^2 - \left(2 \sigma_1^4 +  \frac{a^4}{16}\right) x + \sigma_1^6 = 0
\end{equation}
Note that the larger solution to this equation is given by:
\[ x = \frac{2 \sigma_1^4 +  \frac{a^4}{16} + \sqrt{\left(2 \sigma_1^4 +  \frac{a^4}{16}\right)^2 - 4 \sigma_1^8}}{2 \sigma_1^2} = \frac{2 \sigma_1^4 +  \frac{a^4}{16} + \frac{a^2}{2}\sqrt{\sigma_1^4 + \frac{a^4}{64}}}{2 \sigma_1^2}.\]
Therefore for all $x > \frac{2 \sigma_1^4 +  \frac{a^4}{16} + \frac{a^2}{2}\sqrt{\sigma_1^4 + \frac{a^4}{64}}}{2 \sigma_1^2}$, $\sigma_1^2 (x - \sigma_1^2)^2 - x \ \frac{a^4}{16} > 0$.
Therefore, for
\[\sigma_2^2 > \frac{2 \sigma_1^4 +  \frac{a^4}{16} + \frac{a^2}{2}\sqrt{\sigma_1^4 + \frac{a^4}{64}}}{2 \sigma_1^2} \implies   \frac{\sigma_2^2}{\sigma_1^2} \ \left(\frac{a^2/4}{\sigma_2^2 - \sigma_1^2} \right)^2 < 1,\]
which gives 
\[\frac{4  \sigma_2^4}{b^T D b}  \phi^2 \left(\frac{ a b_1}{2 \sqrt{b^T D b}} \right)  \leq \frac{2}{\pi} \sigma_2^2.\]
Therefore, when $\sigma_2^2 > \frac{2 \sigma_1^4 +  \frac{a^4}{16} + \frac{a^2}{2}\sqrt{\sigma_1^4 + \frac{a^4}{64}}}{2 \sigma_1^2}$, for any $b_1$ satisfying Equation (\ref{eqn::spliti4}),
\[ W(b) \geq  \sum_{i = 1}^d \sigma_i^2 + \frac{a^2}{4} - \frac{2}{\pi} \sigma_2^2 = W_2^*.\]

If we additionally have that
\[ \sigma_2^2 > \frac{\pi}{2}\left( \sqrt{\frac{2}{\pi}} \ \sigma_1 \ e^{-\frac{a^2}{8 \ \sigma_1^2}} + \frac{a}{2} \ P \left(|Z| < \frac{a}{2 \ \sigma_1}\right) \right)^{2}, \]
then 
\[W_2* < W_1^* < W_{i_0}^*, \ \text{ for } \ i_0 \geq 3. \]
Therefore, considering all the three possibilities in Expression (\ref{eqn::sigma2big}), if
\[ \sigma_2^2 > \max\left\lbrace \frac{2 \sigma_1^4 +  \frac{a^4}{16} + \frac{a^2}{2}\sqrt{\sigma_1^4 + \frac{a^4}{64}}}{2 \sigma_1^2}, \frac{\pi}{2}\left( \sqrt{\frac{2}{\pi}} \ \sigma_1 \ e^{-\frac{a^2}{8 \ \sigma_1^2}} + \frac{a}{2} \ P \left(|Z| < \frac{a}{2 \ \sigma_1}\right) \right)^{2} \right\rbrace,\]
then the minimum within sum of squares is given by $W_2^*$ which is achieved for the unique $b^*$ such that $b_2^* = 1$ and $b_j^* = 0$ if $j \neq 2$ and the unique optimal separating hyperplane is given by $\mathcal{H}(b) = \{ y \in \mathbb{R}^d: y_2 = 0\}$.

$\Box$

\subsection{Proof of Lemma~\ref{lemma::Gamma alternate}}
\label{sec:GammaAlt}
This proof is analogous to the proof of the matrix being positive definite for the null case as shown in Lemma~\ref{lemma::verify}. We find the matrix $G$ in the two different cases and show that it is positive definite. 
\begin{enumerate}
\item When condition (\ref{eqn::firstdimsplit}) is true, Theorem \ref{thm::optimalsplit} gives us the unique optimum as $\mu_1^* = - \mu_2^* = \left(\E[X_1|X_1 > 0], 0, \ldots, 0 \right)$, $M_{12} = \{ x = (x_1, \ldots, x_d) \in \mathbb{R}^d : x_1 = 0 \}$, $\mathbb{P}(A_1) = \mathbb{P}(A_2) = 0.5$
and $r_{12} = 2 \E[X_1|X_1 > 0]$, where 
\[\E[X_1|X_1 > 0] = \sqrt{\frac{2}{\pi}} \ \sigma_1 \ e^{-\frac{a^2}{8 \sigma_1^2}} + \frac{a}{2} \ \mathbb{P}\left(|Z| < \frac{a}{2 \sigma_1}\right). \]
From the proof of lemma \ref{lemma::verify}, we know that for $x \in M_{12}$,
\[
(x - \mu_1^*) (x - \mu_1^*)^T =  \left(\begin{array}{c c c c}
\mu_{11}^{*2} &  - \mu^*_{11} x_2 & \hdots &  - \mu^*_{11} x_d\\
- \mu^*_{11} x_2 & x_2^2 & \hdots &  x_2 x_d\\
\vdots & \vdots & \ddots & \vdots \\
- \mu^*_{11} x_d &  x_2 x_d & \hdots &  x_d^2
\end{array} \right),
\]
\[
(x - \mu_1^*) (x - \mu_2^*)^T = \left(\begin{array}{c c c c}
- \mu_{11}^{*2} &  - \mu^*_{11} x_2 & \hdots &  - \mu^*_{11} x_d\\
- \mu^*_{21} x_2 & x_2^2 & \hdots & x_2 x_d\\
\vdots & \vdots & \ddots & \vdots \\
- \mu^*_{21} x_d & x_2 x_d & \hdots & x_d^2
\end{array} \right).
\]
As before, $\mathbb{I}_{M_{12}} = \mathbb{I}_{\{X_1 = 0\}}$. Therefore,
\[
 \mu_{11}^{*2} \int_{M_{12}} f(x) \ d\sigma(x) = \E[X_1|X_1 > 0]^2 \ \frac{e^{-a^2/8 \sigma_1^2}}{\sqrt{2 \pi } \sigma_1}.
\]
 For $2 \leq j \leq d$,
\begin{align*}
 \mu^*_{11} \ \int_{M_{12}} x_j \ f(x) \ d\sigma(x)  &= \mu^*_{11} \ \mathbb{E}[X_j] \ \frac{e^{-a^2/8 \sigma_1^2}}{\sqrt{2 \pi } \sigma_1} = 0,\\
 \mu^*_{21} \ \int_{M_{12}} x_j \ f(x) \ d\sigma(x) &= \mu^*_{21} \ \mathbb{E}[X_j] \ \frac{e^{-a^2/8 \sigma_1^2}}{\sqrt{2 \pi } \sigma_1} = 0,\\
\int_{M_{12}} x_j^2 \ f(x) \ d\sigma(x) &=  \mathbb{E}[X_j^2] \ \ \frac{e^{-a^2/8 \sigma_1^2}}{\sqrt{2 \pi } \sigma_1} = \frac{\sigma_j^2}{\sqrt{2 \pi } \sigma_1} \ e^{-a^2/8 \sigma_1^2}.
\end{align*}
Let $i \neq j,$ and $ i,j \in \{ 2, \ldots, d\}$,
\[\int_{M_{12}} x_i \ x_j \ f(x) \ d\sigma(x) =  \mathbb{E}[X_i] \ \mathbb{E}[X_j] \ \ \frac{e^{-a^2/8 \sigma_1^2}}{\sqrt{2 \pi } \sigma_1}  = 0.\] 
Then the matrix $G$ can be derived as,
\begin{align*}
G_{22} = G_{11} &= \mathbf{I}_d - \frac{1}{ \E[X_1|X_1 > 0]}  \ \frac{e^{-a^2/8 \sigma_1^2}}{\sqrt{2 \pi } \sigma_1}  \left(\begin{array}{c c c c}
\E[X_1|X_1 > 0]^2& 0 & \hdots & 0 \\
0 & \sigma_2^2 & \hdots & 0 \\
\vdots & \vdots & \ddots & \vdots \\
0 & 0 & \hdots & \sigma_d^2
\end{array} \right),\\
G_{21} = G_{12} &=  \frac{1}{ \E[X_1|X_1 > 0]}  \ \frac{e^{-a^2/8 \sigma_1^2}}{\sqrt{2 \pi } \sigma_1}  \left(\begin{array}{c c c c}
\E[X_1|X_1 > 0]^2& 0 & \hdots & 0 \\
0 & - \sigma_2^2 & \hdots & 0 \\
\vdots & \vdots & \ddots & \vdots \\
0 & 0 & \hdots & - \sigma_d^2
\end{array} \right). 
\end{align*}
\citet{boyd2004convex} now gives that the symmetric matrix $G$ is positive definite if and only if $G_{11}$ and $G / G_{11}$ (the Schur complement of $G_{11}$ in $G$) are both positive definite. Let us first look at the diagonal entries of $G_{11}$ and define the following:
\begin{equation}
m_1 = \frac{e^{-a^2/8 \sigma_1^2}}{\sqrt{2 \pi } \sigma_1} \  \E[X_1|X_1 > 0], \ m_j = \frac{\sigma_j^2}{ \E[X_1|X_1 > 0]}  \ \frac{e^{-a^2/8 \sigma_1^2}}{\sqrt{2 \pi } \sigma_1}, \ j \neq 1.
\end{equation}
Then the diagonal entries of $G_{11}$ are given by $1 - m_j$, for $j =1, \ldots, d$. Next note that
\[ \frac{1}{\sqrt{2 \pi}} \ \frac{a}{\sigma_1} e^{-a^2/8 \sigma_1^2} \leq \P\left(|Z| < \frac{a}{2 \sigma_1}\right) \leq \frac{1}{\sqrt{2 \pi}} \ \frac{a}{\sigma_1}\]
and therefore we can get the following bounds on $\E[X_1 | X_1 > 0]$.
\begin{equation}
\E[X_1 | X_1 > 0] \leq \sqrt{\frac{2}{\pi}} \ \frac{1}{\sigma_1} \left( \sigma_1^2 + \frac{a^2}{4} \right), \ \ \E[X_1 | X_1 > 0] \geq \sqrt{\frac{2}{\pi}} \ \frac{e^{-a^2/8 \sigma_1^2}}{\sigma_1} \left( \sigma_1^2 + \frac{a^2}{4} \right)
\end{equation}

Using these inequalities and observing that $e^x \geq 1 + x$ for $x > 0$ along with the assumption that $\sigma_1^2 + \frac{a^2}{4} > \sigma_j^2$ for $j \neq 1$, one can easily verify that $1 - m_j > 0$ for all $j = 1, \ldots, d$. Therefore,  $G_{11}$ is a positive definite matrix. To show $G / G_{11}$ is also positive definite first we simplify it.
\[
G / G_{11} = G_{22} - G_{21} \left[ G_{11} \right]^{-1} G_{12}.
\]
Since all of them are diagonal matrices, $G / G_{11}$ is also a diagonal matrix with the $j^{th}$ entry given by $1 - m_j - \frac{m_j^2}{1 - m_j} = \frac{1 - 2 m_j}{1 - m_j}$. Since, we have already verified that $1 - m_j > 0$, we just have to verify that $1 - 2 m_j$ is also greater than $0$. This can again be easily verified with the properties as mentioned above. That is, by using the inequalities and the assumption. Therefore, $G / G_{11}$ is also a positive definite matrix, which implies $G$ itself is a positive definite matrix. 

\item When condition (\ref{eqn::seconddimsplit}) is true, Theorem \ref{thm::optimalsplit} gives us the unique optimum as $\mu_1^* = - \mu_2^* = \left(0, \E[X_2|X_2 > 0], 0, \ldots, 0 \right)$, $M_{12} = \{ x = (x_1, \ldots, x_d) \in \mathbb{R}^d : x_2 = 0 \}$, $\mathbb{P}(A_1) = \mathbb{P}(A_2) = 0.5$
and $r_{12} = 2 \E[X_2|X_2 > 0]$, where 
\[\E[X_2|X_2 > 0] = \sqrt{\frac{2}{\pi}} \ \sigma_2. \]
Analogous to previous part we can show that for $x \in M_{12}$,
\[
(x - \mu_1^*) (x - \mu_1^*)^T =  \left(\begin{array}{c c c c c}
 x_1^2 &  - \mu^*_{11} x_1  & x_1 x_3 & \hdots &  - \mu^*_{11} x_d\\
- \mu^*_{11} x_1 & \mu^{*2}_{11} & - \mu^*_{11} x_3& \hdots &  - \mu^*_{11} x_d\\
 x_1 x_3 &  - \mu^*_{11} x_3  &  x_3^2 & \hdots &  x_3 x_d\\
\vdots & \vdots & \vdots & \ddots & \vdots \\
x_1 x_d &  - \mu^*_{11} x_d & x_3 x_d & \hdots &  x_d^2
\end{array} \right),
\]
\[
(x - \mu_1^*) (x - \mu_2^*)^T =  \left(\begin{array}{c c c c c}
 x_1^2 &  - \mu^*_{21} x_1  & x_1 x_3 & \hdots &  - \mu^*_{11} x_d\\
- \mu^*_{11} x_1 & - \mu^{*2}_{11} & - \mu^*_{11} x_3& \hdots &  - \mu^*_{11} x_d\\
 x_1 x_3 &  - \mu^*_{21} x_3  &  x_3^2 & \hdots &  x_3 x_d\\
\vdots & \vdots & \vdots & \ddots & \vdots \\
x_1 x_d &  - \mu^*_{21} x_d & x_3 x_d & \hdots &  x_d^2
\end{array} \right).
\]
Now, $\mathbb{I}_{M_{12}} = \mathbb{I}_{\{X_2 = 0\}}$. Therefore,
\[
 \mu_{11}^{*2} \int_{M_{12}} f(x) \ d\sigma(x) = \frac{2 \sigma_2^2}{\pi} \frac{1}{\sqrt{2 \pi } \sigma_2} = \sqrt{\frac{2}{\pi^3}} \sigma_2.
\]
 For $j \neq 2$,
\begin{align*}
 \mu^*_{11} \ \int_{M_{12}} x_j \ f(x) \ d\sigma(x)  &= \mu^*_{11} \ \mathbb{E}[X_j] \ \frac{1}{\sqrt{2 \pi } \sigma_2} = 0,\\
 \mu^*_{21} \ \int_{M_{12}} x_j \ f(x) \ d\sigma(x) &= \mu^*_{21} \ \mathbb{E}[X_j] \ \frac{1}{\sqrt{2 \pi } \sigma_2} = 0,\\
\int_{M_{12}} x_j^2 \ f(x) \ d\sigma(x) &=  \mathbb{E}[X_j^2] \ \ \frac{1}{\sqrt{2 \pi } \sigma_2}.
\end{align*}
Therefore for $j \geq 3$,
\[ \int_{M_{12}} x_j^2 \ f(x) \ d\sigma(x) =  \frac{\sigma_j^2}{\sqrt{2 \pi } \sigma_2},\]
and for $j = 1$,
\[ \int_{M_{12}} x_j^2 \ f(x) \ d\sigma(x) =  \frac{\sigma_1^2 + \frac{a^2}{4}}{\sqrt{2 \pi } \sigma_2}.\]
Let $i \neq j,$ and $ i,j \in \{ 1, 3, \ldots, d\}$,
\[\int_{M_{12}} x_i \ x_j \ f(x) \ d\sigma(x) =  \mathbb{E}[X_i] \ \mathbb{E}[X_j] \ \ \frac{1}{\sqrt{2 \pi } \sigma_2}  = 0.\] 
Then the matrix $G$ can be derived as,
\begin{align*}
G_{22} = G_{11} &= \mathbf{I}_d - \frac{1}{2 \sigma_2^2}  \left(\begin{array}{c c c c c}
\sigma_1^2 + \frac{a^2}{4 }& 0 & 0 &  \hdots & 0 \\
0 & \frac{2 \sigma_2^2}{\pi} & 0 & \hdots & 0 \\
0 & 0 &\sigma_3^2& \hdots & 0 \\
\vdots & \vdots & \vdots & \ddots & \vdots \\
0 & 0 & 0 & \hdots & \sigma_d^2
\end{array} \right),\\
G_{21} = G_{12} &=  \frac{1}{2 \sigma_2^2} \left(\begin{array}{c c c c c}
- \sigma_1^2 - \frac{a^2}{4 }& 0 & 0 &  \hdots & 0 \\
0 & \frac{2 \sigma_2^2}{\pi} & 0 & \hdots & 0 \\
0 & 0 & - \sigma_3^2& \hdots & 0 \\
\vdots & \vdots & \vdots & \ddots & \vdots \\
0 & 0 & 0 & \hdots & - \sigma_d^2
\end{array} \right). 
\end{align*}
\citet{boyd2004convex} now gives that the symmetric matrix $G$ is positive definite if and only if $G_{11}$ and $G / G_{11}$ (the Schur complement of $G_{11}$ in $G$) are both positive definite. Since $\sigma_2^2 > \sigma_1^2 + \frac{a^2}{4}$ under the assumption (\ref{eqn::seconddimsplit}) and $\sigma_2^2 > \sigma_j^2$ for every $j \geq 3$, all the diagonal elements of $G_{11}$ are strictly positive and therefore $G_{11}$ is a positive definite matrix. Now as observed in the previous part, $G / G_{11}$ is given by
\[
G / G_{11} = G_{22} - G_{21} \left[ G_{11} \right]^{-1} G_{12},
\]
which ends up being a diagonal matrix with the $j^{th}$ entry given by $\frac{1 - 2 m_j}{1 - m_j}$, where in this part,
\[m_1 = \frac{\sigma_1^2 + \frac{a^2}{4}}{2 \sigma_2^2}, \ m_2 = \frac{1}{\pi}, \ m_j = \frac{\sigma_j^2}{2 \sigma_2^2}, \ j \geq 3.\]
$1 - m_j$ are the diagonal entries of $G_{11}$ which we have verified are greater than zero. Again since  $\sigma_2^2 > \sigma_1^2 + \frac{a^2}{4}$ and $\sigma_2^2 > \sigma_j^2$ for every $j \geq 3$, $1 - 2 m_j$ is  greater than $0$ for every $j$ and therefore, $G_0 / G_{11}$ is also a positive definite matrix, which implies $G$ itself is a positive definite matrix. 
\end{enumerate}
$\Box$

\subsection{Proofs of Additional Lemmas Supporting Theorem~\ref{thm::optimalsplit}}
\label{sec:AdditionalProofs}
\subsubsection{Proof of Lemma \ref{lemma::4.1}}
Let $Z$ be a random variable generated from $N(0,1)$. Then,
\begin{align*}
\mathbb{E}\left[ Y | Y > 0 \right] &= 2 \mathbb{E}[Y \mathbb{I}_{\{ Y > 0 \}}] \\
&= \int_{0}^\infty x f(x) dx +  \int_{0}^\infty x g(x) dx \\
&= \frac{1}{\sqrt{2 \pi}\sigma_1} \left[ \int_{0}^\infty x e^{-\frac{1}{2 \sigma_1^2} (x + a/2)^2} dx +  \int_{0}^\infty x e^{-\frac{1}{2 \sigma_1^2} (x - a/2)^2} dx \right]\\
&= \frac{1}{\sqrt{2 \pi}\sigma_1} \left[ \int_{0}^\infty \left(x + \frac{a}{2} \right) e^{-\frac{1}{2 \sigma_1^2} (x + a/2)^2} dx +  \int_{0}^\infty \left(x - \frac{a}{2} \right) e^{-\frac{1}{2 \sigma_1^2} (x - a/2)^2} dx \right] + \frac{a}{2} \mathbb{P}\left(|Z| < \frac{a}{2 \sigma_1}\right)\\
&= \frac{\sigma_1}{\sqrt{2 \pi}} \left[\left. e^{-\frac{1}{2 \sigma_1^2} (x + a/2)^2} \right\vert_{\infty}^0 +  \left. e^{-\frac{1}{2 \sigma_1^2} (x - a/2)^2} \right\vert_{\infty}^0 \right] + \frac{a}{2} \mathbb{P}\left(|Z| < \frac{a}{2 \sigma_1}\right)\\
&= \sqrt{\frac{2}{\pi}} \ \sigma_1 \ e^{-\frac{a^2}{8 \sigma_1^2}} + \frac{a}{2} \ \mathbb{P}\left(|Z| < \frac{a}{2 \sigma_1}\right). \ 
\end{align*}
$\Box$

\subsubsection{Proof of Lemma \ref{lemma::WSSb}}

The within sum of squares can be written as:
\begin{align*}
W(b) &= P(b^T X > 0) E[||X - E[X| b^T X > 0]||^2 | b^T X > 0] \\
& \ \ \ \ \ \ \ \ +  P(b^T X < 0) E[||X- E[X| b^T X < 0]||^2 | b^T X < 0]\\
&= E\left[\left. ||X- E[X| b^T X > 0]||^2 \right\vert b^T X > 0 \right], \ \ \ \ \ (\text{Since, } - X \overset{d}{=} X).
\end{align*}

Since $ b_j = 0 \ \forall \ j \geq 3$, and $X_j$ for $j \geq 3$ is independent of $X_1$ and $X_2$,
\begin{align*}
W(b) &= E\left[\left. ||X- E[X| b^T X > 0]||^2 \right\vert b^T X > 0 \right]\\
&= E\left[\left. (X_1 - E[X_1| b^T X > 0])^2 \right\vert b^T X > 0 \right] + E\left[\left. (X_2 - E[X_2| b^T X > 0])^2 \right\vert b^T X > 0 \right] + \sum_{j = 3}^d \sigma_j^2,
\end{align*}
since for $j \geq 3$, $E[X_j| b^T X > 0] = E[X_j] = 0$ and $E[X_j^2| b^T X > 0] = E[X_j^2] = \sigma_j^2$. Therefore,
\begin{equation}
\label{eqn::WSSb}
W(b) = E\left[\left. (X_1 - E[X_1| b^T X > 0])^2 \right\vert b^T X > 0 \right] + E\left[\left. (X_2 - E[X_2| b^T X > 0])^2 \right\vert b^T X > 0 \right] + \sum_{j = 3}^d \sigma_j^2.
\end{equation}

We know that,
\[ E\left[\left. (X_1 - E[X_1| b^T X > 0])^2 \right\vert b^T X > 0 \right] = E\left[\left. X_1^2 \right\vert b^T X > 0 \right]  - \left(E[X_1| b^T X > 0]\right)^2, \]
and similarly,
\[ E\left[\left. (X_2 - E[X_2| b^T X > 0])^2 \right\vert b^T X > 0 \right] = E\left[\left. X_2^2 \right\vert b^T X > 0 \right]  - \left(E[X_2| b^T X > 0]\right)^2. \]

Since $-X \overset{d}{=} X$, we can write,
\begin{align*}
E\left[X_1^2 \right] &= P\left(b^T X > 0\right) E\left[\left. X_1^2 \right\vert b^T X > 0 \right] + P\left(b^T X < 0\right) E\left[\left. X_1^2 \right\vert b^T X < 0 \right] \\
&= P\left(b^T X > 0\right) E\left[\left. X_1^2 \right\vert b^T X > 0 \right] + P\left(b^T X < 0\right) E\left[\left. X_1^2 \right\vert b^T X > 0 \right]  \\
&= E\left[\left. X_1^2 \right\vert b^T X > 0 \right] .
\end{align*}
Hence,
\begin{equation}
\label{eqn::1stdim2ndmoment}
E\left[\left. X_1^2 \right\vert b^T X > 0 \right] = E\left[X_1^2 \right] = \sigma_1^2 + \frac{a^2}{4},
\end{equation}
and following similar arguments,
\begin{equation}
\label{eqn::2nddim2ndmoment}
E\left[\left. X_2^2 \right\vert b^T X > 0 \right] = E\left[X_2^2 \right] = \sigma_2^2.
\end{equation}

To find the conditional first moments, for simplicity of notation, let us define $f$ to be the pdf of $N(-\theta_1, D)$ and $g$ to be the pdf of $N(\theta_1, D)$. Let us define a latent variable $Q \sim \mbox{Ber}(0.5)$ and define $Y \sim f$ if $Q = 0$ and $Y \sim g$ if $Q = 1$. Then $ X \overset{d}{=} Y$ and by the law of total expectation, 
\begin{align*}
E\left[\left. X_1 \right\vert b^T X > 0 \right] &= E\left[\left. Y_1 \right\vert b^T Y > 0 \right]\\
&= E\left[ \left. E[Y_1 | b^TY > 0, Q] \right\vert b^TY > 0\right]\\
&= P\left(Q = 0| b^TY > 0 \right) \ E_f[ Y_1 | b^TY > 0, Q = 0] \\
& \ \ \ \ \ \ +  P\left(Q = 1| b^TY > 0 \right) \ E_g[ Y_1 | b^TY > 0, Q = 1]\\
&= \alpha \ E_f[ Y_1 | b^TY > 0] + (1 - \alpha) \ E_g[ Y_1 | b^TY > 0],
\end{align*}
where $\alpha = P\left(Q = 0| b^TY > 0 \right)$, $E_f$ is the expectation when the distribution of $Y$ has a pdf $f$ and $E_g$ is the expectation when the pdf is $g$.

Similarly,
\begin{align*}
E\left[\left. X_2 \right\vert b^T X > 0 \right] &= \alpha \ E_f[ Y_2 | b^TY > 0] + (1 - \alpha) \ E_g[ Y_2 | b^TY > 0]\\
\end{align*} 

To simplify further, we consider each of these terms separately. We first start with $E_f[ Y_1 | b^TY > 0]$ and to compute it we define random variable $V$ such that when $Y \sim f$, that is, $Y \sim N(-\theta_1, D)$,
\[V = \left( \begin{array}{c}
V_1 \\
V_2
\end{array}\right) := \left( \begin{array}{c}
\frac{Y_1 + a/2}{\sigma_1} \\
\frac{b_1 Y_1 + b_2 Y_2 + a b_1/2}{\sqrt{b^T D b}}
\end{array}\right)
 \sim N\left( \left( \begin{array}{c}
0 \\
0
\end{array}\right), \left( \begin{array}{c c}
1 & \frac{b_1 \sigma_1}{\sqrt{b^T D b}} \\
\frac{b_1 \sigma_1}{\sqrt{b^T D b}} & 1
\end{array}\right)  \right).  \]

Note that $b^TY = b_1Y_1 + b_2 Y_2 > 0 $ is equivalent to $V_2 > \frac{ a b_1}{2 \sqrt{b^T D b}}$. In order to find $E\left[V_1\left\vert V_2 > \frac{ a b_1}{2 \sqrt{b^T D b}}\right. \right]$ we use moments derived for truncated bivariate normal distribution, presented in \citet{rosenbaum1961moments}. We get
\begin{align*}
E\left[V_1 \left\vert V_2 > \frac{ a b_1}{2 \sqrt{b^T D b}}\right. \right] &= \frac{b_1 \sigma_1}{\sqrt{b^T D b}} \left( \frac{\phi \left(\frac{ a b_1}{2 \sqrt{b^T D b}} \right)}{P\left(V_2 > \frac{ a b_1}{2 \sqrt{b^T D b}} \right)} \right)\\
&=\frac{b_1 \sigma_1}{\sqrt{b^T D b}} \left( \frac{\phi \left(\frac{ a b_1}{2 \sqrt{b^T D b}} \right)}{1 - \Phi \left(\frac{ a b_1}{2 \sqrt{b^T D b}} \right)} \right).
\end{align*}

Note that
\[
E\left[V_1 \left\vert V_2 > \frac{ a b_1}{2 \sqrt{b^T D b}}\right. \right] = E_f\left[\left. \frac{Y_1 + a/2}{\sigma_1} \right\vert b^TY > 0\right ].
\]
Therefore,
\begin{equation}
E_f\left[\left. Y_1 \right\vert b^TY > 0\right ] = - \frac{a}{2} + \frac{b_1 \sigma_1^2}{\sqrt{b^T D b}} \left( \frac{\phi \left(\frac{ a b_1}{2 \sqrt{b^T D b}} \right)}{1 - \Phi \left(\frac{ a b_1}{2 \sqrt{b^T D b}} \right)} \right).
\label{eqn::tm1}
\end{equation}

Similarly in order to find $E_f[ Y_2 | b^TY > 0]$, we define random variable $V^*$ such that when $Y \sim f$, that is, $Y \sim N(-\theta_1, D)$,
\[V^* = \left( \begin{array}{c}
V_1^* \\
V_2
\end{array}\right) := \left( \begin{array}{c}
\frac{Y_2}{\sigma_2} \\
\frac{b_1 Y_1 + b_2 Y_2 + a b_1/2}{\sqrt{b^T D b}}
\end{array}\right)
 \sim N\left( \left( \begin{array}{c}
0 \\
0
\end{array}\right), \left( \begin{array}{c c}
1 & \frac{b_2 \sigma_2}{\sqrt{b^T D b}} \\
\frac{b_2 \sigma_2}{\sqrt{b^T D b}} & 1
\end{array}\right)  \right).  \]

Then again using moments derived for truncated bivariate normal distribution, presented in \citet{rosenbaum1961moments}. We get
\begin{align*}
E\left[V_1^* \left\vert V_2 > \frac{ a b_1}{2 \sqrt{b^T D b}}\right.\right]
&=\frac{b_2 \sigma_2}{\sqrt{b^T D b}} \left( \frac{\phi \left(\frac{ a b_1}{2 \sqrt{b^T D b}} \right)}{1 - \Phi \left(\frac{ a b_1}{2 \sqrt{b^T D b}} \right)} \right).
\end{align*}
Again note that,
\[
E\left[V_1^* \left\vert V_2 > \frac{ a b_1}{2 \sqrt{b^T D b}}\right.\right] = E_f\left[\left. \frac{Y_2 }{\sigma_2} \right\vert b^TY > 0\right ].
\]
Therefore,
\begin{equation}
E_f\left[\left. Y_2 \right\vert b^TY > 0\right] = \frac{b_2 \sigma_2^2}{\sqrt{b^T D b}} \left( \frac{\phi \left(\frac{ a b_1}{2 \sqrt{b^T D b}} \right)}{1 - \Phi \left(\frac{ a b_1}{2 \sqrt{b^T D b}} \right)} \right).
\label{eqn::tm3}
\end{equation}

In order to find $E_g\left[\left. Y_1 \right\vert b^TY > 0\right]$ and $E_g\left[\left. Y_2 \right\vert b^TY > 0\right]$, note that $f$ is the density of $N(-\theta_1, D)$ and $g$ is the density of $N(\theta_1, D)$, where $\theta = (a/2, 0, \ldots, 0)$. So just replacing $a$ with $-a$ in equations (\ref{eqn::tm1}) and (\ref{eqn::tm3}) will give us the corresponding expectations when $Y \sim g$. So we get
\begin{align}
E_g\left[\left. Y_1 \right\vert b^TY > 0\right] &= \frac{a}{2} + \frac{b_1 \sigma_1^2}{\sqrt{b^T D b}} \left( \frac{\phi \left(\frac{ a b_1}{2 \sqrt{b^T D b}} \right)}{\Phi \left(\frac{ a b_1}{2 \sqrt{b^T D b}} \right)} \right)\\
E_g\left[\left. Y_2 \right\vert b^TY > 0\right] &= \frac{b_2 \sigma_2^2}{\sqrt{b^T D b}} \left( \frac{\phi \left(\frac{ a b_1}{2 \sqrt{b^T D b}} \right)}{ \Phi \left(\frac{ a b_1}{2 \sqrt{b^T D b}} \right)} \right)
\end{align}

To find $\alpha$ we note that,
\begin{align*}
\alpha &= P\left(Q = 0| b^TY > 0 \right) = \frac{P\left(Q = 0, b^TY > 0 \right)}{P\left(Q = 0, b^TY > 0 \right) + P\left(Q = 1, b^TY > 0 \right)}\\
&= \frac{P_f\left(b^TY > 0 \right)}{P_f\left(b^TY > 0 \right) + P_g\left(b^TY > 0 \right)}.
\end{align*}
Now if $Y \sim g$, then $-Y \sim f$. So $P_g\left(b^TY > 0 \right) =  P_f\left(b^TY < 0 \right) = 1 - P_f\left(b^TY > 0 \right).$ Now if $Y \sim f$, then $b^T Y \sim N(-ab_1/2, b^T D b)$. Therefore,
\begin{equation}
\alpha = P_f\left(b^TY > 0 \right) = P_f\left(\frac{b^TY + ab_1/2}{\sqrt{b^TDb}} > \frac{ab_1/2}{\sqrt{b^TDb}} \right) = 1 - \Phi\left(\frac{ab_1}{2 \sqrt{b^TDb}} \right).
 \label{eqn::splitalpha}
\end{equation}

Using all of the above equations we get:
\begin{align*}
E\left[\left. X_1 \right\vert b^T X > 0 \right] &= \alpha E_f\left[\left. Y_1 \right\vert b^T Y > 0 \right] + (1 - \alpha) E_g\left[\left. Y_1 \right\vert b^T Y > 0 \right]\\
&=  \left( 2  \Phi\left( \frac{ab_1}{2 \sqrt{b^TDb}}\right)  - 1 \right) \frac{a}{2} + \frac{2 b_1 \sigma_1^2}{\sqrt{b^T D b}}  \phi \left(\frac{ a b_1}{2 \sqrt{b^T D b}} \right).
\end{align*}

Similarly,
\begin{align*}
E\left[\left. X_2 \right\vert b^T X > 0 \right] &= \alpha E_f\left[\left. Y_2 \right\vert b^T Y > 0 \right] + (1 - \alpha) E_g\left[\left. Y_2 \right\vert b^T Y > 0 \right]\\
&= \frac{2 b_2 \sigma_2^2}{\sqrt{b^T D b}}  \phi \left(\frac{ a b_1}{2 \sqrt{b^T D b}} \right).
\end{align*}

Plugging the expressions for $E\left[\left. X_1 \right\vert b^T X > 0 \right]$ and $E\left[\left. X_2 \right\vert b^T X > 0 \right]$ along with equations (\ref{eqn::1stdim2ndmoment}) and (\ref{eqn::2nddim2ndmoment}) into equation (\ref{eqn::WSSb}), we get that:
\begin{equation*}
W(b) = \sum_{j = 1}^d \sigma_j^2 + \frac{a^2}{4} - \left[ \left( 2  \Phi\left( \frac{ab_1}{2 \sqrt{b^TDb}}\right)  - 1 \right) \frac{a}{2} + \frac{2 b_1 \sigma_1^2}{\sqrt{b^T D b}}  \phi \left(\frac{ a b_1}{2 \sqrt{b^T D b}} \right)\right]^2 - \frac{4 b_2^2 \sigma_2^4}{b^T D b}  \phi^2 \left(\frac{ a b_1}{2 \sqrt{b^T D b}} \right).
\end{equation*}

$\Box$

\subsubsection{Proof of Lemma \ref{lemma::WSSbi1}}

The within sum of squares can be written as:
\begin{align*}
W(b) &= P(b^T X > 0) E[||X - E[X| b^T X > 0]||^2 | b^T X > 0] \\
& \ \ \ \ \ \ \ \ +  P(b^T X < 0) E[||X- E[X| b^T X < 0]||^2 | b^T X < 0]\\
&= E\left[\left. ||X- E[X| b^T X > 0]||^2 \right\vert b^T X > 0 \right], \ \ \ \ \ (\text{Since, } - X \overset{d}{=} X)\\
&= E\left[\left. ||X- E[X| X_i > 0]||^2 \right\vert b^T X_i > 0 \right]\\
&= E\left[\left(X_i - E[X_i| X_i > 0]\right)^2| X_i  > 0\right] + \sum_{j \neq i} E[X_j^2].
\end{align*}

Recall that $X_i \sim N(0, \sigma_i^2)$, therefore $E[X_i| X_i > 0] = \sqrt{\frac{2}{\pi}} \sigma_i$. This implies that the within sum of squares is given by,
\begin{align*}
W(b) &= Var( X_i | X_i > 0) +  E[X_1^2] + \sum_{j \neq i, j \neq 1} E[X_j^2]\\
& = \sigma_i^2 - \frac{2}{\pi} \sigma_i^2 + \sigma_1^2 + \frac{a^2}{4} + \sum_{j \neq i, j \neq 1} \sigma_j^2 \\
&= \sum_{j = 1}^d \sigma_j^2 +  \frac{a^2}{4}  - \frac{2}{\pi} \sigma_i^2.
\end{align*}

$\Box$

\subsubsection{Proof of Lemma \ref{lemma::1Nprojection}}

 First, for any point $v = (v_1, \ldots, v_d) \in \mathbb{R}^d$, its projection, $u = (u_1, \ldots, u_d)$, onto the hyperplane $\mathcal{H}(b) = \{ y \in \mathbb{R}^d: b^T y = 0\}$ is given by the following equations:
\[ u_i = v_i + b_i t, \text{ for some } t, \ \forall \ i \text{ and } b^Tu = 0, \]
solving which gives us that for every $i$,
\[u_i =  v_i - b_i \sum_{i = 1}^d b_i v_i = v_i - b_i (b^T v).\]

Now for simplicity, define $Z_{1i} = Y_i - b_i (b^T Y)$, the projection of $Y$ onto the plane and $Z_2 =  b^TY$. Therefore, the $i^{th}$ coordinate of the projection of $E[Y|b^T Y > 0]$ is given by:
\begin{align*}
\mathcal{P}_i &=  E\left[Y_i | Z_2 > 0\right] - b_i \left(b^T E\left[Y| Z_2 > 0\right] \right)\\
&=  E\left[Y_i - b_i (b^T Y) | Z_2 > 0\right] \\
&= E\left[Z_{1i} | Z_2 > 0\right]\\
&= \frac{E\left[Z_{1i} I\{Z_2 > 0\}\right]}{P(Z_2 > 0)}\\
&= \frac{E\left[E\left[Z_{1i} I\{Z_2 > 0\}|Z_2\right]\right]}{P(Z_2 > 0)}\\
&= \frac{E\left[I\{Z_2 > 0\} E\left[Z_{1i}|Z_2\right]\right]}{P(Z_2 > 0)}\\
&=E\left[\left.E\left[Z_{1i}|Z_2\right]\right\vert Z_2 > 0 \right]\\
&=E\left[\left.E[Z_{1i}] + \frac{Cov(Z_{1i}, Z_2)}{Var(Z_2)} \left( Z_2 - E[Z_2]\right)\right\vert Z_2 > 0 \right]\\
&=E[Z_{1i}] + \frac{Cov(Z_{1i}, Z_2)}{Var(Z_2)} \left( E\left[\left.Z_2\right\vert Z_2 > 0 \right] - E[Z_2]\right).
\end{align*}

Note that we can do this because $Y$ is a multivariate normal random variable and therefore $Z_{1i}$ and $Z_2$ are jointly normal. Now since $Y \sim N(\theta_1, D)$, where $\theta_1 = (a/2, 0, \ldots, 0)$,  and $D$ is a diagonal matrix,
\[E[Z_2] = E\left[b^TY \right] = b^T E\left[Y \right] =  \frac{a b_1}{2},\]
\[ E[Z_{1i}] = E\left[Y_i - b_i (b^T Y)\right] = E\left[Y_i - b_i Z_2\right] = E\left[Y_i\right] - b_i E\left[Z_2\right] = \frac{a}{2} \mathbb{I}\{i = 1\} - \frac{a b_1 b_i}{2},\]
\[ Var(Z_2) = Var\left(b^TY\right) = b^T D \ b,\]
 and
\[Cov(Z_{1i}, Z_2) =  Cov\left(Y_i - b_i Z_2, Z_2\right) = Cov\left(Y_i, Z_2\right) - b_i Var\left(Z_2\right) = b_i Var(Y_i) - b_i \left(b^T D \ b\right).\]
Therefore the projection is given by:
\[ \mathcal{P}_i  = \frac{a}{2} \mathbb{I}\{i = 1\} - \frac{a b_1 b_i}{2} + \frac{b_i Var(Y_i) - b_i \left(b^T D b\right)}{b^T D b} \ \left(E\left[\left.Z_2\right\vert Z_2 > 0 \right] - \frac{a b_1}{2}\right).\]
$\Box$

\subsubsection{Proof of Lemma~\ref{lemma::lowerboundW}}
\label{app:lowerboundW}
Our goal is to lower bound the term:
\begin{align*}
T := \left[ \frac{a}{2} \mathbb{P}(|Z| \leq u) + \sqrt{\frac{2}{\pi}} \sigma_1 \exp(-u^2/2) \right]^2,
\end{align*}
where $u = a/(2\sigma_1)$.
Defining,
\begin{align*}
R := \left[ \frac{a}{2} \mathbb{P}(|Z| \leq u) + \sqrt{\frac{2}{\pi}} \sigma_1 \exp(-u^2/2) \right],
\end{align*}
we see that if we can lower bound $R$ with $k$, i.e., find $k$ such that $R \geq k > 0$, then $k^2$ is a lower bound on $T^2$. 
We consider two cases:

\noindent {\bf Case when $0 \leq u \leq 2$:}
We first claim that the following hold for all $u \geq 0$:
\begin{align*}
\exp(-u^2/2) &\geq 1 - \frac{u^2}{2} + \frac{u^4}{8} - \frac{u^6}{48} + \frac{u^8}{384} - \frac{u^{10}}{3840} \\
\mathbb{P}(|Z| \leq u) &\geq \sqrt{\frac{2}{\pi}} \left[ u - \frac{u^3}{6} +  \frac{u^5}{40} - \frac{u^7}{336} + \frac{u^9}{3456} - \frac{u^{11}}{42240} \right],
\end{align*}
and both lower bounds are positive for $0 \leq u \leq 2.$
The first bound follows from the Taylor expansion of $\exp(x)$. For the second we use that:
\begin{align*}
\mathbb{P}(|Z| \leq u) &= \sqrt{\frac{2}{\pi}} \int_0^u \exp(-x^2/2) dx \\
&\geq \sqrt{\frac{2}{\pi}}\int_0^u \left(1-\frac{x^2}{2} + \frac{x^4}{8} - \frac{x^6}{48} + \frac{x^8}{384} - \frac{x^{10}}{3840}  \right) dx.
\end{align*}
The fact that the bounds are positive for the specified range can be directly verified.

\noindent Now, we can simply plug-in these estimates to obtain the following:
\begin{align*}
\frac{-2}{\pi} \left( \sigma_1^2 + \frac{a^2}{4} \right) + T &\geq \frac{-2}{\pi} \left( \sigma_1^2 + \frac{a^2}{4} \right) +\frac{a^2}{2\pi} \left[ \frac{1}{u} + \frac{u}{2} - \frac{u^3}{24} + \frac{u^5}{240} - \frac{u^7}{2688} + \frac{u^9}{34560} - \frac{u^{11}}{42240}\right]^2, \\
&= \frac{a^2u^2}{2\pi} \left[ \frac{1}{6} - \frac{u^2}{30} + \frac{13 u^4}{2520} - \frac{u^6}{1512} + 
\frac{797 u^{8}}{26611200} - \frac{233 u^{10}}{7983360} + \frac{42067 u^{12}}{17882726400} \right. \\
&~~~~~ \left. - \frac{559 u^{14}}{2554675200} + \frac{1697 u^{16}}{91968307200} - \frac{u^{18}}{729907200} + \frac{u^{20}}{1784217600} \right].
\end{align*}
and using the fact that $u \leq 2$ to bound the negative terms (and dropping some positive terms) we obtain that,
\begin{align*}
\frac{-2}{\pi} \left( \sigma_1^2 + \frac{a^2}{4} \right) + T &\geq \frac{a^2 u^2}{2\pi} \left[ \frac{1}{6} - \frac{u^2}{30} + \frac{19u^4}{7560} - \frac{6929 u^8}{79833600}   \right] \\
&\geq \frac{a^2 u^2}{2\pi} \left[ \frac{1}{6} - \frac{u^2}{30}   \right] \geq \frac{a^2 u^2}{60\pi},
\end{align*}
as desired. 

\vspace{0.1cm}

\noindent  {\bf Case when $u \geq 2$: } Observe that if $u^2 \geq 4$, then:
\begin{align*}
\frac{-2}{\pi} \left( \sigma_1^2 + \frac{a^2}{4} \right) + T \geq  \frac{a^2}{40}. 
\end{align*}
Notice that since $u^2 \geq 4$, we obtain that,
\begin{align*}
\frac{2}{\pi} \left( \sigma_1^2 + \frac{a^2}{4} \right) \leq \frac{5a^2}{8\pi}.
\end{align*}
We also notice that we can verify numerically that,
\begin{align*}
T \geq \frac{a^2}{4} (\mathbb{P}(|Z| \leq 2))^2 \geq \frac{9a^2}{40}.
\end{align*}
Putting these two bounds together yields the desired result.

$\Box$

\section{Proof of Theorem~\ref{thm::power1}}
\label{sec:power}
Throughout this proof we use $c, C, c_1, C_1, \ldots$ to denote positive constants whose value may change from line to line.
Recall, that in studying the power of SigClust we suppose that,
we observe samples:
\begin{align}
\label{eqn:model_repeat}
\{X_1,\ldots,X_n\} \sim \frac{1}{2} N(-\theta_1, D) +
\frac{1}{2} N(\theta_1, D)
\end{align}
where $\theta_1 = (a/2, 0, \ldots, 0) \in \mathbb{R}^d$ and $a > 0$. Furthermore, $D$ is a diagonal
matrix with elements $\Sigma_{jj} = \sigma_j^2$, such that $\sigma_1^2, \sigma_2^2 > \sigma_3^2 \geq \ldots \geq \sigma_d^2$. 
Recall that our goal is to show that when condition \eqref{eqn::firstdimsplit},
\[
\sigma_2^2 < \sigma_1^2 + \frac{a^2}{4}
\]
holds, SigClust is asymptotically consistent, and when condition \eqref{eqn::seconddimsplit},
\[
\sigma_2^2 > \max\left\lbrace \frac{2 \sigma_1^4 +  \frac{a^4}{16} + \frac{a^2}{2}\sqrt{\sigma_1^4 + \frac{a^4}{64}}}{2 \sigma_1^2}, \frac{\pi}{2} \ \kappa^{2} \right\rbrace
\]
holds, SigClust is asymptotically inconsistent. Before we embark on the proof of the theorem we first recollect that Theorem~\ref{thm::optimalsplit} gave a characterization of the population-level optimal symmetric $2$-means solution in this model. Under the model in~\eqref{eqn:model_repeat} described above, the population-level optimal $2$-means solution is unique and is given by \eqref{eqn:optsplit1} and \eqref{eqn:optsplit2}. 

Let us now first derive the power of the test in terms of the limiting distribution of the statistic under the null and alternate. We let $\mathbb{P}_0$ denote the Gaussian distribution with
mean 0, diagonal covariance matrix:
\begin{align*}
D_0 = \left[ \begin{matrix} 
\sigma_1^2 + \frac{a^2}{4} & 0 & 0 \ldots & 0 \\
0 & \sigma_2^2 & 0 &\ldots & 0 \\
\vdots \\
0 & 0 & 0 & \ldots & \sigma_d^2 
\end{matrix} \right].
\end{align*}
and use $\mathbb{P}_1$ to denote the distribution in~\eqref{eqn:model_repeat}. 
We let $W_0(\vecmu_0)$ denote the population optimal $2$-means value under $\mathbb{P}_0$, and 
let 
\begin{align*}
\tau_0^2 = \sum_{i = 1}^2 \mathbb{P}_0(A_i) \mathbb{E}_{X \sim \mathbb{P}_0}[\|X- \mathbb{E}[X| X \in A_i]\|^4 | X \in A_i] - [W_0(\vecmu_0)]^2.
\end{align*}
Similarly, we let $W_1(\vecmu_1)$ be the population optimal $2$-means value under $\mathbb{P}_1$, and 
let 
\begin{align*}
\tau_1^2 = \sum_{i = 1}^2 \mathbb{P}_1(A_i) \mathbb{E}_{X \sim \mathbb{P}_1}[\|X- \mathbb{E}[X| X \in A_i]\|^4 | X \in A_i] - [W_1(\vecmu_1)]^2.
\end{align*}
With this notation in place the following result characterizes the power of SigClust. We let $\Phi$ denote the standard normal CDF.
\begin{lemma}
\label{lem:sigpower}
SigClust has power:
\begin{align*}
\text{Power}_n(a) = \Phi\left(\frac{\tau_0 \Phi^{-1}(\alpha)}{\tau_1} + \sqrt{n} \frac{W_0(\vecmu_0) - W_1(\vecmu_1)}{\tau_1}\right).
\end{align*}
\end{lemma}
\noindent We prove this result in Appendix~\ref{app:sigpower}, but note that it follows from straightforward calculations based on Lemma~\ref{lem:bock} and Theorem~\ref{thm::newPollard}.
As a consequence of this result, we have the following characterization of SigClust:
\begin{lemma}
\label{lem:conditions}
Suppose that for some constant $C > 0$,
\begin{align}
\label{eqn:conscone}
\frac{\tau_0}{\tau_1} &\leq C~~~~\text{and}, \\
\label{eqn:consctwo}
\sqrt{n} \frac{W_0(\vecmu_0) - W_1(\vecmu_1)}{\tau_1} &\rightarrow \infty,~~~~\text{as}~n \rightarrow \infty,
\end{align}
then SigClust is asymptotically consistent. On the other hand if, 
\begin{align}
\label{eqn:consicone}
\frac{\tau_0}{\tau_1} &\leq C~~~~\text{and}, \\
\label{eqn:consictwo}
W_1(\vecmu_1) &= W_0(\vecmu_0)
\end{align}
then SigClust is asymptotically inconsistent.
\end{lemma}
\noindent 
This Lemma provides sufficient conditions for consistency and inconsistency respectively and we proceed to verify these conditions in the sequel.
The proof of this Lemma is straightforward and is omitted. 

To find the expression for $W_0(\vecmu_0)$, note that in~\eqref{eqn:muopt} we had calculated the population optimal within sum of squares for regular $2-$means clustering. But now since the test statistic considers a symmetric version of $2-$means clustering we need a version of Lemma~\ref{lem:bock} for the within sum of squares for symmetric $2-$means clustering,
\[ W_n^{(0)}(t) = \frac{1}{n}\sum_{i = 1}^n \min \{||X_i -  t||^2, ||X_i +  t||^2\},\]
as given by \eqref{eqn::symmsampleWSS}. This is easily provided by an analogous version of Theorem~\ref{thm::newPollard} for a single Normal distribution as follows:
\begin{lemma}
Let the data be generated from $N(0, D_0)$, as defined above, and $\tau_0$ and $W_0(\vecmu_0)$ be as given by~\eqref{eqn:tausq} and~\eqref{eqn:muopt}. Then as $n \to \infty$, 
\[\sqrt{n}( W_n^{(0)}(\mathbf{b_n}^{(0)}) - W_0(\mu_0)) \rightsquigarrow N(0, \tau_0^{2}),\] 
where $W_n^{(0)}(\mathbf{b_n}^{(0)}) = \min_t W_n^{(0)}(t)$, the minimum within sum of squares for symmetric $2-$means clustering
\label{lemma::PollardNull}
\end{lemma}

We skip the proof of this lemma as it follows exactly along the lines of the proof of Theorem~\ref{thm::newPollard} along with the observation that the unique $ \mu_0$ that minimizes the within sum of squares for regular $2-$means is itself symmetric and hence it also minimizes the symmetric version. Additionally the positive definiteness of the corresponding matrix $G_0$ has already been shown in Lemma~\ref{lemma::verify}. 

So given the expressions for $\tau_0$ and $W_0(\vecmu_0)$ in~\eqref{eqn:tausq} and~\eqref{eqn:muopt}, it now remains to calculate $\tau_1$ and $W_1(\vecmu_1)$ to analyze the power of SigClust.
The following Lemma builds on Thorem~\ref{thm::optimalsplit} to calculate these quantities. We analyze two cases which depend on whether the optimal population-level split occurs along the first or second coordinate. 
\begin{lemma}
\label{lem:t1w1}
There are universal constants $0 < c \leq C$ such that:
\begin{enumerate}
\item If \eqref{eqn::firstdimsplit} holds,
then:
\begin{align*}
W_1(\vecmu_1) &= \sum_{j = 1}^d \sigma_j^2 +  \frac{a^2}{4}  - \kappa^2. \\
c &\leq \tau_1^2 \leq C.
\end{align*}
\item If \eqref{eqn::seconddimsplit} holds, then:
\begin{align*}
W_1(\vecmu_1) &= W_0(\vecmu_0), \\
c &\leq \tau_1^2 \leq C.
\end{align*}
\end{enumerate}
\end{lemma}
\noindent We prove this result in Appendix~\ref{app:t1w1}. To complete the proof of the Theorem we need to put together Lemmas~\ref{lem:conditions} and~\ref{lem:t1w1} to show the consistency and inconsistency of SigClust in different regimes. 

We note that using~\eqref{eqn:needtwo} and the result of Lemma~\ref{lem:t1w1} that both $\tau_0^2$ and $\tau_1^2$ are bounded by constants (recall that we take $\{a, \sigma_1^2,\ldots,\sigma_d^2\}$ to be fixed) as $n \rightarrow \infty$ verifying Conditions~\eqref{eqn:conscone} and~\eqref{eqn:consicone}.
Thus, Lemmas~\ref{lem:conditions} and~\ref{lem:t1w1} directly yield the inconsistency of SigClust when condition \eqref{eqn::seconddimsplit} holds. On the other hand, in order to establish consistency when \eqref{eqn::firstdimsplit} holds, to verify Condition~\eqref{eqn:consctwo} we note that for some constant $c > 0$,
\begin{align*}
\sqrt{n} \frac{W_0(\vecmu_0) - W_1(\vecmu_1)}{\tau_1} &\geq c \sqrt{n}\left[W_0(\vecmu_0) - W_1(\vecmu_1)\right] \\
&= c \sqrt{n} \underbrace{\left[ \kappa^2 - \frac{2}{\pi} \max \left\{\left( \sigma_1^2 + \frac{a^2}{4} \right), \sigma_2^2 \right\}\right]}_{T}, \\
\end{align*}
so to complete the proof of the Theorem it suffices to lower bound the term $T > 0$ as $n \rightarrow \infty,$ when \eqref{eqn::firstdimsplit} holds. Clearly in this regime $\kappa^2 -   2\sigma_2^2/\pi > 0$ so Lemma~\ref{lemma::lowerboundW} as stated in Appendix~\ref{sec:AlternativeSplit}, completes the proof of our Theorem.

\subsection{Proof of Lemma~\ref{lem:sigpower}}
\label{app:sigpower}

Under the null, the distribution of the statistic follows from Theorem~\ref{thm:null} and Lemma~\ref{lemma::PollardNull}. Concretely, for
\begin{align}
\label{eqn:needone}
W_0(\vecmu_0) &=  \widetilde{\sigma}^2   - \frac{2}{\pi} \max\left\lbrace \left( \sigma_1^2 + \frac{a^2}{4} \right), \sigma_2^2 \right\rbrace, \\
\label{eqn:needtwo}
\tau_0^2 &= 2\sum_{i = 2}^d \sigma_i^4 + 2\left( \sigma_1^2 + \frac{a^2}{4} \right)^2  - \frac{16}{\pi^2} \left[\max\left\lbrace \left( \sigma_1^2 + \frac{a^2}{4} \right), \sigma_2^2 \right\rbrace\right]^2,
\end{align}
we have by a combination of Theorem~\ref{thm:null} and Lemma~\ref{lemma::PollardNull} that we would expect under the null that, 
\begin{align*}
\sqrt{n} \Big(T_n^{(0)} - \frac{W_0(\vecmu_0)}{\widetilde{\sigma}^2}\Big) \rightsquigarrow 
N\Big(0, \left[\frac{\tau_0}{\widetilde{\sigma}^2}\right]^2\Big), \ \ \text{as} \ \ n \to \infty.
\end{align*}
Thus, we reject at level $\alpha$, if:
\begin{align*}
\sqrt{n} \Big(T_n^{(0)} -  \frac{W_0(\vecmu_0)}{\widetilde{\sigma}^2}\Big) \leq \frac{\tau_0 \Phi^{-1}(\alpha)}{\widetilde{\sigma}^2}.
\end{align*}
Under the alternate we can once again use Theorem~\ref{thm::newPollard} to obtain that,
\begin{align}
\label{eqn:limit_alt}
\sqrt{n} \Big(T_n^{(0)} -  \frac{W_1(\vecmu_1)}{\widetilde{\sigma}^2}\Big) \rightsquigarrow N\Big(0, \left[\frac{\tau_1}{\widetilde{\sigma}^2}\right]^2\Big),
\end{align}
where $W_1(\vecmu_1)$ denotes the optimal $2$-means objective under the alternate, and 
\begin{align*}
\tau_1^2 =  \sum_{i = 1}^2 \mathbb{P}_1(A_i) \mathbb{E}_{X \sim \mathbb{P}_1} [\|X- \mathbb{E}[X| X \in A_i]\|^4 | X \in A_i] - [W_1(\vecmu_1)]^2,
\end{align*}
where $\{A_1,A_2\}$ denotes the Voronoi partition induced by $\vecmu_1$.
Accordingly letting $\mathbb{P}_1$ denote the distribution in~\eqref{eqn:model_repeat} we have that, 
\begin{align*}
\text{Power}_n(a) &= \mathbb{P}_1\left(\sqrt{n} \Big(T_n^{(0)} -  \frac{W_0(\vecmu_0)}{\widetilde{\sigma}^2}\Big) \leq \frac{\tau_0 \Phi^{-1}(\alpha)}{\widetilde{\sigma}^2}\right) \\
&= \mathbb{P}_1 \left( \frac{\sqrt{n} \widetilde{\sigma}^2}{\tau_1} \Big(T_n^{(0)} -  \frac{W_1(\vecmu_1)}{\widetilde{\sigma}^2}\Big) \leq \frac{\tau_0 \Phi^{-1}(\alpha)}{\tau_1} + \sqrt{n} \frac{W_0(\vecmu_0) - W_1(\vecmu_1)}{\tau_1}\right) \\
&\stackrel{\text{(i)}}{=} \Phi\left(\frac{\tau_0 \Phi^{-1}(\alpha)}{\tau_1} + \sqrt{n} \frac{W_0(\vecmu_0) - W_1(\vecmu_1)}{\tau_1}\right),
\end{align*}
where (i) follows from~\eqref{eqn:limit_alt}.

\subsection{Proof of Lemma~\ref{lem:t1w1}}
\label{app:t1w1}
We divide our analysis into two cases, according to the optimal $2$-means solution.

\noindent {\bf When condition \eqref{eqn::firstdimsplit} holds: } In this case, the population optimal $2$-means split is along the first coordinate. The expression for $W_1(\vecmu_1)$ follows from \eqref{eqn:w1mu}, and it only remains to bound $\tau_1^2$. 
To lower bound $\tau_1^2$ we note that,
\begin{align*}
\tau_1^2 &= \sum_{i = 1}^2 \mathbb{P}_1(A_i) \mathbb{E}_{X\sim\mathbb{P}_1}[\|X- \mathbb{E}[X| X \in A_i]\|^4 | X \in A_i] - [W_1(\vecmu_1)]^2 \\
&= 3 \sum_{j=2}^d \sigma_j^4 + \sum_{i,j \neq 1, j \neq i} \sigma_i^2 \sigma_j^2 + \mathbb{E}[(X_1 - \kappa)^2 | X_1 \geq 0] \sum_{j \neq 1} \sigma_j^2 \\
&~~~~~~+ \mathbb{E}[(X_1 - \kappa)^4 | X_1 \geq 0] -  \left[ \mathbb{E}[(X_1 - \kappa)^2| X_1 \geq 0] + \sum_{j = 2}^d \sigma_j^2 \right]^2 \\
&=  2 \sum_{j =2}^d \sigma_j^4 + \text{var}\left((X_1 - \kappa)^2| X_1  \geq 0\right).
\end{align*}
Using the fact that the variances and $a$ are all fixed and bounded above and below we obtain that for two universal
constants $0 < c \leq C$,
\begin{align*}
c \leq \tau_1^2 \leq C.
\end{align*}

\vspace{0.2cm}

\noindent {\bf When condition \eqref{eqn::seconddimsplit} holds: } 
In this case, the population optimal $2$-means split is along the second coordinate. The expression for $W_1(\vecmu_1)$ follows from \eqref{eqn:w1mu2}, and once again it only remains to bound $\tau_1^2$.
In this case,
\begin{align*}
\tau_1^2 &= \sum_{i = 1}^2 \mathbb{P}_1(A_i) \mathbb{E}_{X\sim\mathbb{P}_1}[\|X- \mathbb{E}[X| X \in A_i]\|^4 | X \in A_i] - [W_1(\vecmu_1)]^2.
\end{align*}
Noting that,
\begin{align*}
\mathbb{E}[X_1^2] = \sigma_1^2 + \frac{a^2}{4},~~~\mathbb{E}[X_1^4] = \frac{a^4}{16} + 3 \sigma_1^4 + 3 \sigma_1^2 \frac{a^2}{2},
\end{align*}
we obtain
\begin{align*}
\tau_1^2 &=  \left[\frac{a^4}{16} + 3 \sigma_1^4 + 3 \sigma_1^2 \frac{a^2}{2}\right] + 2 \sum_{j = 3}^d \sigma_j^4 + 
\text{var}[(X_2 - \sqrt{2}{\pi} \sigma_2)^4 | X_2 \geq 0] -  \left[\sigma_1^2 + \frac{a^2}{4}\right]^2 \\
&= 2 \sum_{j \neq 2} \sigma_j^4 + \text{var}[(X_2 - \sqrt{2}{\pi} \sigma_2)^4 | X_2 \geq 0] + \sigma_1^2 a^2.
\end{align*}
Once again using the fact that the variances and $a$ are all fixed and bounded above and below we obtain that for two universal
constants $0 < c \leq C$,
\begin{align*}
c \leq \tau_1^2 \leq C,
\end{align*}
as desired.

\clearpage


\section{Proof of our main results for \riftspace}

In this Appendix, we collect the proofs of the main results for the \riftspaces in our paper. In Sections~\ref{sec:BE1} and~\ref{sec:BE2} we consider the limiting distributions of the \riftspace statistic, and its $\ell_2$ counterpart under the null and prove Theorems~\ref{thm::BE1} and~\ref{thm::BE2}. In Section~\ref{sec:riftpower} we consider Thorem~\ref{thm::goodpower} and analyze the power of the \riftspace and finally in Section~\ref{sec:validMixture} we consider Theorem~\ref{thm::valid} where we verify the validity of the modified \riftspace to test for mixtures of two Normals. 

\subsection{Proof of Theorem \ref{thm::BE1}}
\label{sec:BE1} 
In the following proof all probabilities and expectations are taken conditioned on $\mathcal{D}_1$.
By the Berry-Esseen theorem, given $\mathcal{D}_1$, 
\[ \sup_t|\mathbb{P}(\sqrt{n}(\hat\Gamma - \Gamma)\leq t) - \mathbb{P}(Z \leq t)| \leq \frac{C_0 \rho}{\tau^3 \sqrt{n}},\]
where $C_0$ is a constant,
\begin{equation}\label{eq::tau}
\rho = \mathbb{E}\left[\left\vert \tilde R_i  - \Gamma \right\vert^3 \right] \ \ \ \ \text{and} \ \ \ \ 
\tau^2 =  \mathbb{E}\left[\left( \tilde R_i  - \Gamma \right)^2 \right].
\end{equation}

Now $\Gamma = \mathbb{E}\left[\tilde R_i \right]$, therefore,
\[
\tau^2 =  {\rm Var}\left( \tilde R_i\right) = \text{Var}\left( R_i + \delta{Z_i} \right) \geq \delta^2 \text{Var}(Z_i) = \delta^2 >0.
\]
Now note that,
\begin{align*}
|R_i| &= \left\vert \log \left( \frac{\hat{p}_2(X_i)}{\hat{p}_1(X_i)} \right) \right\vert =\left\vert \log \hat{p}_2(X_i) - \log {\hat{p}_1(X_i)} \right\vert  \\
&\leq \max_{x} \left\lbrace  \log \hat{p}_2(x) - \log {\hat{p}_1(x)},  \log \hat{p}_1(x) - \log {\hat{p}_2(x)}\right\rbrace  \\
&\leq \max_{x, i \in \{1,2 \}} \left\lbrace  \log \hat{f}_i(x) - \log {\hat{p}_1(x)},  \log \hat{p}_1(x) - \log {\hat{f}_i(x)}\right\rbrace.
\end{align*}
Then we get,
\begin{align*}
|R_i| \leq \max_{i \in \{1,2 \}} & \left\lbrace  \frac{1}{2} \log \left(\frac{|\hat\Sigma|}{|\hat\Sigma_i|} \right) + \frac{1}{2} \left(\hat\mu^T \hat\Sigma^{-1} \hat\mu - \left( \hat\Sigma^{-1} \hat\mu - \hat\Sigma_i^{-1} \hat\mu_i\right)^T \left(\hat\Sigma^{-1} - \hat\Sigma_i^{-1} \right)^{-1}\left( \hat\Sigma^{-1} \hat\mu - \hat\Sigma_i^{-1} \hat\mu_i\right) \right),\right.  \\
& \  \left.\frac{1}{2} \log \left(\frac{|\hat\Sigma_i|}{|\hat\Sigma|} \right) +  \frac{1}{2} \left(\hat\mu_i^T \hat\Sigma_i^{-1} \hat\mu_i - \left( \hat\Sigma_i^{-1} \hat\mu_i - \hat\Sigma^{-1} \hat\mu\right)^T \left(\hat\Sigma_i^{-1} - \hat\Sigma^{-1} \right)^{-1}\left( \hat\Sigma_i^{-1} \hat\mu_i - \hat\Sigma^{-1} \hat\mu\right) \right)\right\rbrace.
\end{align*}
Now we notice that since for $i = 1,2$, $\mu, \hat\mu_i \in \mathcal{A}$ and the eigenvalues of $\hat\Sigma, \hat\Sigma_i$ lie in a bounded set, there exists a constant $k \geq 0$ such that 
\begin{align*}
    &\hat\mu^T \hat\Sigma^{-1} \hat\mu - \left( \hat\Sigma^{-1} \hat\mu - \hat\Sigma_i^{-1} \hat\mu_i\right)^T \left(\hat\Sigma^{-1} - \hat\Sigma_i^{-1} \right)^{-1}\left( \hat\Sigma^{-1} \hat\mu - \hat\Sigma_i^{-1} \hat\mu_i\right) \leq k,\\
    &\hat\mu_i^T \hat\Sigma_i^{-1} \hat\mu_i - \left( \hat\Sigma_i^{-1} \hat\mu_i - \hat\Sigma^{-1} \hat\mu\right)^T \left(\hat\Sigma_i^{-1} - \hat\Sigma^{-1} \right)^{-1}\left( \hat\Sigma_i^{-1} \hat\mu_i - \hat\Sigma^{-1} \hat\mu\right) \leq k.
\end{align*}

Also note that,
\[ \left\vert \log \left(\frac{|\hat\Sigma|}{|\hat\Sigma_i|} \right)\right\vert \leq d\log \left(\frac{c_2}{c_1} \right) \ \ \text{and} \ \ 
\left\vert \log \left(\frac{|\hat\Sigma_i|}{|\hat\Sigma|} \right)\right\vert \leq d\log \left(\frac{c_2}{c_1} \right).\]

Therefore
\[ |R_i| \leq \frac{1}{2} \left(d \log \left(\frac{c_2}{c_1} \right) + k\right) = C_1.\]
So we can also say,
\[|\Gamma| = \left\vert \mathbb{E}\left[R_i \right] \right\vert \leq C_1.\]

Then,
\begin{align*}
\rho = \mathbb{E}\left[\left\vert \tilde R_i  - \Gamma \right\vert^3 \right] &\leq 
\mathbb{E}\left[\left( |\tilde R_i|  + |\Gamma| \right)^3 \right]\\
&\leq \mathbb{E}\left[\left( 2 C_1  + \delta |Z_i| \right)^3 \right]\\
&= 8C_1^3 + 12 C_1^2 \delta \mathbb{E}[|Z_i|] + 6 C_1\delta^2 \mathbb{E}[Z_i^2] + \delta^3 \mathbb{E}[|Z_i|^3]\\
&= 8C_1^3 + \delta \left[ 12 C_1^2 \sqrt{\frac{2}{\pi}} + 6 C_1\delta + 2 \sqrt{\frac{2}{\pi}} \delta^2  \right].
\end{align*}
Therefore,
\[\frac{C_0 \rho}{\tau^3} \leq  \frac{C_0 }{\delta^3} \left[8C_1^3 + \delta \left( 12 C_1^2 \sqrt{\frac{2}{\pi}} + 6 C_1\delta + 2 \sqrt{\frac{2}{\pi}} \delta^2  \right)\right].\]
Hence,
\[\sup_t|\mathbb{P}(\sqrt{n}(\hat\Gamma - \Gamma)\leq t) - \mathbb{P}(Z \leq t)| \leq \frac{C}{\sqrt{n}},\]
where $C = \frac{C_0 }{\delta^3} \left[8C_1^3 + 
\delta \left( 12 C_1^2 \sqrt{\frac{2}{\pi}} + 6 C_1\delta + 2 \sqrt{\frac{2}{\pi}} \delta^2  \right)\right]$.
Since the upper bound does not depend on ${\cal D}_1$, the result holds unconditionally as well. $\Box$

\subsection{Proof of Theorem \ref{thm::goodpower}}
\label{sec:riftpower}
Let $p\in {\cal P}_2- {\cal P}_1$.
Conditional on ${\cal D}_1$,
$\mathbb{E}[\hat\Gamma | {\cal D}_1] = \Gamma = K(p,\hat p_1)-K(p,\hat p_2)$.
There exists $\gamma>0$ such that
$K(p,p_1)\geq \gamma>0$ for all $p_1\in {\cal P}_1$.
It follows from the law of large numbers,
with probability 1,  that
$\liminf_{n\to\infty}K(p,\hat p_1)> \gamma/2$.
Since $\hat p_2$ is consistent,
$K(p,\hat p_2) = o_\mathbb{P}(1)$.
Thus,
with probability 1,  
$\Gamma > \gamma/2$ for all large $n$.
Also, with probability 1,  
$\hat\tau/\tau = 1+o(1)$.
Combining these facts with the Berry-Esseen result, we have that
\begin{align*}
\mathbb{P}\left(\hat\Gamma > \frac{z_\alpha \hat\tau}{\sqrt{n}} \Biggm| {\cal D}_1\right) &=
\mathbb{P}\left( \frac{\sqrt{n}(\hat\Gamma - \Gamma)}{\tau} > (1+o(1))z_\alpha - 
\frac{\sqrt{n}\Gamma}{\tau} \Biggm| {\cal D}_1\right)\\
&=\mathbb{P}(Z > (1+o(1))z_\alpha - \sqrt{n}\Gamma/\tau | {\cal D}_1) + \frac{C}{\sqrt{n}}\\
& \geq
\mathbb{P}(Z > (1+o(1))z_\alpha - \sqrt{n}\gamma/(2\tau) ) + \frac{C}{\sqrt{n}}\\
\end{align*}
where 
$Z\sim N(0,1)$ and
$C$ is a constant that does not depend on ${\cal D}_1$.
It follows that
$\mathbb{P}(\hat\Gamma > z_\alpha \hat\tau/\sqrt{n})\to 1$. $\Box$

\vspace{1cm}

\subsection{Proof of Theorem \ref{thm::BE2}}
\label{sec:BE2}
In the following proof all probabilities and expectations are taken conditioned on $\mathcal{D}_1$.
By Berry-Esseen theorem, given $\mathcal{D}_1$, 
\[ 
\sup_t|\mathbb{P}(\sqrt{n}(\hat\Theta - \Theta)\leq t) - \mathbb{P}(Z \leq t)| \leq 
\frac{C_0  \mathbb{E}\left[\left\vert \tilde U_i  - \Theta \right\vert^3 \right]}{a^3 \sqrt{n}},
\]
where $C_0$ is a constant and $a^2 =  \mathbb{E}\left[\left( \tilde U_i  - \Theta \right)^2 \right]$.
Let
\begin{equation}\label{eq::a}
a^2 =  {\rm Var}\left( \tilde U_i\right).
\end{equation}
Now $\Theta = \mathbb{E}\left[U_i \right] = \mathbb{E}\left[\tilde U_i \right]$, therefore,
\[
a^2 =  {\rm Var}\left( \tilde U_i\right) = \text{Var}\left( U_i + \delta{Z_i} \right) \geq \delta^2 \text{Var}(Z_i) = \delta^2.
\]
Now note that,
\begin{align*}
|U_i| &= \left\vert \hat p_1 (X_i) - \hat p_2 (X_i) \right\vert \\
&\leq \left\vert \hat p_1 (X_i) \right\vert + \left\vert \hat p_2 (X_i) \right\vert \\
&\leq \frac{1}{(2 \pi)^{d/2} |\hat \Sigma |^{1/2}} + \max_{ i \in \{1,2 \}} \frac{1}{(2 \pi)^{d/2} |\hat \Sigma_i |^{1/2}} \leq \frac{2}{(2 \pi c_1)^{d/2}} = C_2.
\end{align*}

Therefore we also have that,
\[|\Theta| = \left\vert \mathbb{E}\left[U_i \right] \right\vert \leq C_2.\]

Then following the same arguments as before while finding a bound for $\rho$, we can see that
\begin{align*}
\mathbb{E}\left[\left\vert \tilde U_i  - \Theta \right\vert^3 \right] 
&\leq 8C_2^3 + \delta \left[ 12 C_2^2 \sqrt{\frac{2}{\pi}} + 6 C_2\delta + 2 \sqrt{\frac{2}{\pi}} \delta^2  \right].
\end{align*}
Therefore,
\[\frac{C_0  \mathbb{E}\left[\left\vert \tilde U_i  - \Theta \right\vert^3 \right]}{a^3 } \leq  \frac{C_0 }{\delta^3} \left[8C_2^3 + \delta \left( 12 C_2^2 \sqrt{\frac{2}{\pi}} + 6 C_2\delta + 2 \sqrt{\frac{2}{\pi}} \delta^2  \right)\right].\]
Hence,
\[
\sup_t|\mathbb{P}(\sqrt{n}(\hat\Gamma - \Gamma)\leq t) - \mathbb{P}(Z \leq t)| \leq \frac{\tilde C}{\sqrt{n}},
\]
where 
$\tilde C = \frac{C_0 }{\delta^3} 
\left[8C_2^3 + \delta \left(  12 C_2^2 \sqrt{\frac{2}{\pi}} + 6 C_2\delta + 2 \sqrt{\frac{2}{\pi}}
\delta^2 \right)\right]$. $\Box$

\subsection{Proof of Theorem \ref{thm::valid}}
\label{sec:validMixture}
Suppose $H_0$ is true.
Crucially, the error from the Berry-Esseen theorem does not depend on ${\cal D}_1$.
The unconditional type I error of the split test is thus
\begin{align*}
\mathbb{P}\left(\hat\Gamma > \frac{z_\alpha \hat\tau}{\sqrt{n}}\right) &=
\mathbb{P}\left( \frac{\sqrt{n}(\hat\Gamma - \Gamma)}{\hat\tau} > z_\alpha - \frac{\sqrt{n}\Gamma}{\hat\tau}\right)\\
&=
\mathbb{E}\left[
\mathbb{P}\left( \frac{\sqrt{n}(\hat\Gamma - \Gamma)}{\hat\tau} > z_\alpha - \frac{\sqrt{n}\Gamma}{\hat\tau}\,
\Biggm| \, {\cal D}_1 \right)\right]\\
& =
\mathbb{E}\left[
\mathbb{P}\left( Z > z_\alpha - \frac{\sqrt{n}\Gamma}{\tau}\, \Biggm| \, {\cal D}_1 \right)\right] + O(n^{-1/2})\\
&=
\mathbb{E}\left[ \overline{\Phi}\left(z_\alpha - \frac{\sqrt{n}\Gamma}{\tau} \right)\right]  + O(n^{-1/2})
\end{align*}
where $Z\sim N(0,1)$,
$\overline\Phi = 1-\Phi$
and $\Phi$ is the normal cdf.

Recall that when we fit $H_1$, we constrain the solution to satisfy
$K(p,\hat p_2) > \Delta$.
Under $H_0$,
$\mathbb{P}(A_n)\to 1$ where $A_n$ is the event:
$$
A_n = \Biggl\{  K(p,\hat p_1) < \Delta + \frac{\tau \delta_n}{\sqrt{n}} \Biggr\}
$$
and $\delta_n$ is any sequence such that $\delta_n = o(1)$.
(In fact, $\Delta$ can also be taken to be a slowly decreasing sequence.)
On the event $A_n$ we have
that
\begin{align*}
z_\alpha - \frac{\sqrt{n}\Gamma}{\tau} &=
z_\alpha - \frac{\sqrt{n} (K(p,\hat p_1) - K(p,\hat p_2))   }{\tau}\\
& \geq
z_\alpha - \frac{\sqrt{n} (K(p,\hat p_1) - \Delta_n)}{\tau} \geq z_\alpha - \delta_n.
\end{align*}
So
\begin{align*}
\mathbb{E}\left[ \overline{\Phi}\left(z_\alpha - \frac{\sqrt{n}\Gamma}{\tau} \right)\right] &=
\mathbb{E}\left[ \overline{\Phi}\left(z_\alpha - \frac{\sqrt{n}\Gamma}{\tau} \right) \mathbb{I}_{A_n}\right] +
\mathbb{E}\left[ \overline{\Phi}\left(z_\alpha - \frac{\sqrt{n}\Gamma}{\tau} \right) \mathbb{I}_{A_n^c}\right]\\
& \leq
\mathbb{E}\left[ \overline{\Phi}\left(z_\alpha -\delta_n \right) \mathbb{I}_{A_n}\right] +
\mathbb{P}(A_n^c)\\
&= \alpha + o(1).
\end{align*}
$\Box$

\clearpage
\end{appendices}

\end{document}